\documentclass[12pt]{article}

\usepackage{geometry}
\usepackage{fullpage}                    %
\usepackage[parfill]{parskip}
\usepackage{setspace}
\usepackage{lscape}
\usepackage{changepage}
\usepackage[hang,flushmargin,bottom]{footmisc}
\usepackage[ascii]{inputenc}
\usepackage[T1]{fontenc}
\usepackage{lmodern}
\usepackage[english]{babel}

\usepackage{amsmath}
\usepackage{amssymb}
\usepackage{amsthm}
\usepackage{mathtools}
\usepackage{bbm}
\usepackage{dsfont}
\usepackage{accents}
\usepackage{upgreek}
\usepackage{cancel}
\usepackage{esint}

\usepackage{array}
\usepackage{booktabs}
\usepackage{longtable}
\usepackage{multirow}
\usepackage{tabularx}                    %
\usepackage{makecell}

\usepackage{bigdelim}
\usepackage{colortbl}
\newcolumntype{H}{>{\setbox0=\hbox\bgroup}c<{\egroup}@{}}

\usepackage{graphicx}
\usepackage{float}
\usepackage{caption}
\usepackage[flushleft]{threeparttable}
\makeatletter
\@ifundefined{showcaptionsetup}{}{%
  \PassOptionsToPackage{caption=false}{subfig}}
\makeatother
\usepackage{subfig}

\usepackage{tikz}
\usetikzlibrary{
  calc,
  matrix,
  positioning,
  shapes,
  shapes.geometric,                      %
  arrows,
  arrows.meta                            %
}
\usepackage{tkz-graph}

\tikzset{
  pipebox/.style={
    rectangle, draw, thick, rounded corners=1pt,
    minimum width=3.6cm, minimum height=1.05cm,
    align=center, font=\small
  },
  databox/.style={
    ellipse, draw, thick,
    minimum width=3cm, minimum height=0.9cm,
    align=center, font=\footnotesize
  },
  bootbox/.style={
    rectangle, draw, thick, rounded corners=1pt,
    minimum width=2.8cm, minimum height=1cm,
    align=center, font=\small,
    fill=gray!8
  },
  outlabel/.style={font=\large},
  arr/.style={-Stealth, thick},
  bootarr/.style={-Stealth, thick, dashed},
}

\usepackage{enumitem}
\usepackage{multicol}
\usepackage{xcolor}
\usepackage{color}
\usepackage{blindtext}
\usepackage{listings}
\usepackage{comment}
\usepackage{datetime}

\usepackage{titlesec}
\usepackage{sectsty}
  \sectionfont{\fontsize{16}{20}\selectfont}
  \subsectionfont{\fontsize{14}{15}\selectfont}
\makeatletter
\@ifundefined{counterwithout}{\usepackage{chngcntr}}{}
\makeatother
\counterwithin{figure}{section}
\counterwithin{table}{section}
\renewcommand\thesection{\arabic{section}}
\renewcommand\thesubsection{\arabic{section}.\arabic{subsection}}
\usepackage[backref=page]{hyperref}
\usepackage{cleveref}
\usepackage[authoryear]{natbib}
\usepackage{bibentry}
\hypersetup{
    colorlinks=true,
    citecolor=blue,
    linkcolor=blue,
    filecolor=magenta,
    urlcolor=blue,
}

\theoremstyle{remark}

\usepackage{epigraph}
\usepackage{etoolbox}
\usepackage{catchfilebetweentags}
\makeatletter
\patchcmd{\CatchFBT@Fin@l}{\endlinechar\m@ne}{}
  {}{\typeout{Unsuccessful patch!}}
\makeatother

\AtBeginDocument{}
\AtBeginDocument{}
\AtBeginDocument{}
\AtBeginDocument{}

\providecommand\abstractname{Abstract}
\def\abstract{}

\renewenvironment{abstract}{%
  \centering\small
  \textbf\abstractname
  \list{}{\leftmargin1cm \rightmargin\leftmargin}
  \item\relax
}{%
  \endlist \par\bigskip
}

\providecommand{\tofill}[1]{}

\makeatletter
\newcommand{\distas}[1]{\mathbin{\overset{#1}{\kern\z@\sim}}}
\newsavebox{\mybox}\newsavebox{\mysim}
\newcommand{\distras}[1]{%
  \savebox{\mybox}{\hbox{\kern3pt$\scriptstyle#1$\kern3pt}}%
  \savebox{\mysim}{\hbox{$\sim$}}%
  \mathbin{\overset{#1}{\kern\z@\resizebox{\wd\mybox}{\ht\mysim}{$\sim$}}}%
}
\makeatother

\AtBeginDocument{%
  \renewcommand{\doi}[1]{}%
  \renewcommand{\url}[1]{}%
}



\def\mydate{\leavevmode\hbox{\monthname\ \the\year}}
\def\twodigits#1{\ifnum#1<10 0\fi\the#1}

\author{ 
  Remy Levin\\
  \textit{University of Connecticut}
  \and
  Daniela Vidart\\
  \textit{University of Connecticut  and NBER}
  }
\title{The Yeoman's Portfolio: Measuring Historical Risk Preferences Using Crop Choice\thanks{Email:  remy.levin@uconn.edu and daniela.vidart@uconn.edu. We would like to thank Leah Boustan, Katherine Hauck, David Lagakos, Omer Moav, Allen Peters, Steve Ross, and Wayne Sandholtz for their insightful feedback on this paper. We would also like to thank Desmond Ang, Price Fishback, Paul Rhode, and Ken Sylvester for kindly sharing data.}}
\date{August 2026}

\begin{document}

\maketitle

\begin{abstract}

We design a method for measuring the risk preferences of agents in the deep past. The method combines a structural model of crop choice as a portfolio allocation with machine-learning prediction of expected crop returns, using historic agronomic and climate data. We estimate county-level risk preferences for the United States and farmer-level preferences in Kansas from 1889 to 1929. More risk averse farmers leveraged less, were less likely to purchase novel WWI Liberty Bonds, and were more likely to participate in local risk-sharing institutions. We show that higher risk aversion predicts slower tractor adoption and farm mechanization during the 1920s.

\bigskip
\bigskip



\end{abstract}

\newpage

\section{Introduction}

Risk preferences are a central driver of economic decision-making. Understanding how they vary across populations and over time is essential for explaining patterns of economic behavior and their evolution. While contemporary data have enabled economists to measure individual risk preferences with increasing precision, evidence on historical preferences is practically nonexistent. This scarcity reflects a fundamental measurement problem: widely used methods for eliciting risk preferences are modern innovations, with no analog in the historical record. As a result, we know very little about the distribution of risk preferences in the past, how they changed over time, and their role in shaping long-run economic development.

In this paper, we design a new method for measuring the risk preferences of historical populations using agricultural production decisions. Our method centers on farmers' decisions of which crops to plant, a choice that is akin to allocating a portfolio of land to safe and risky assets. We formalize this insight in a parsimonious structural model in which an expected-utility-maximizing farmer allocates land across crops under exogenous price and yield risk. The model produces a tractable estimating equation in which the coefficient of relative risk aversion can be recovered from data on cropland shares, expected crop revenues, the variance-covariance matrix of those revenues, and the returns to scale of production, without imposing a functional form on utility. 

We operationalize the framework in the United States between 1889 and 1929. We combine county-level agricultural census data on acreage and yields for seven major crops, a harmonized panel of state-level crop prices, and high-resolution historical climate data. The centerpiece of the empirical methodology is the construction of expected revenues. We use machine learning to predict local crop yields from climate histories, construct annual county-level revenue panels between census waves, and use their moments as empirical proxies for farmer expectations about risks and returns.

The result is the first dataset of risk preferences for a historical population in the literature. We estimate county-level preferences for farmers in 94.5\% of counties in the contiguous United States, in six waves from 1889 to 1929. This represents approximately 40\% (1889) to 25\% (1929) of the U.S. population in this period. We also estimate preferences for a sample of approximately 2,500 farmers in Kansas in six waves from 1895 to 1930, using a unique farm-level panel. This allows us to validate our methodology at the individual level.

\begin{figure}[tbp]\centering
\caption{Estimated median risk aversion, 1889 -- 1929}\label{fig:rc5medmap}
\vspace{0.15cm}

\includegraphics[width=.82\textwidth]{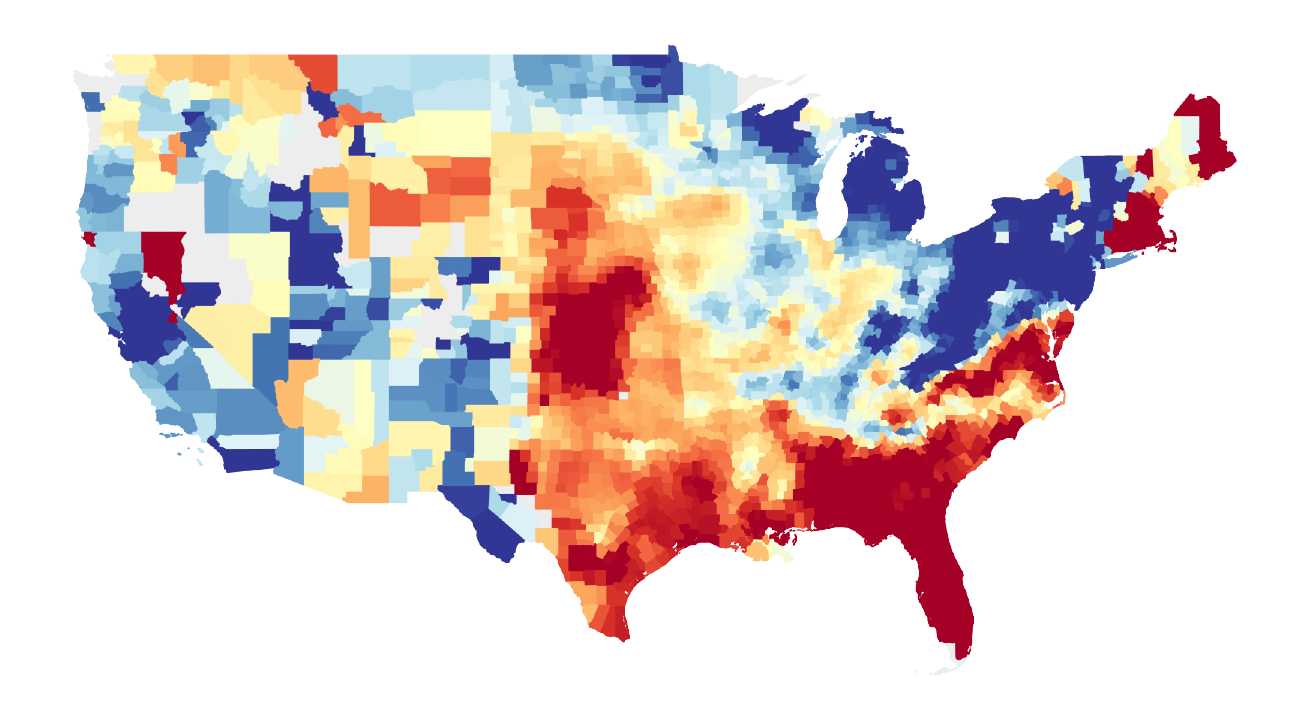}

\vspace{0.15cm}
\includegraphics[width=0.72\textwidth]{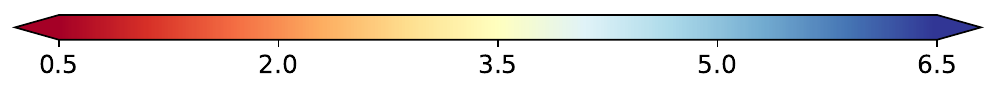}
\vspace{0.1cm}

{\footnotesize\raggedright \textbf{Notes:} This figure shows the estimated risk aversion coefficient $\widehat{\rho}$ across counties, taking the median across waves in which it is observed. Blue indicates more risk averse, red more risk-seeking, and grey no estimate.\par}
\end{figure}

We find that the median county exhibits considerable risk aversion ($\widehat{\rho} = 3.23$), and that 60.7\% of counties lie in the range of values commonly used in macroeconomic models ($\widehat{\rho} \in [1,5]$). Risk aversion displays significant local heterogeneity alongside distinct regional patterns: higher risk aversion in the Mid-Atlantic, Upper Midwest and West Coast, and more risk-seeking in the South and the Great Plains (\Cref{fig:rc5medmap}). The estimates also move over time in a way that tracks the macroeconomic history of the period. Median risk aversion rises between 1889 and 1909, amid the volatility and deflation of the turn of the century, and falls during the 1910s and 1920s, alongside the wartime commodity boom and subsequent agricultural speculation.

We design four empirical exercises to validate the method's basic assumptions. First, we show that crops behave like financial assets: higher expected revenues come with greater variance, creating a portfolio-style risk--return tradeoff. Second, using the Kansas farmer panel, we show that farmers respond to this tradeoff like investors, shifting acreage towards crops whose expected revenue rises over time, and away from crops that become riskier. Third, observed crop portfolios lie close to the efficient frontier. The median county achieves 95\% of its maximum risk-adjusted return, with 81\% of county-waves attaining at least 80\%, matching or exceeding efficiency levels documented for modern households and retail investors. Fourth, we show that the township-level preference estimates in Kansas closely track the median farm-level estimates within each township, indicating that the aggregate measure represents the underlying distribution of individual preferences.

We next validate the risk aversion estimates against contemporaneous risk-taking in other domains. We find that more risk averse counties had a lower proportion of mortgaged farms, indicating that farmers in these counties took on less financial leverage. We also find that more risk averse counties in 1919 subscribed at lower rates to the Fourth Liberty Bond Drive, a federal program that offered a novel and risky financial instrument aimed at funding the war effort during World War I (\cite{hilt2020}). At the individual level, using our Kansas data, we show that more risk averse farmers were more likely to join fraternal organizations that insured members against death, disability, and family hardship.

Finally, we show that risk preferences help explain the adoption of tractors, the defining agricultural technology of the period (\cite{olmstead2000}). Tractors required large, indivisible investments with uncertain returns, and more risk averse counties adopted them more slowly. A one-standard-deviation increase in $\widehat{\rho}$ predicts 0.15 to 0.20 standard deviations fewer tractors per farm up to a decade later. We find similar patterns for other risky technologies, including commercial fertilizer and motor trucks, but not for draft animals, the technology being replaced by tractors, or for household amenities such as electric lighting and running water, whose availability was determined largely off the farm. Thus, estimated risk aversion specifically predicts the adoption of risky productive investments.

Methodologically, this paper builds on and advances the literature on the estimation of risk preference distributions from observational data. This growing set of studies employs structural methods to estimate preferences for populations of agents, using models of insurance choice (\cite{CicchettiDubin1994}, \cite{cohenEstimatingRiskPreferences2007}, \cite{Handel2013}, \cite{BarseghyanMolinari2013}, \cite{Barseghyan2016}, \cite{BarseghyanMolinari2021}), financial portfolio allocation (\cite{Chiappori2011}, \cite{Bucciol2011}, \cite{paravisiniRiskAversionWealth2017}, \cite{CalvetCampbellGomesSodini2026}, \cite{EganMacKayYang2025}), behavior in betting markets (\cite{ChiapporiSalanie2019}, \cite{AndrikogiannopoulouPapakonstantinou2020}) and game shows (\cite{FullenkampTenorioBattalio2003}, \cite{BombardiniTrebbi2012}), consumption under risk-sharing (\cite{Chiappori2014}), and agricultural production (\cite{Antle1987}). Most closely related to our paper are studies that use models of crop choice as a portfolio allocation to back out the risk preferences of farmers (\cite{BarShira1997}, \cite{Kurosaki2002}, \cite{KoundouriLaukkanen2009}, \cite{AllenAtkin2022}).  

The literature in this area has so far progressed by inverting ever-more sophisticated models of risky decisions using richer and richer choice data. We turn this paradigm on its head by designing a method that is explicitly aimed at recovering preferences in data-poor historical environments. To do this, we make several methodological innovations. Our procedure for constructing expected returns solves the problem of how to recover agents' risk-return environment when we only observe some of its elements (here yields) intermittently and after choices are made. We recover local heterogeneity in risk preferences without specifying a global utility function or a low-dimensional relationship between risk aversion and characteristics that can be hard to observe historically.\footnote{\cite{BarShira1997} impose a functional form for ARA ($A(w) = \alpha w^\beta$). The level parameter varies with farmer characteristics, but the wealth elasticity $\beta$ is common across farmers. \cite{Kurosaki2002} parameterize risk aversion as a function of land, livestock, and education, and their final RRA also varies with proximity to subsistence consumption. \cite{KoundouriLaukkanen2009} specify ARA as a common quadratic function of expected profit. Farm-year heterogeneity comes from evaluating that common function at different profit levels. \cite{AllenAtkin2022} initially estimate common effective risk aversion and then allow it to vary parametrically with district bank access. All four studies use complete panel datasets in which the outcomes needed to construct the relevant return distributions are observed in each period.} Our validation pipeline is also unusually extensive, including direct tests of the model's assumptions about agents' behavior, comparisons with risk-taking in other domains, and evidence that the aggregate measure (all we are likely to see in historical data) is representative of individual preferences. This is particularly important when trying to understand the behavior of people far removed from us by time and culture (\cite{HenrichHeineNorenzayan2010}). The result is a portable methodology that here delivers the first estimates of risk preferences for any population observed before 1970. 

This methodology creates a host of opportunities to further our understanding of economic history by observing its behavioral underpinnings. In our setting, the most prominent example of this is the long-standing debate over the causes of the slow and uneven diffusion of the tractor in the United States during the early twentieth century. Prior work has attributed this pattern to Southern labor and tenancy institutions (\cite{Whatley1985}), financing barriers (\cite{Clarke1991}), high fixed costs relative to farm scale (\cite{OlmsteadRhode2001}), limited early labor savings (\cite{MartiniSilberberg2006}), low initial tractor quality and high costs relative to human and animal power (\cite{ManuelliSeshadri2014}), and the limited scope of tasks that early tractors could perform (\cite{Gross2018}). A common theme throughout these studies is that tractor adoption in the 1920s was a very risky proposition for farmers, due to the high costs and uncertain returns of the technology.

Our analysis indicates that heterogeneity in farmer risk aversion helps explain the diffusion pattern of tractors in this period. This finding lends empirical content to earlier sociological accounts of ``farmer conservatism'' (\cite{AnkliOlmstead1981}), and quantifies a behavioral determinant of capital deepening that can be studied alongside the institutional, financial, and technological forces emphasized in prior work. It also places the economic history of American mechanization in direct conversation with a central finding in contemporary development economics: risk averse farmers may forgo profitable but uncertain innovations in favor of safer, lower-yielding alternatives (\cite{Rosenzweig1993}, \cite{Dercon1996}, \cite{BanerjeeDuflo2005}). More generally, it suggests that preference data of this kind has the potential to turn behavioral conjectures elsewhere in economic history into testable hypotheses.

Perhaps the broadest contribution of this paper is to the literature on long-run growth. Classical theorists (\cite{Knight1921}, \cite{Schumpeter1934}) and economic historians (\cite{Landes1949}, \cite{McCloskey1976}, \cite{Rosenberg1996}) have long recognized the central role that risk-taking has played in the making of the wealth of nations. The modern growth literature has developed these ideas along two parallel lines. One takes preferences as given, and examines how institutions for managing risk helped unlock economic growth (\cite{North1991}, \cite{AcemogluZilibotti1997}, \cite{Levine2005}). The other places the endogeneity of preferences at the center of the analysis, and studies how risk attitudes coevolved with development through natural selection (\cite{Galor2012}) and cultural transmission (\cite{Doepke2014}, \cite{Klasing2014}, \cite{ChakrabortyThompsonYehoue2016}, \cite{Doepke2017}).

Our method is a blueprint for creating data that can advance both traditions. For the institutions literature, it makes preferences observable alongside other initial conditions, allowing behavioral heterogeneity to be integrated into the study of institutional development much as cultural differences have been (\cite{Greif1994}, \cite{StulzWilliamson2003}). For the coevolution literature, our method provides a way of observing preference dynamics \textit{as they unfolded in history}. Empirical work in this area has so far been limited to testing historical theories against present-day cross-sectional differences in preferences (\cite{Dohmen2012}, \cite{Bouchouicha2019}, \cite{Becker2020}) or short-run changes in the recent past (\cite{DoepkeKlasing2026}). But just as the fossil record is indispensable for studying biological evolution, historical measures of preferences are crucial for developing theories of preference evolution. In our setting, they provide direct evidence for a central coevolutionary proposition: risk preferences moved alongside economic growth, and these changes in turn shaped structural transformation by altering the pace of technological adoption.

The remainder of the paper is organized as follows. \Cref{sec: conceptual} explains the conceptual framework underlying our methodology and previews its mechanics. \Cref{sec: setting} provides context on the historical setting of the analysis. In \Cref{sec: model} we present the structural model, and in \Cref{sec: data} our data. \Cref{sec: methodology} describes the empirical methodology in detail. \Cref{sec: validation} presents the four analyses in our validation pipeline. In \Cref{sec: descriptive results} we present the main results, including descriptive results for the estimated preference distribution and the relationship of the measure with risk-taking in other domains. \Cref{sec: structural transformation} presents our results on the effects of estimated risk aversion on agricultural technological adoption. \Cref{sec: conclusion} concludes.

\section{Conceptual framework and method overview}
\label{sec: conceptual}

If making sense of the past is like mapping a foreign country, then recovering past preferences is like imaging the surface of another planet. To succeed, one would need something akin to a \textit{behavioral telescope}: a tool for inferring motives from choices observed at a great distance. Constructing this telescope and tuning it to risk preferences is the object of this paper. Its design rests on a series of conceptual choices about what it should observe, how its signals should be interpreted, and where we should aim it first.

The natural starting point for the design is the suite of tools used to measure risk preferences today. One option is experimental measures using objective income lotteries, but these cannot be deployed retroactively. The other alternative is methods that take observational data on risky decisions --- like insurance choices, portfolio allocations, and agricultural production  --- fit them to a structural model of the choice environment, and back out the agent's risk preference parameter. This is the basic architecture we employ.

The next decision is which choice domain to use. The ideal candidate would involve high-stakes risky decisions, undertaken by a large share of the population, that can be observed in the historical record. Agriculture is the clear winner. Farmers face significant climate, pest, disease, and price risk, and their production decisions determine their livelihoods and often their subsistence. From the Neolithic Revolution until recently the vast majority of humans worked in agriculture, allowing us in principle to measure the preferences of much of the population. And since agriculture had long been the dominant sector of the global economy, records on its inputs and outputs often exist in written and archaeological sources.

Within agriculture, we focus on how farmers allocate land across crops at planting. Development and agricultural economists have long treated this as a portfolio problem (\cite{Anderson1977}, \cite{Fafchamps1992a}, \cite{Rosenzweig1993}): arable land is capital, crops are assets with stochastic returns, and allocations reflect expected returns and risk aversion. We model risk as uninsurable price and yield fluctuations, the two principal sources of agricultural uncertainty (\cite{Moschini1999}), and include a reduced-form parameter linking per-acre yields to production scale to capture scale-dependent technology and the differential use of marginal land.

The resulting model is rich enough to capture the relevant risk-taking margin, yet parsimonious enough for historical data. Estimation requires four inputs: cropland allocations and observed yields at each measurement date, price histories, and high-frequency climate histories. Repeated yield observations allow us to account for local heterogeneity, crop suitability, and historical farming technology during the period we study, rather than relying on present-day proxies like FAO-GAEZ, which may contain substantial non-classical measurement error (\cite{Rhode2024}, \cite{araujoPotatoPotahtoFAOGAEZ2026a}).

Our method works as follows. The model contains four objects: the crop allocation $C$, expected crop returns $R$, scale parameter $\alpha$, and the relative risk aversion coefficient $\rho$. We observe $C$ directly for each location and measurement period. We estimate $\alpha$ from the relationship between per-acre yields and production scale across the panel, controlling for location and time fixed effects. The main empirical task is constructing $R$, the mean vector and variance-covariance matrix of expected per-acre crop revenues in each location and period. We assume that agents form expectations from local revenue histories, and use the moments of realized returns in the decade before each crop allocation as measures of $R$. However, because yields are observed only at measurement dates, we need a way to recover them between measurements. To do this, we use machine learning to predict measured per-acre yields as a function of climate, locality fixed effects, and time fixed effects. We then combine the fitted climate response and estimated fixed effects with higher-frequency climate histories to impute yields between measurements. Multiplying predicted yields by prices produces a revenue panel from which we construct $R$. We then recover $\rho$ using the observed $C$ and the estimated $\alpha$ and $R$.

With the telescope in hand, the final decision is when and where to aim it first. We choose the United States at the turn of the twentieth century, for four reasons. First, data availability: the U.S. Census of Agriculture recorded county-level cropland allocations and crop yields each decade from 1879 to 1929, as well as in 1924. State-level prices and high-frequency climate data are also available. Second, model fit: farms during this period grew a diverse menu of crops, giving us meaningful variation in portfolio allocations across space and time. Agriculture employed a large share of the population, was predominantly small-scale and commercial rather than subsistence, and faced substantial risks that remained largely unbuffered by government intervention until the New Deal. The price and yield fluctuations we observe, therefore, closely approximate the risks shaping farmers' choices. Third, verifiability: the setting is sufficiently well studied that we can validate the method and estimates against other data on risk-related behavior and farm-level decisions. Finally, the period is important in its own right. It marks the opening decades of America's ``special century'' of economic growth (\cite{gordonRiseFallAmerican2016}) and a period of rapid structural transformation, especially in agriculture, that our results can help illuminate.

\section{Historical setting: U.S. agriculture, 1879--1929}
\label{sec: setting}

From the 1880s through the 1920s American agriculture was labor-intensive, local, and central to rural life. Farms were generally small and diversified \citep{cain2018}. Although the U.S. underwent significant urbanization over this half century, much of the population still lived in rural areas (64.9\% in 1890, 43.8\% in 1930) and on farms (39.3\% in 1890, 24.8\% in 1930), and a large share of workers were in agricultural occupations (42.6\% in 1890, 21.4\% in 1930) (\cite{Katz2014}, \cite{Kim2004}, \cite{uscensusbureau1975}). 

On the macroeconomic front, the 1880's were marked by a slow recovery from the panic of 1873, a severe financial crisis that resulted in a global depression lasting until 1879. In the United States, the crisis was triggered by railroad overexpansion and financial speculation, and amplified by the federal demonetization of silver and the adoption of a \textit{de facto} gold standard. The latter led to a large contraction in the money supply, deflation, and a strong dollar, which hurt the profits of farmers who were largely debtors and exporters. 

Macroeconomic upheaval continued throughout the 1890s, and triggered the Panic of 1893 and the Panic of 1896. Together these led to a severe economic depression that significantly raised unemployment and poverty nationally. Political upheaval followed the economic upheaval, culminating in the elections of 1896 and 1900, which the Democratic presidential nominee William Jennings Bryan strongly contested by running on a populist platform of abandoning the gold standard to help farmers (\cite{Frieden1997}).

The turn of the twentieth century, and especially the 1910s, saw a regime change in the macroeconomy and the agricultural sector. Inflation returned and farm profits soared, especially during World War I, when national and global demand for agricultural products outstripped supply. This resulted in substantial increases in commodity prices, which in turn led to an expansion in speculative activity in the sector. Prices collapsed when European agriculture rebounded faster than expected after the war, exposing farmers with highly leveraged land purchases and leaving many regions in a prolonged farm crisis, even as incomes partially recovered from the 1921 trough (\cite{Rajan2015}). 

Accompanying the economic volatility were significant climatic and ecological shocks. The most important ones on this long list were the major drought that hit the Great Plains and the western states in the 1890s; the massive floods that inundated large swathes of the Midwest and the East in 1913 and the southern Mississippi river basin in 1927; and the cotton boll weevil, a pest that entered Texas in 1892, spread to the entire cotton belt by 1922, and devastated yields throughout much of the American south (\cite{Lange2009}). 

Another defining feature of farmers' experience during this period was rapid technological change in agriculture. Biological innovation expanded the range and productivity of crops such as wheat and corn, while improvements in commercial fertilizers, soil amendments, and seed varieties altered the returns to cultivation in many regions. At the same time, a wave of new powered machinery, including trucks, tractors, and automobiles began to transform farm power \citep{olmstead2000,OlmsteadRhode2008,Olmstead2018}.

\begin{figure}[!ht]\centering
\caption{The diffusion of the tractor and the shift from animal to mechanical power}\label{fig:tractordiffusion}
\vspace{0.15cm}

\resizebox{\linewidth}{!}{%
\begin{tabular}{@{}c@{\hspace{1.5em}}c@{}}
\subfloat[County share of farms with tractors, 1929]{\includegraphics[width=0.48\textwidth]{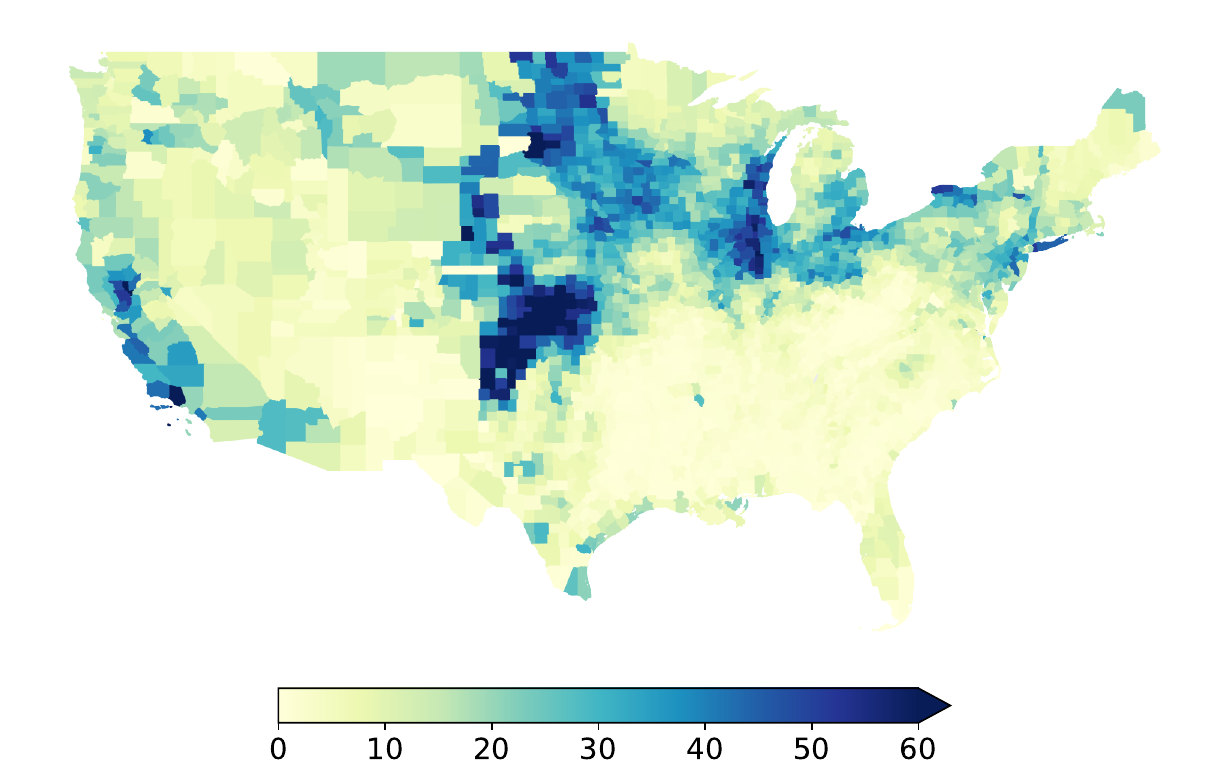}} &
\subfloat[Tractors vs.\ draft animals, 1920--1940]{\includegraphics[width=0.48\textwidth]{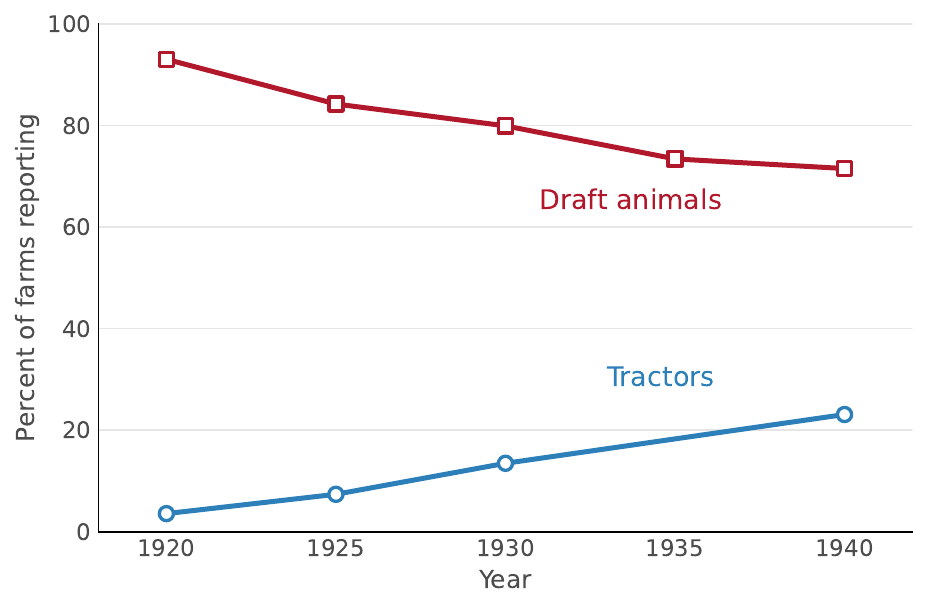}} \\[0.5cm]
\subfloat[Tractor with planter, Ralls, Texas, 1939]
{\includegraphics[height=5.5cm]{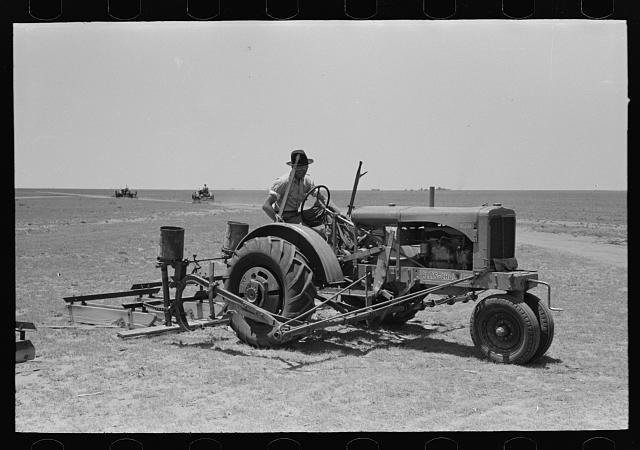}} &
\subfloat[Horse and tractor, Jasper County, Iowa, 1940]
{\includegraphics[height=5.5cm]{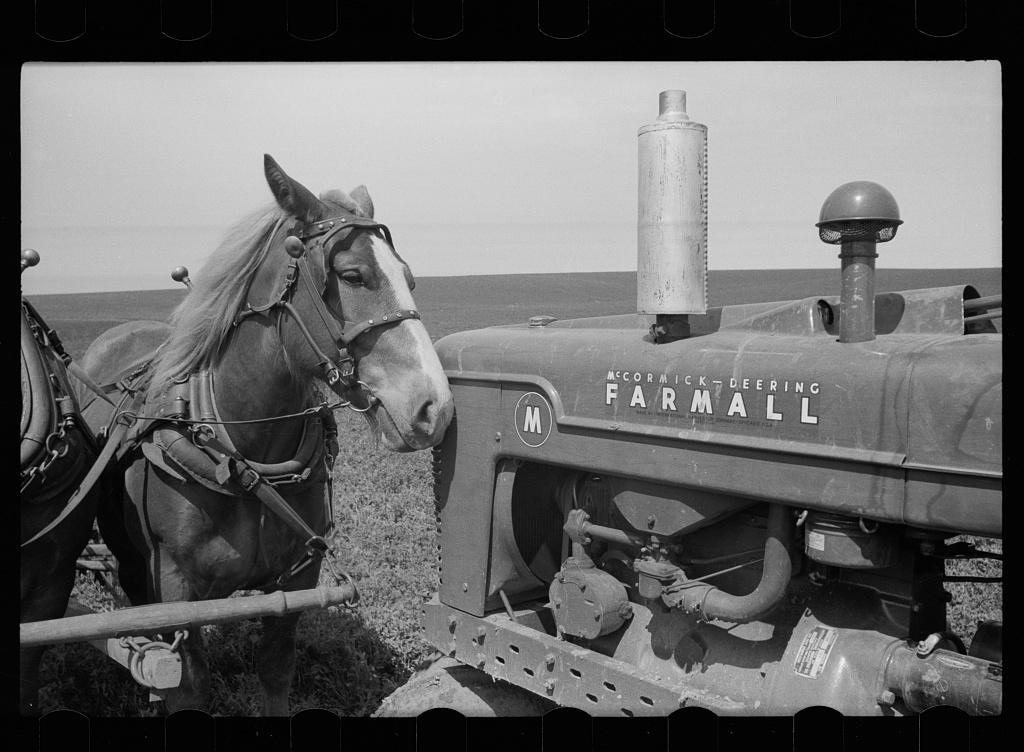}}
\end{tabular}}
\vspace{0.1cm}

{\footnotesize\raggedright \textbf{Notes:} This figure illustrates the diffusion of the tractor in American agriculture. Panel (a) maps the 1930 county-level share of farms operating a tractor (farms reporting a tractor as a share of all farms). Panel (b) plots the share of U.S.\ farms reporting tractors versus horses or mules, 1920--1940 (the 1920 horses-or-mules value is the authors' extrapolation). Panel (c) shows a tractor with planter and go-devil attached on a large farm near Ralls, Texas, photographed by Russell Lee in May 1939. Panel (d) shows a horse and tractor in Jasper County, Iowa, photographed by John Vachon in May 1940. Sources: 1930 U.S.\ Census of agriculture (panel a); \cite{olmstead2000}, Table~1 (panel b); Library of Congress Farm Security Administration/Office of War Information Black-and-White Negatives collection (panels c and d).\par}
\end{figure}

The most visible of these changes was the diffusion of the tractor. As can be seen in \Cref{fig:tractordiffusion}, tractors were nearly absent from U.S. farms in 1910 but spread rapidly after World War~I, while the horses and mules that had long supplied farm power began a steady decline. This shift from animal to mechanical power was highly uneven across the country: by 1930 adoption was well advanced in parts of the Corn Belt and the Great Plains but had barely begun across much of the South. Photos from the period illustrate the coexistence of these two power systems, with animal and mechanical power appearing side by side in some places while tractors were already integrated into field operations in others.

Behind the diffusion pattern of the tractor was a costly and uncertain adoption decision faced by individual farmers. The tractor was expensive, indivisible, and unfamiliar, and its purchase represented a substantial and often irreversible capital commitment. A large literature has emphasized farm scale, credit access, market conditions, and technical improvements as determinants of uneven tractor diffusion \citep{olmstead2000,OlmsteadRhode2008}. Our results below show that farmer risk preferences also predict the pattern and speed of tractor adoption and technological diffusion throughout the 1920s.

\section{Model}
\label{sec: model}

We formalize the farmer's crop-choice decision as a standard portfolio allocation problem. The farmer, an expected utility maximizer, allocates arable land across crops, and faces exogenous risk over the price and yield (and therefore revenue) generated by each crop. We also allow for heterogeneous returns to scale in yields by crop. This enables us to control in the empirical analysis for non-linear deterministic factors, like crop-specific production technologies, that can affect crop returns alongside stochastic prices and yields.

\subsection{The yeoman's portfolio problem}

The farmer decides what to plant, given uncertainty over both prices and yields, which realize for each crop after the planting decision is made. We assume that the decision is static, and that land is separable from other factors of production. Let $c$ denote revenue from the modeled crop portfolio, interpreted as crop income or consumption in the period. The farmer's problem is then
\begin{equation}
\max_{c,\{l_i\}^N_{i=1}}\mathbb{E}\left[U(c)\right]
\quad \text{s.t.} \quad
c=\sum^N_{i=1}p_iq_i\left(l_i\bar{L}\right)^{\alpha_i}
\quad \text{and} \quad
\sum^N_{i=1}l_i=1,
\label{eq: farmerproblem}
\end{equation}

where $U$ is a twice continuously differentiable utility function, $N$ is the total number of crops available to the farmer, and $\bar{L}$ is the total amount of land. $l_i$ denotes the share of land dedicated to crop $i$, while $p_i$ is the crop's price and $q_i$ is a stochastic per-acre crop-specific productivity shifter. The parameter $\alpha_i$ governs crop-specific returns to scale in land, allowing per-acre yields to vary nonlinearly with planted acreage and absorbing crop-specific differences in the mapping from land allocations to output. Total output from planting $l_i\bar{L}$ acres of crop $i$ is therefore $q_i(l_i\bar{L})^{\alpha_i}$, and the corresponding observed per-acre yield is $\tilde{q}_i = q_i(l_i\bar{L})^{\alpha_i-1}$. The stochastic component of revenue, $p_iq_i$, captures exogenous price and yield risk and is separable from the deterministic scale component, $(l_i\bar{L})^{\alpha_i}$.

\subsection{The estimating equation}

To solve the farmer's problem, we take first-order conditions and difference them against a numeraire crop (crop one). Applying a standard linearization yields the well-known mean-variance solution to the portfolio problem:
\begin{equation}
\begin{aligned}
\alpha_i\mathbb{E}(p_i\tilde{q}_i) - \alpha_1\mathbb{E}(p_1\tilde{q}_1) = \frac{-U''(\mathbb{E}(c))}{U'(\mathbb{E}(c))}\mathbb{E}(c)\; 
\times \Biggl( &\;\frac{\alpha_i\sum^N_{j=1}\text{Cov}(p_j\tilde{q}_j,p_i\tilde{q}_i)(l_j\bar{L})}{\sum^N_{j=1}\mathbb{E}(p_j\tilde{q}_j)(l_j\bar{L})} \\[.4cm]
&-  \;\frac{\alpha_1\sum^N_{j=1}\text{Cov}(p_j\tilde{q}_j,p_1\tilde{q}_1)(l_j\bar{L})}{\sum^N_{j=1}\mathbb{E}(p_j\tilde{q}_j)(l_j\bar{L})} \Biggl),
\end{aligned}
\label{eq: main model}
\end{equation}
where $\mathbb{E}(c)=\sum^N_{j=1}\mathbb{E}(p_j\tilde{q}_j)(l_j\bar{L})$ is expected total revenue. \Cref{app: irs} provides the full derivation.

\Cref{eq: main model} is the basis for the estimating equation that we bring to the data in \Cref{sec: model estimation}. The mean term on the left serves as the dependent variable, and the variance-covariance term on the right as the independent variable. The coefficient of relative risk aversion, $\rho \equiv \frac{-U''(\mathbb{E}(c))}{U'(\mathbb{E}(c))}\mathbb{E}(c)$, is the regression coefficient, capturing an effective  portfolio-specific measure of curvature over crop income. This is the standard object identified in revealed-preference settings, which recover curvature over the payoff relevant to the observed choice (\cite{cohenEstimatingRiskPreferences2007}, \cite{Chiappori2011}, \cite{CalvetCampbellGomesSodini2026}).

Note that $\rho$ can be recovered without imposing a functional form on utility, since we never have to compute moments of $U$ directly. Thus, estimating $\rho$ in this model requires only information on cropland shares $l_i$, total land $\bar{L}$, returns-to-scale parameters $\alpha_i$, and expected per-acre revenues and their covariances, constructed from per-acre yields $\tilde{q}_i$ and prices $p_i$.

\section{Data}
\label{sec: data} 

In this section, we describe the data used in our empirical exercise to recover $\rho$ from the model in the previous section.

\subsection{The U.S. census of agriculture}
\label{Subsec: AgCensus data}

Our data on total farmland, planted acreage, and crop yields come from the U.S. Census of Agriculture, hereafter the Ag Census, a comprehensive national farm census administered by the Census Bureau from 1840 onward. The survey was administered decennially through 1920 and quinquennially from 1925 onward. The census was conducted during the census year, but output and acreage data referred to the previous calendar year. For example, 1899 information was collected in the 1900 wave, so we refer to estimates from this wave as estimates for 1899. Although the underlying data were collected at the farm level, the farm-level records are not readily available. We therefore use the county-level data compiled by \cite{Haines2018}. Since county boundaries changed over time, we harmonize counties to 1900 boundaries and link other years using the crosswalk built by \cite{Eckert2020b}. Throughout the analysis, we restrict our attention to the contiguous United States.

\begin{figure}[t]\centering
\caption{Core-crop coverage of improved farmland}\label{fig: cropcoverage}
\vspace{0.15cm}

\subfloat[Average across waves, by county]{\includegraphics[width=0.53\textwidth]{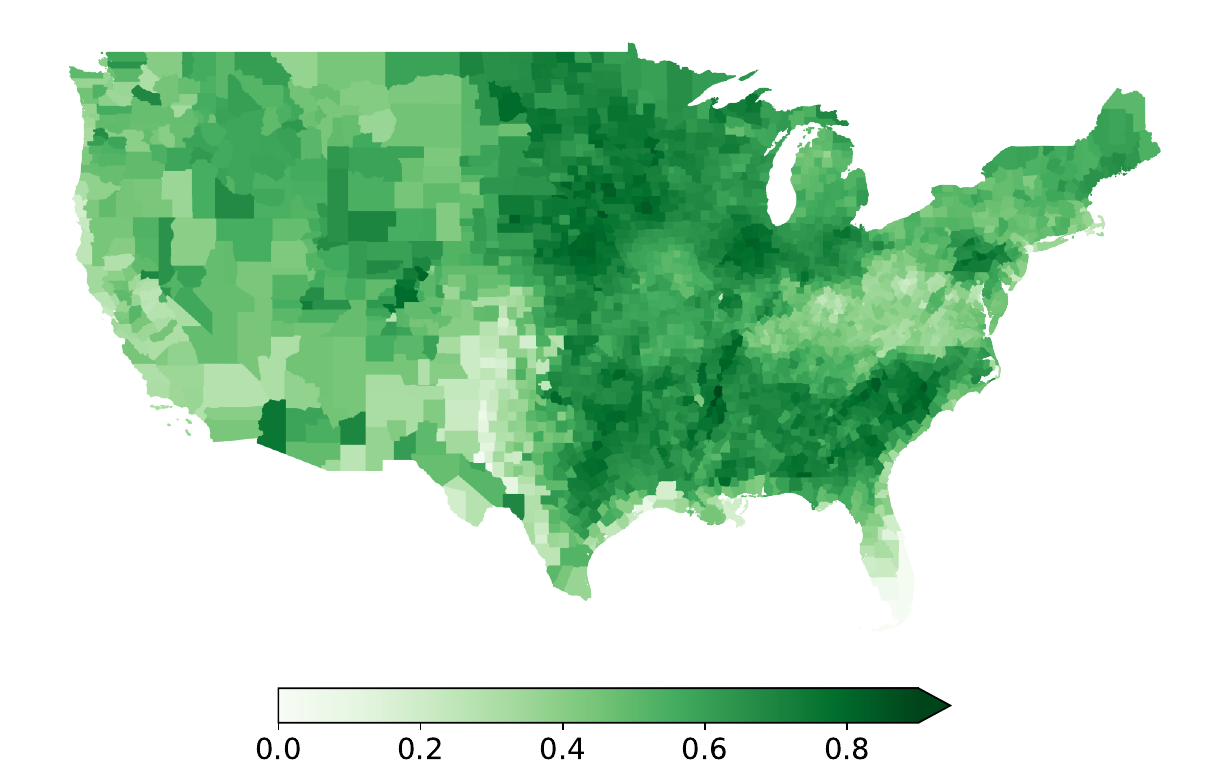}}\hfill
\subfloat[Mean and median by wave]{\includegraphics[width=0.45\textwidth]{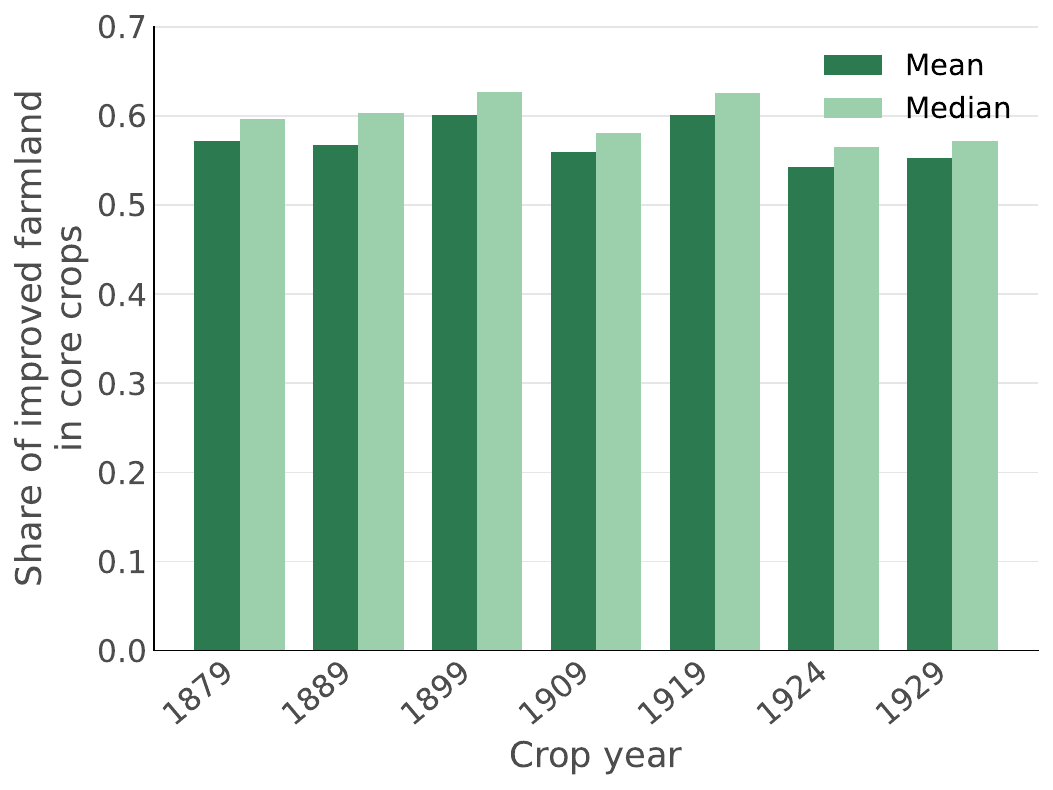}}
\vspace{0.1cm}

{\footnotesize\raggedright \textbf{Notes:} This figure shows crop-core coverage, the share of a county's improved farmland planted in the seven core crops (corn, cotton, wheat, hay, oats, barley, rye), across space and time. Panel~(a) maps each county's average coverage across the waves in which it is observed, while panel~(b) gives the cross-county mean and median coverage in each agricultural census year.\par}
\end{figure}

Between 1840 and 1870, the Ag Census collected crop output but not planted acreage information. Starting in the 1880 wave, it also collected planted acreage for a set of core crops that remain available and comparable over time: corn, oats, wheat, cotton, hay, rye, and barley. We harmonize the survey information for these crops across waves and focus our analysis on the 1880--1930 period. These core crops account for roughly 60\% of improved land in every wave and encompass the majority of improved land in most counties, as shown in \Cref{fig: cropcoverage} and \Cref{fig: cropcoveragewave}.\footnote{Improved land is defined in the Ag Census as all land regularly tilled or mowed, land pastured and cropped in rotation, land lying fallow, land in gardens, orchards, vineyards, and nurseries, and land occupied by farm buildings. Unimproved land encompasses any land which is not  improved or is forested. Some examples of unimproved land are woodlands (natural or planted, and which may later be used for firewood), brush land, rough or stony land, swamp, etc. The estimating equation defines the farmer's portfolio over this core set of crops, so our object of interest is the curvature of utility over core-crop income (see \Cref{sec: model}). In \Cref{sec: results correlates} we verify that the estimates are unrelated to how much of a county's land this set covers.}

For each core crop, we define observed per-acre yield as
\[
\tilde{q}_{i,c,t}=\frac{Q_{i,c,t}}{L_{i,c,t}},
\]
where $Q_{i,c,t}$ and $L_{i,c,t}$ are the total output and acreage planted with crop $i$ in county $c$ and year $t$, respectively. We winsorize the per-acre yield distribution at the 1st and 99th percentiles within each crop and year to limit the influence of outliers and misreported values.

We then construct the land share of each crop as
\[
l_{i,c,t}=\frac{L_{i,c,t}}{\sum^N_{j=1}L_{j,c,t}},
\]
which measures acreage in crop $i$ as a fraction of total core-crop acreage. This maps to the model in \Cref{sec: model} by identifying crop $i$'s planted acres, $l_{i,c,t}\bar{L}_{c,t}$, with observed acreage $L_{i,c,t}$, and total land, $\bar{L}_{c,t}\equiv\sum^N_{j=1}L_{j,c,t}$, with the acreage the county devotes to the core crops. The shares $l_{i,c,t}$ therefore sum to one within each county-year. \Cref{tab: acressumstats} reports the mean and median share of land allocated to each of the core crops in different waves.

\subsection{Cultivation clusters}

Using information on the acreage shares for different crops across counties in each year, we construct cultivation clusters that group counties with similar crop portfolios. These clusters provide a coarse summary of agricultural environments and are used in the pipeline below to estimate cluster-level returns to scale parameters for each crop that help account for systematic differences in cultivation technology and resources across regions.

We apply a $k$-means clustering procedure to county crop-share vectors in each census year to form four broad staple-crop clusters: a corn/hay cluster (the Corn belt), a cotton/corn cluster (the Cotton belt in the deep south and Texas), a hay/oats cluster (the Hay belt in the northeast and the arid west), and a hay/wheat cluster (the Wheat belt in the Great Plains). \Cref{app: cultclusters} provides further details on the construction of the cultivation clusters.

\Cref{fig: clusters4} maps these four clusters in each of the census years. \Cref{fig: clustercentroids} reports the associated cultivation mixes of these clusters in each census year, showing that each cluster has dominant crops but remains diversified, and that these mixes change over time.

\subsection{Price data}
\label{sec: price data}

We compile a panel of state-level annual prices for our core set of crops from 1879 to 1929 using several sources. Our main source is the state-level price histories constructed by \cite{cooley1977} for 1866--1930, which report the average price received by farmers in each state using USDA data. Because this dataset contains gaps for some crops, years, and states, we supplement it with four national-level datasets, detailed in \Cref{tab:pricedata}. We harmonize these sources by filling missing crop-state-year observations in \cite{cooley1977} using national price movements and cross-crop differences between state and national prices. \Cref{App: prices} describes the harmonization procedure in detail. We deflate all prices in the harmonized panel to 1913 dollars using the Consumer Price Index.

\Cref{tab:pricestats} reports summary statistics for the price series by crop and decade, and \Cref{fig:pricetseries} plots the state-level price time series for each crop. These series show substantial variation across crops, time, and space, reflecting both dispersion in commodity-crop prices and potential differences in the crop varieties dominant in different states.

\subsection{Climate data}
\label{sec: climate data}

Our source of historical climate data is the NOAA/CIRES/DOE 20th Century Reanalysis version 3 dataset, hereafter 20CRV3. 20CRV3 is a gridded climate dataset that reconstructs historical climate conditions using atmospheric pressure observations, sea-ice observations, and climate-model output \citep{slivinski2019}. The data are available globally at $0.7^\circ$ by $0.7^\circ$ spatial resolution and at 3-hour intervals starting in 1806. We assign each county to its nearest grid cell based on centroid distance, and aggregate the data to daily frequency.\footnote{Support for the Twentieth Century Reanalysis Project version 3 dataset is provided by the U.S. Department of Energy, Office of Science Biological and Environmental Research (BER), by the National Oceanic and Atmospheric Administration Climate Program Office, and by the NOAA Earth System Research Laboratory Physical Sciences Laboratory. NOAA/CIRES/DOE 20th Century Reanalysis (V3) data provided by the NOAA PSL, Boulder, Colorado, USA, from their website at \href{https://psl.noaa.gov}{https://psl.noaa.gov}.}

\subsection{Kansas farm-level dataset}
\label{sec: kansas data}

We complement the national analysis with a farm-level panel from Kansas, using the dataset compiled by \cite{sylvester2002}. These data contain detailed agricultural records for all farms in 24 Kansas townships in 1895, 1905, 1915, 1920, 1925, and 1930. Each wave contains 2,446 farms on average. To our knowledge, this dataset is the only farm-level data currently available for the United States during this period.

Using this farm panel, we also construct a \textit{farmer}-level panel, leveraging farmer name information. We link farmers in the dataset over time using Soundex phonetic encoding of last names combined with a Jaro-Winkler similarity score on given names. With this procedure we are able to pairwise match 33\%--49\% of farmers across waves. Using the same matching procedure, we also link farmers to a separate dataset of membership in fraternal organizations, digitized and compiled by \cite{ang2026}. Specifically, we code a farmer as a lodge member if any death-notice matching their name and town is found.

Information on the sample geographic distribution, cultivation patterns, and the matching procedures is available in \Cref{app: kansas data}.

\section{Methodology}
\label{sec: methodology}

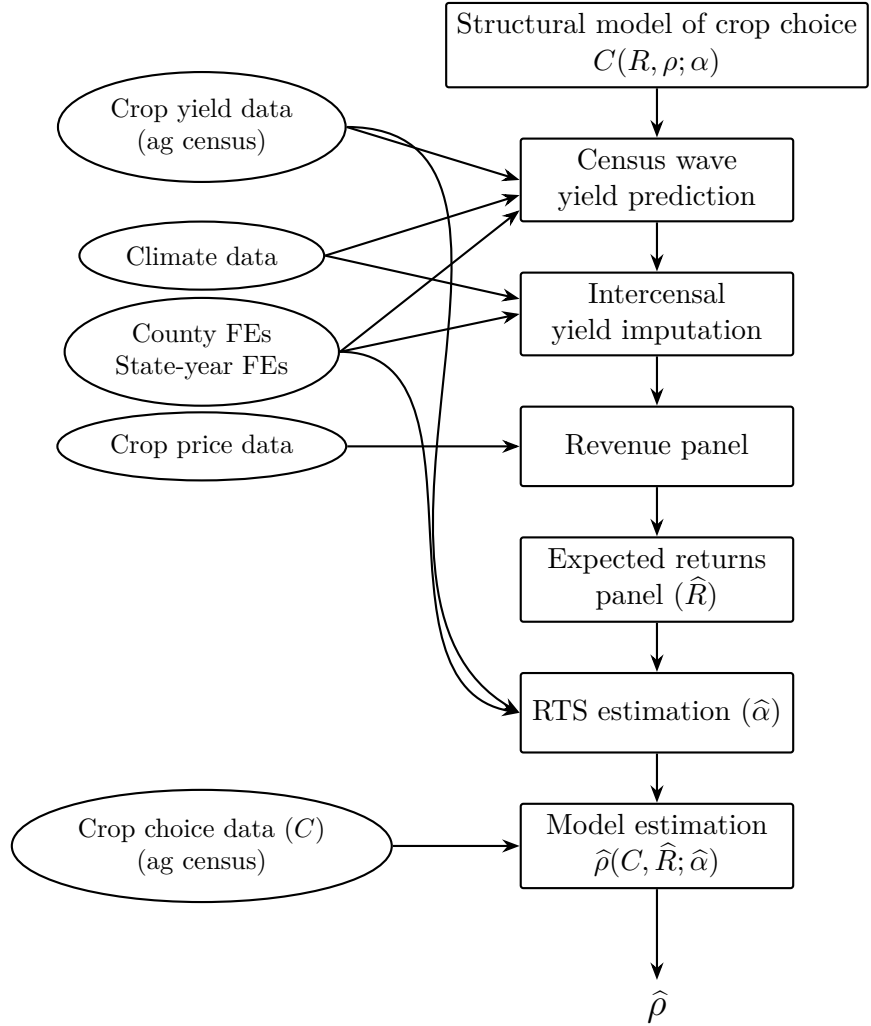
\begin{figure}[!ht]\centering
\caption{Visualization of the empirical methodology}\label{fig: pipeline}
\vspace{.25cm}
\begin{tikzpicture}[node distance=0.65cm]
\node[pipebox] (model) {Structural model of crop choice \\ $C(R, \rho; \alpha)$};
\node[pipebox, below=of model] (yieldpred) {Census wave \\ yield prediction};
\node[pipebox, below=of yieldpred] (yieldimp) {Intercensal \\ yield imputation};
\node[pipebox, below=of yieldimp] (revenue) {Revenue panel};
\node[pipebox, below=of revenue] (beliefs) {Expected returns \\ panel ($\widehat R$)};
\node[pipebox, below=of beliefs] (alpha) {RTS estimation ($\widehat\alpha$)};
\node[pipebox, below=of alpha] (estim) {Model estimation \\ $\widehat\rho(C,\widehat R; \widehat{\alpha})$};
\node[outlabel, below=1.2cm of estim] (output) {$\widehat\rho$};
\node[databox] (cropyield) at ($(yieldpred.west) + (-4.2, 0.7)$) {Crop yield data \\ (ag census)};
\node[databox] (climate) at ($(yieldpred.west) + (-4.2, -1.0)$) {Climate data};
\node[databox] (fes) at ($(yieldimp.west) + (-4.2, -0.5)$) {County FEs \\ State-year FEs};
\node[databox] (price) at ($(revenue.west) + (-4.2, 0)$) {Crop price data};
\node[databox] (cropchoice) at ($(estim.west) + (-4.2, 0)$) {Crop choice data ($C$) \\ (ag census)};
\draw[arr] (model) -- (yieldpred);
\draw[arr] (yieldpred) -- (yieldimp);
\draw[arr] (yieldimp) -- (revenue);
\draw[arr] (revenue) -- (beliefs);
\draw[arr] (beliefs) -- (alpha);
\draw[arr] (alpha) -- (estim);
\draw[arr] (estim) -- (output);
\draw[arr] (cropyield.east) -- (yieldpred.west);
\draw[arr] (climate.east) -- ($(yieldpred.west) + (0, -0.2)$);
\draw[arr] (climate.east) -- ($(yieldimp.west) + (0, 0.2)$);
\draw[arr] (fes.east) -- ($(yieldpred.west) + (0, -0.4)$);
\draw[arr] (fes.east) -- (yieldimp.west);
\draw[arr] (price.east) -- (revenue.west);
\draw[arr] (cropchoice.east) -- (estim.west);
\draw[arr] (cropyield.east) to[out=0,in=150] (alpha.west);
\draw[arr] (fes.east) to[out=0,in=170] (alpha.west);
\end{tikzpicture}
\vspace{.4cm}

{\footnotesize\raggedright \textbf{Notes:} This figure shows the empirical pipeline for the analysis. Data inputs (ovals, left) feed into the processing chain (rectangles, center), producing the final
risk-aversion estimate $\widehat{\rho}$.\par}
\end{figure}

This section describes how we take the model of \Cref{sec: model} to the data. \Cref{fig: pipeline} maps the pipeline and serves as a guide to this section.

\subsection{Census-wave yield prediction}
\label{sec: yield prediction}

We first estimate the relationship between observed per-acre yields, climate, and county- and time- fixed effects in census years. We start by describing the prediction model and evaluating its performance.

We build the climate predictors from the 20CRV3 daily climate data. For each county and Ag Census year, we compute ten agronomically-informed base statistics: four for precipitation (total, maximum daily, wet days, and longest dry spell), and six for temperature (mean, maximum, minimum, standard deviation, frost days, and hot days). We compute these statistics at three temporal frequencies --- 10-day periods (dekads), months, and 3-month seasons --- for a two year lag window ending in December of the cultivation year (e.g. for the 1900 census, spanning the cultivation year 1899 and prior year 1898). This allows the prediction model to select agronomically-relevant climate moments, timescales, and lags for each crop. In total, we cover $104$ temporal windows for each of the 10 variables, resulting in $1{,}040$ candidate climate features per county-year. \Cref{tab: climatefeatures} summarizes the base climate statistics and temporal window construction.

Per-acre yields are driven by both climate and non-climate factors. Because we have repeat observations of yields in each county over time, we can capture persistent local factors (such as soil quality and crop suitability) \textit{at the time of cultivation} by including county fixed effects in the prediction exercise. These substitute for present-day proxies like the crop suitability index in FAO-GAEZ commonly used in other analyses. The relationship between these and historic yields, especially in the deep past, is unclear, and they can suffer from significant non-classical measurement error (\cite{Rhode2024}, \cite{araujoPotatoPotahtoFAOGAEZ2026a}). County fixed effects also capture other persistent local predictors outside of crop suitability that can affect yields, such as local persistent cultivation practices. We also include year- and state-year fixed effects to capture the effects of contemporaneous national and regional shocks.

Our sample period spans 50 years in which significant technological change takes place in agriculture. Predicting yields from the entire sample period simultaneously risks contaminating predictions in one period with factors from another. To ensure that we are staying within one technological regime when predicting yields, we estimate separate models for each adjacent pair of census waves, $(1880,1890)$, $(1890,1900)$, $(1900,1910)$, $(1910,1920)$, $(1920,1925)$, and $(1925,1930)$. In other words, we fit a separate prediction model for each crop and wave pair. Because of this approach and the inclusion of the fixed effects, when we use the within-census predictions to impute the full yield panel, we can only do so for the years between census waves. This means that we can compute expected revenues (and risk aversion) only starting in the second wave for which we have acreage data (1889).

For the prediction exercise, we residualize yields and all climate features on the fixed effects. We then use six machine learning methods to predict yields from climate: three linear shrinkage models (LASSO, Ridge regression, Elastic nets), and three tree-based methods (gradient boosting machine, random forest, and quantile gradient boosting). These approaches are complementary. The linear shrinkage models efficiently recover smooth, additive climate responses spread across many correlated windows. Within the linear family, the three estimators differ in their sparsity assumptions. The tree-based methods capture threshold effects and interactions that the linear models cannot, such as frost and heat cutoffs or the interaction of precipitation timing with temperature. Within this family, the three methods trade off variance against bias in different ways. This diversity of prediction errors across methods allows us to improve on each individual member by using the set as a whole.

For each crop and census-wave pair, we combine the six methods into a stacked ensemble. Each method is fit on the fixed-effect residuals and evaluated using county-aware five-fold cross-validation, producing an out-of-fold prediction for every county-year observation. The stacking weights are then chosen to minimize the squared error of the weighted combination of out-of-fold predictions, subject to being non-negative and summing to one. These constraints ensure that the ensemble weights reward out-of-sample performance and limit extrapolation when the model is applied to intercensal years. The final yield prediction additively combines the fixed-effect component (county, year, and state-by-year effects) with the stacked climate prediction of the residual. \Cref{app: yield prediction} describes the implementation details. 

We assess performance by the out-of-fold $R^2$ of the stacked predictions. The cross-validation is built around the panel structure: each county's two wave-pair observations are assigned to different folds, and all residualization and prediction models are refit within each training fold. County fixed effects for held-out observations are recovered from the county's observation in the other wave, mirroring the imputation setting in which census anchors are observed and only intercensal variation is predicted out of sample. \Cref{fig: oofr2} reports the average out-of-fold $R^2$ of the stacked predictions for each crop and \Cref{tab: climateuplift} separates the contribution of the climate features from the fixed-effect baseline. Predictive performance is strong for most crops, especially for more-widely grown grains, with $R^2$ values ranging from 0.78 to 0.63 for corn, oats, and wheat, and weaker for sparse crops like barley and rye.

\subsection{Intercensal yield imputation}
\label{sec: yield imputation}
With the prediction models in hand, we impute per-acre yields for every county, crop, and year between census waves. For an intercensal year $\tau$ in census wave pair $p\equiv(p_s,p_e)$, let $f_\tau=(\tau-p_s)/(p_e-p_s)$ denote its relative position between the two census waves. The baseline imputation linearly interpolates the fixed-effect yield component for crop $i$ and county $c$ between the two census anchors, then adds the stacked climate-predicted residual for year $\tau$:\\
\[
\widehat{\tilde q}_{i,c,\tau}=
(1-f_\tau)\widehat{\text{FE}}_{i,c,p_s}
+
f_\tau\widehat{\text{FE}}_{i,c,p_e}
+
\hat\varepsilon^{\text{stack}}_{i,c,\tau}.
\]

Here, $\widehat{\text{FE}}_{i,c,p}$ is the county-level fixed-effect yield component, incorporating county, year, and state-by-year fixed effects, and $\hat\varepsilon^{\text{stack}}_{i,c,\tau}$ is the climate-predicted residual from the stacked model.\footnote{When the climate model performs poorly out of sample for a crop in a given cultivation cluster and census wave, we drop the climate-based component. This correction is described in detail in \Cref{app: yieldpredictionfallbackadd}, and affects only about 1\% of planted acreage.} Imputed yields are floored at zero. \Cref{fig: imputedpanels} plots the resulting national-average yield trajectories by crop, showing substantial variation across crops and over time.

\subsection{Revenue panel}
\label{sec: revenue panel}

\begin{figure}[t]
\begin{center}
\caption{Imputed average per-acre revenue by crop, 1878--1929}\label{fig: revenuepanels}
\vspace{.1cm}
\includegraphics[width=\textwidth]{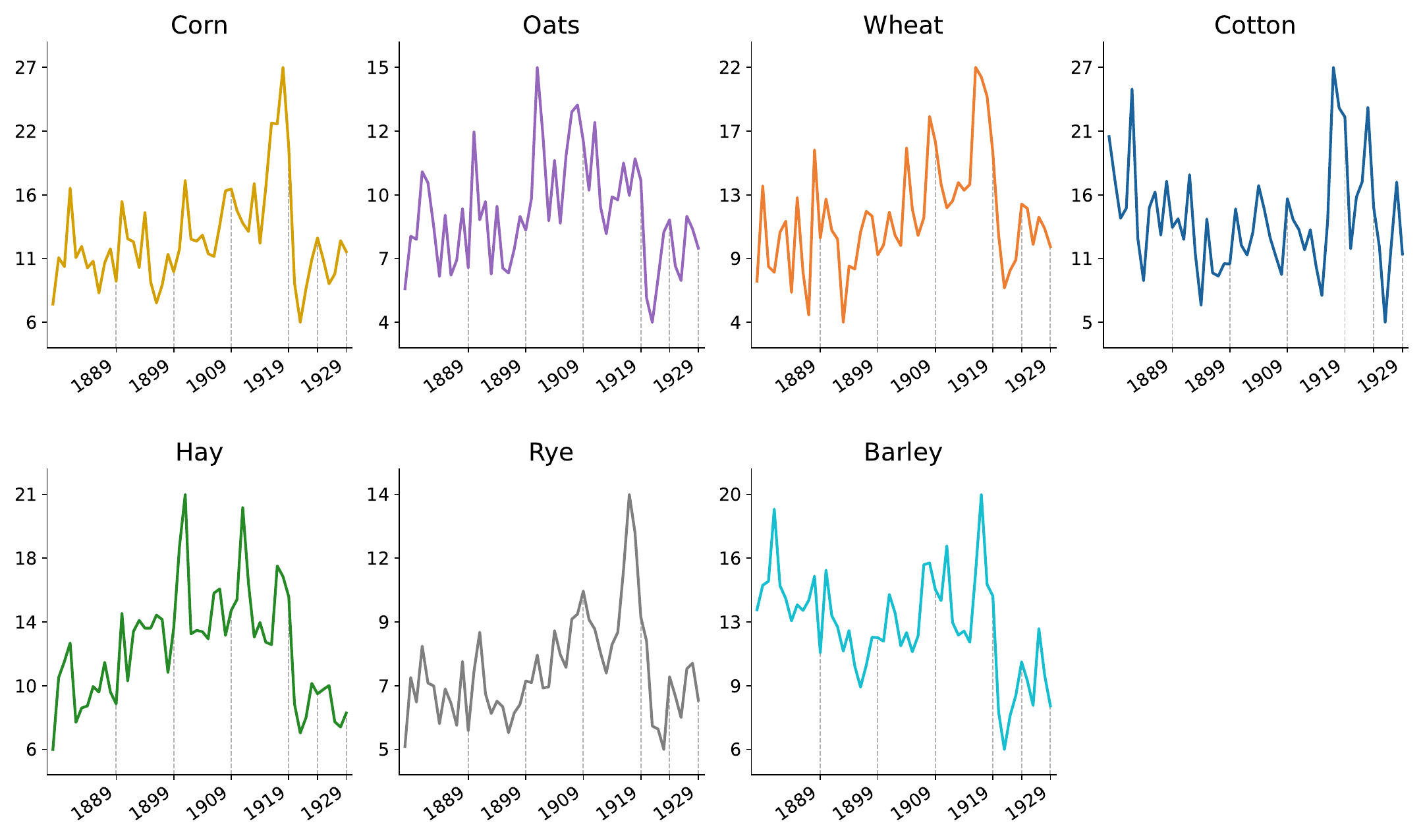}
\end{center}
\smallskip\footnotesize{\textbf{Notes}: These figures show the average revenue (imputed yield times deflated state-year price), by crop and year (in 1913 dollars). Light dashed vertical lines mark the census years.}
\end{figure}

We then combine the imputed yields with prices to form the revenue panel used in the rest of the analysis. For each county $c$, crop $i$, and year $t$, per-acre revenue is
\[
R_{i,c,t}=p_{i,s(c),t}\;\times\;\widehat{\tilde q}_{i,c,t},
\]
where $p_{i,s(c),t}$ is the state-level price of crop $i$ in state $s(c)$ and year $t$, deflated to $1913$ dollars, and $\widehat{\tilde q}_{i,c,t}$ is the imputed per-acre yield. \Cref{fig: revenuepanels} plots the resulting per-acre revenue trajectories. These vary widely across crops and over time, but share some comovement from national price swings such as the World War I agricultural price boom and subsequent bust.

\subsection{Constructing the expected returns panel}
\label{sec: returns}

We next construct the expected returns used in the model, namely the the means and variance-covariance matrices of county-level per-acre revenues. For each county $c$ and census year $t$, we compute these moments using realized revenues over the preceding ten years. Let $\hat\mu(i,c,t)$ and $\hat\Sigma(ij,c,t)$ denote the empirical mean and covariance with crop $j$ for crop $i$ in county $c$ and year $t$ that enter the model. We compute
\begin{align*}
\hat\mu(i,c,t) &= \frac{1}{h_{i,c,t}}\sum_{\tau=t-10}^{t-1}  R_{i,c,\tau}\;\text{ and }\;\hat\Sigma(ij,c,t) = \frac{1}{h_{ij,c,t}-1}\sum_{\tau\in\mathcal{T}_{ij,c,t}}
\bigl( R_{i,c,\tau}-\bar R^{(ij)}_{i,c,t}\bigr)
\bigl( R_{j,c,\tau}-\bar R^{(ij)}_{j,c,t}\bigr),
\end{align*}
where $R_{i,c,\tau}$ is the per-acre revenue (in 1913 dollars) of crop $i$ in county $c$ and year $\tau$, and $h_{i,c,t}$ is the number of years in which crop $i$ is observed in the ten-year window before census year $t$. $h_{ij,c,t}$ is the number of years in which both crops $i$ and $j$ are observed, and $\mathcal{T}_{ij,c,t}$ is the set of these years in which both crops are observed, $h_{ij,c,t}=|\mathcal{T}_{ij,c,t}|$. $\bar R^{(ij)}_{i,c,t}$ is the mean of crop $i$'s revenue over those common years (so each covariance entry is centered on the pair's common support). We require at least five years of data for each mean and covariance. Because the revenue panel begins in 1878, the ten-year window is complete only from 1889 onward. The expected-returns panel, and therefore our main analysis, covers the six census years $t\in\{1889,1899,1909,1919,1924,1929\}$.

\subsection{Returns to scale estimation}
\label{sec: RTS}

We next estimate the returns-to-scale parameters $\alpha_i$. Since the observed per-acre yield is $\tilde q_{i,c,t}=q_{i,c,t}(l_{i,c,t}\bar{L}_{c,t})^{\alpha_i-1}$, the elasticity of observed per-acre yield with respect to own-crop planted acreage pins down $\alpha_i-1$. We allow this elasticity to vary by crop, census wave, and cultivation cluster, capturing regional variation in the scale-yield relationship. Specifically, for county $c$ growing crop $i$ in census year $t$ and located in cultivation cluster $k$, we estimate\\
\begin{equation}
\ln\tilde q_{i,c,t} = \beta_{i,k,t}\ln (l_{i,c,t}\bar{L}_{c,t}) + \delta_c + \delta_{s,t} + \varepsilon_{i,c,t},
\label{eq: alphaest}
\end{equation}
where $\beta_{i,k,t}=\alpha_{i,k,t}-1$. We estimate \Cref{eq: alphaest} separately for each crop, wave, and cultivation cluster, pooling each wave with the preceding decennial census, and assigning each county to its cluster in the later wave $t$. We recover $\widehat\alpha_{i,k,t}=1+\widehat\beta_{i,k,t}$. \Cref{app: RTS} contains further details.

We present the estimated returns-to-scale parameters for each crop and census year, averaged across cultivation clusters, in \Cref{fig: alphacropseries}. The estimates sit close to one for nearly every crop and year, indicating returns to scale that are close to constant. \Cref{fig: alphacluster} shows the corresponding estimates separately by cultivation cluster.

\subsection{Model estimation}
\label{sec: model estimation}

With expected returns, cropland shares, and the returns-to-scale parameters in hand, we estimate the coefficient of relative risk aversion in each county and census year using \Cref{eq: main model}. To do this, we need to select a numeraire crop (crop 1), since the model differences crop returns against those of the risky numeraire. We choose corn as that numeraire, because it has the broadest geographic and temporal coverage (between 92\% and 99\% of counties grew corn between 1879 and 1929), and because corn also has the least noisy yield predictions (\Cref{sec: yield prediction}).

For crop $i$ grown in county $c$ and census year $t$ located in cultivation cluster $k$, define
\begin{equation}    Y_{i,c,t}=\hat\alpha_{i,k(c,t),t}\times\hat\mu(i,c,t)-\hat\alpha_{\text{corn},k(c,t),t}\times\hat\mu(\text{corn},c,t)
\end{equation}
and
\begin{equation}
    X_{i,c,t}= 
\frac{\hat\alpha_{i,k(c,t),t}\sum_j (l_{j,c,t}\bar L_{c,t})\hat\Sigma(ji,c,t)-\hat\alpha_{\text{corn},k(c,t),t}\sum_j (l_{j,c,t}\bar L_{c,t})\hat\Sigma(j,\text{corn},c,t)}
{\sum_j (l_{j,c,t}\bar L_{c,t})\hat\mu(j,c,t)},
\end{equation}
where $l_{j,c,t}\bar L_{c,t}$ is the acreage planted with crop $j$ as measured in the Ag Census, $\hat\mu$ and $\hat\Sigma$ are the expected-return moments constructed in \Cref{sec: returns}, and $\hat\alpha$ are the returns to scale parameters estimated in \Cref{sec: RTS}. With this notation, the estimating equation, \Cref{eq: main model}, becomes
\begin{equation}
    Y_{i,c,t}=\rho_{c,t}X_{i,c,t}+\varepsilon_{i,c,t}.
\end{equation}
We then estimate $\rho_{c,t}$ as the slope of  a no-intercept ordinary-least-squares regression of $Y_{i,c,t}$ on $X_{i,c,t}$ across the crops grown in county $c$ in census year $t$. 

Because each county-year contains at most six observations, one for each core crop minus corn, to ensure consistency and maximize power we estimate risk aversion using a local space-time county cropshare pool, rather than a single county-year regression. For each county-year, we run a kernel-weighted regression that combines observations from a county's six nearest neighboring counties (the optimal spatial bandwidth) and two nearest census waves. Specifically, we estimate the risk aversion in a county observed in 1909 from cropland allocations in that county and its six nearest neighbors in 1909, 1899, and 1919 --- 21 counties in total, each with up to six observations. County-years in edge waves (1889, 1929) are pooled with the single wave closest to them (1899, 1924). Spatial weights decline with distance between county centroids, while time weights decline with elapsed years. The resulting county-year estimate is the slope from this local regression. This procedure borrows strength from geographically and temporally nearby observations while still allowing risk aversion to vary flexibly across counties and over time. \Cref{app: rho} provides details on it.

In total, we are able to estimate risk aversion for 2,808 distinct counties, 94.5\% of the 2,973 counties that ever appear in the Ag Census.\footnote{A county-year receives a risk-aversion estimate when we can form its own first-order-condition moments, which requires the following: (i) it grows corn, the numeraire, whose expected revenue and return covariances enter every other crop's first-order condition; (ii) it grows at least one other crop from the core set, which supplies the first-order condition(s) from which risk aversion is recovered; and (iii) it has at least five years of revenue history in the preceding decade from which to form the expected-return and covariance moments.}  Of those 2,808, 2,321 counties (82.7\%) have estimates in all six waves, with the median county measured in 5.66 waves. Coverage is highest for later waves: 2,339 (85.4\%) in 1889; 2,556 (87.8\%) in 1899; 2,692 (93.6\%) in 1909; 2,774 (96.4\%) in 1919; 2,782 (96.9\%) in 1924; and 2,752 (95.7\%) in 1929. This set comprises almost all counties across 40 states, but excludes some counties in the western states.

\subsection{Kansas farm- and township-level risk aversion}
\label{sec: kansas RA}

For our Kansas analyses, we produce risk aversion estimates at the farm and the aggregate township levels using the same procedure as that outlined in this section. We focus on the six core crops cultivated in Kansas in the period: barley, corn, hay, oats, rye, and wheat. As in the main analysis, we set corn as the numeraire crop. We construct expected revenues and revenue covariances for each crop, township, and year using the per-acre yield histories in the county corresponding to the township, combined with Kansas crop prices from the main price dataset. Returns-to-scale parameters are assigned from the township's cultivation cluster in the nearest available census wave. \Cref{app: kansas methodology} provides additional details, and \Cref{tab: kansassamplesize} tabulates the number of farms for which we recover risk aversion in each wave.

\section{Method validation}
\label{sec: validation}

In this section we present four analyses that empirically validate key aspects of our method.

\subsection{Crop returns behave like financial asset returns}
\label{sec: validation returns}

A basic assumption of our method is that crops behave like financial assets, by offering a risk-return tradeoff where crops with higher expected returns also carry higher expected risk. This means that the menu of crops does not have a dominant or dominated options. If some crop offered high return at low risk, every farmer should plant it regardless of preferences, and crop allocations would tell us nothing about risk attitudes.

\begin{figure}[tbp]
\begin{center}
\caption{Expected crop returns: mean--variance positions, averaged across waves}\label{fig: retmvavg}
\includegraphics[width=.85\textwidth]{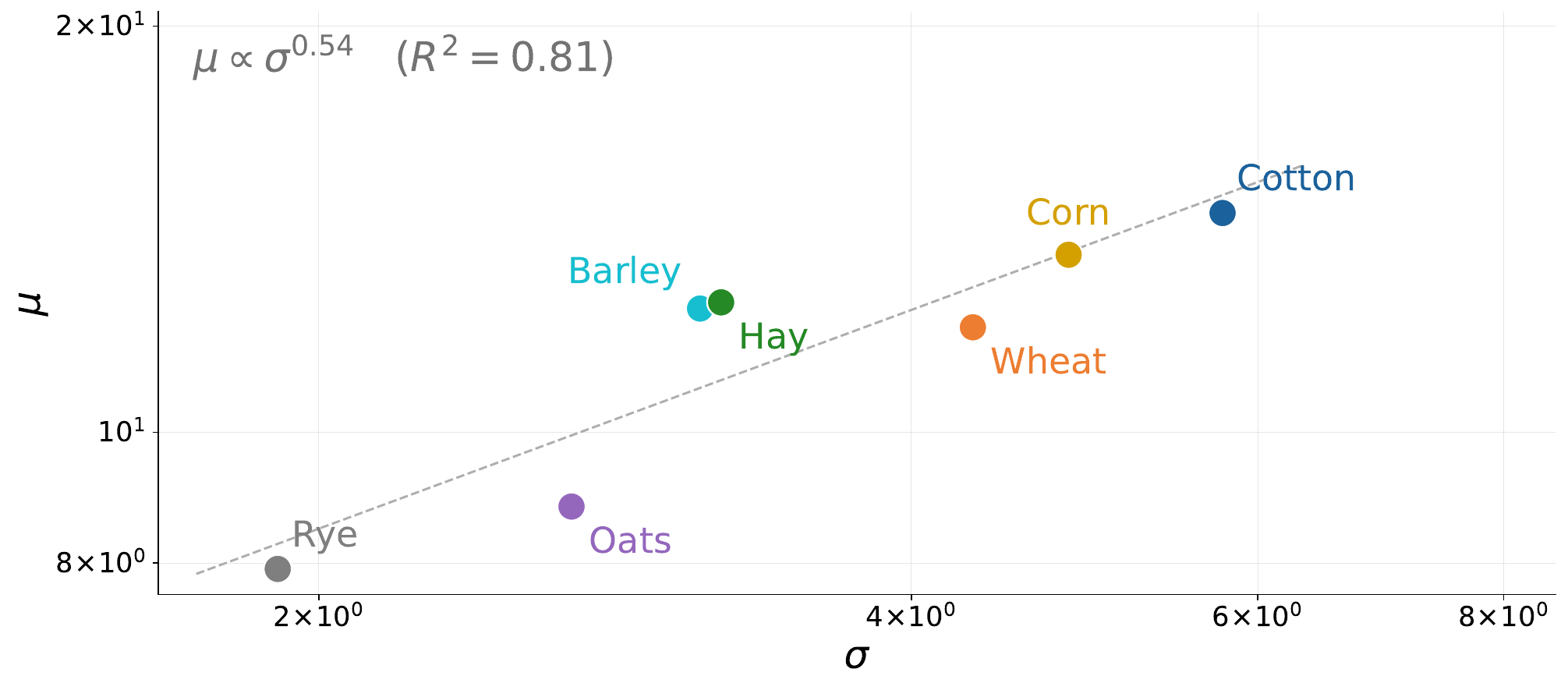}
\end{center}
\smallskip\footnotesize \textbf{Notes:} This figure shows the  mean--variance positions across crops. Each dot is one crop at its expected per-acre return $\mu$ against the standard deviation $\sigma$ of that return, both in real 1913
\$/acre and averaged across the six crop-year waves. Both axes use a common log--log scale, and
the dashed line is the power-law fit, $\mu \propto \sigma^{0.54}$ ($R^2=0.81$).
\end{figure}

We test this premise in our data by plotting the log mean and log standard deviation of each crop's per-acre expected revenue in our constructed returns dataset. \Cref{fig: retmvavg} presents the results averaged across census years. We find that expected returns are increasing in return risk: no crop offers high return at low risk, and none is clearly dominated. The ordering along the line matches the agricultural history of the period, with hardy, low-value small grains like rye and oats at the bottom of the ladder, wheat and corn in the middle, and cotton, which is the classic high-risk high-return cash crop-at the top. \Cref{app: returns} displays the results of this analysis for each census wave separately, which are qualitatively similar.

The slope of the fitted line shows that the menu offers \emph{decreasing returns to risk}: the elasticity of expected return with respect to risk is $0.54$, so doubling a portfolio's return volatility raises its expected return by approximately 45 percent. This is in line with theoretical and empirical evidence for concave returns in financial markets with leverage constraints (\cite{Black1972}, \cite{FrazziniPedersen2014}). We would expect such constraints to bite in our setting, and the empirical returns pattern suggests that it does.

\subsection{Farmers respond to crop risk-return tradeoffs like investors}
\label{sec: kansas expectations validation}

Conditional on crops offering a plausible risk-return tradeoff, the next assumption in the method is that farmers respond to the tradeoff in the returns we construct in the same way as traders allocating a financial portfolio. This is not a foregone conclusion. One might imagine that farmers would make planting decisions based on other criteria like cultural practices or tradition. Alternatively, it is possible that, even if they were responding rationally to risk, the returns measures we construct might not adequately capture the real risks facing farmers. Either case would make it difficult to identify risk preferences in the way that we do from their observed choices.

To test this assumption, we turn to our Kansas farmer-level panel dataset. This offers the strongest test of the assumption that farmers respond to the tradeoff in mean and variance of crop returns like investors, because it speaks to individual farmer behavior, and because we can observe repeated choices by each farmer over time, which means we can see the dynamic adjustment to a changing risk landscape that we expect from a rational investor.

For each farmer-crop-year observation, we regress the planted acreage share on the expected revenue per acre for that crop and on the standard deviation of that revenue, both constructed as described in \Cref{sec: returns}. \Cref{tab: plantingacshare} reports the results. We progressively add fixed effects: crop fixed effects in column (1), crop and year fixed effects in column (2), crop, year, and farmer fixed effects in column (3), and farmer-by-crop and year fixed effects in column (4). The last specification identifies the relationship from changes over time in expected returns and risk for a given farmer-crop pair, absorbing all time-invariant farmer-specific and farmer-crop-specific heterogeneity.

\begin{table}[t]
\caption{Planting share, expected returns, and revenue risk: Kansas farmer panel \label{tab: plantingacshare}}
\begin{footnotesize}
\begin{center}
\begin{tabular}{lcccc}
\hline
\noalign{\vskip 4pt}
 & \multicolumn{4}{c}{Acreage share} \\
\cmidrule(lr){2-5}
 & (1) & (2) & (3) & (4) \\
\midrule
E[revenue/acre] & 0.0131*** & 0.0174*** & 0.0262*** & 0.0128*** \\
 & (0.0017) & (0.0020) & (0.0029) & (0.0026) \\
\addlinespace[2pt]
SD[revenue/acre] & -0.0179*** & -0.0300*** & -0.0310*** & -0.0181*** \\
 & (0.0025) & (0.0039) & (0.0036) & (0.0034) \\
\midrule
Observations & 79,915 & 79,915 & 79,915 & 43,738 \\
$R^2$ & 0.569 & 0.573 & 0.581 & 0.835 \\
\addlinespace[2pt]
Crop FE & X & X & X &  \\
Year FE &  & X & X & X \\
Farmer FE &  &  & X &  \\
Farmer $\times$ Crop FE &  &  &  & X \\
\noalign{\vskip 4pt}
\hline
\end{tabular}

\end{center}
\end{footnotesize}
\vspace{.25cm}
{\footnotesize \textbf{Notes:} This table regresses the within-farmer share of total cropland devoted to a crop on the 10-year mean and the 10-year standard deviation of revenue per acre for that crop, constructed from the imputed county-level yield panel and Kansas state prices. The sample is farmer-crop-year observations for the six core Kansas crops in all Kansas panel waves (six waves, 1895--1930). Column (4) restricts to farmers observed with the same crop in at least two waves. Standard errors clustered at the township level in parentheses. *** p$<$0.01, ** p$<$0.05, * p$<$0.1.}

\end{table}

The two margins have exactly the signs the method predicts in every specification. The coefficient on expected crop returns is positive and highly significant throughout, and its magnitude remains similar across fixed-effects structures, including the farmer-by-crop specification in column (4). The coefficient on the standard deviation of per acre revenue is negative throughout and highly significant: holding a crop's expected revenue fixed, farmers pull acreage away from crops whose revenue has become more volatile. Thus, the relationship between expected returns, expected risk, and planting shares is not driven only by static farmer-crop sorting, such as some farmers persistently favoring corn while others persistently favor wheat. Instead, the estimates reflect within-farmer adjustment over time: when the expected revenue of a crop rises for a given farmer-crop pair, the farmer allocates more acreage to that crop, and when the expected risk rises, they allocate less acreage to that crop. These results also indicate that our empirical measures of expected risk and returns appear to capture the relevant margin of risky choice in the crop allocation decision.

\subsection{Farmers allocate their crop portfolios efficiently}
\label{sec: frontier efficiency}

Having established that crops exhibit a risk-return profile similar to financial assets, and that farmers respond to this profile like investors, the next question is whether farmers are \textit{efficient} in making their crop choice investments. Our method attributes differences in crop portfolios to movements along the efficient investment frontier, driven by risk aversion, so departures from efficiency can add noise to the measures. Here we test farmer efficiency.

To do this, we compare the realized crop portfolio in each county-wave to the maximum-Sharpe portfolio of the same crops, which lies on that county-wave's efficient frontier, and calculate $\varphi = \mathrm{Sharpe}(\text{realized})/\mathrm{Sharpe}(\text{best})$ (\cite{GibbonsRossShanken1989}, \cite{CalvetCampbellSodini2007}). Both Sharpe ratios are evaluated using the county's own expected-return vector and full cross-crop covariance matrix, and the benchmark portfolio is restricted to non-negative acreage shares. To reduce attenuation from noise in constructed returns, we estimate the frontier on a bootstrap resample of belief-window years and evaluate both portfolios on held-out years.\footnote{Because the frontier is optimized over the covariance matrix, we restrict this exercise to crops observed in nearly every window year and apply a small ridge adjustment. \Cref{app: phiconstruction} provides details.} This out-of-sample procedure can generate $\varphi>1$, which we interpret as being indistinguishable from the frontier.\footnote{The maximum-Sharpe portfolio is only one point on the frontier. The optimal portfolio of a farmer with different risk preferences may have a lower Sharpe ratio. Thus, $\varphi$ is a conservative measure of efficiency: values near one indicate little scope for improving risk-adjusted returns, and, for $\varphi\leq1$, $1-\varphi$ provides an upper bound on the share of risk-adjusted return left on the table.}

\begin{figure}[t]\centering
\caption{Out-of-bag Sharpe efficiency of county crop allocations}\label{fig: phiefficiency}
\vspace{.1cm}
\includegraphics[width=0.6\textwidth]{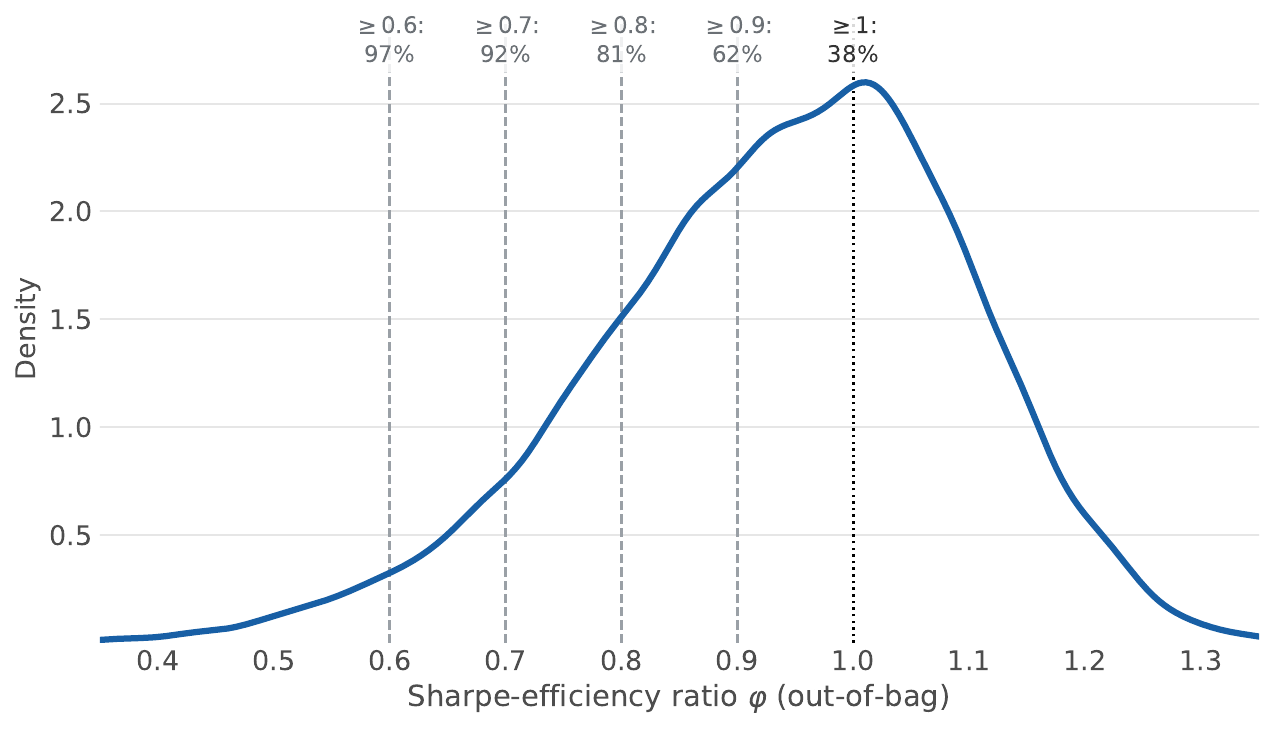}

{\footnotesize\raggedright \textbf{Notes:} Kernel density of the out-of-bag Sharpe-efficiency ratio $\varphi = \mathrm{Sharpe}_{\text{oob}}(\text{realized})/\mathrm{Sharpe}_{\text{oob}}(\text{best})$ across 15,665 county-waves, pooled over the six census waves. The best (maximum-Sharpe) portfolio of each county's own crops is estimated on bootstrap-resampled in-bag years and both portfolios are scored on the held-out years, so the benchmark is not overfit to noise. $\varphi=1$ indicates the efficient frontier, with values greater or equal than 1 corresponding to portfolios where the measure cannot reject full efficiency. Values $1-\varphi < 1$ values less than 1 indicate the fraction of risk-adjusted returns lost to inefficient allocations. \Cref{fig: phiwaves} reports the corresponding densities by census wave.\par}
\end{figure}

\Cref{fig: phiefficiency} presents the distribution of $\varphi$. We find high levels of efficiency in farmer crop portfolio allocations. The pooled median is $0.95$, and $38\%$ of county-waves sit at or beyond $\varphi = 1$. $62\%$ of county-waves achieve at least $90\%$ of their attainable risk-adjusted return, $81\%$ achieve at least $80\%$, $92\%$ achieve at least $70\%$, and $97\%$ achieve at least $60\%$. The pattern is stable over time: per-wave medians range from $0.89$ to $1.00$. For comparison, in studies of modern investors, median values for this measure are comparable or significantly lower (0.85 for Dutch households (\cite{vonGaudecker2015}); 0.73 for Swedish households (\cite{CalvetCampbellSodini2007}); 0.41 for Danish retail investors (\cite{FlorentsenEtAl2019})).\footnote{We choose the most comparable measure to our own in each paper: efficiency relative to the CAPM frontier of their own investable universe, net of fund fees in \cite{vonGaudecker2015}; relative to the highest Sharpe ratio attained by any household in the sample in \cite{CalvetCampbellSodini2007}; and the implied ratio under the CAPM framework with $\sigma_{mkt}/\sigma_p = 15.5/38 = 0.41$ in \cite{FlorentsenEtAl2019}.} Our results suggest that farmers are much closer to professional investors than retail investors in their risk management, which is consistent with their presumed domain knowledge, the high stakes of making errors, and the assumptions of our model.

\subsection{The aggregate risk preference measure is representative}
\label{sec: aggregation}

Finally, we test whether the aggregate risk preference measure we derive is representative of the underlying individual risk preference distribution. Our overarching goal in this paper is to devise a method to measure individual risk preferences in the past. This is not realistic in our setting (or likely in future work delving into the deeper past) given data constraints, as the finest data we can obtain at scale is at the county level. The next best thing would be to provide evidence that the aggregate measure is representative of the underlying individual distribution, as we do here.

To do this we use the Kansas farm-level dataset, where we can estimate risk aversion both at the farm level and at the township level (as each township consists of hundreds of farms, and we observe the universe of farms in each township in the dataset). For each farm $f$ in township $w$ and year $t$, we recover a farm-level risk-aversion coefficient, $\widehat{\rho}_{f,t}$, from the model's first-order condition using that farm's crop allocation. We then recover an analogous township-level estimate, $\widehat{\rho}_{w,t}$, after aggregating crop allocations across all farms in the township-year, analogous to our county-level estimates. To match the spatial pooling used for our county-level estimation, we pool first-order-condition moments locally using each farm's neighbors in the within-township census enumeration order in each year, a standard proxy for geographic proximity in door-to-door historical censuses.\footnote{We pool only within a wave, and not across waves, because individual farms cannot be fully tracked over time due to changes in proprietorship, mergers, and other turnover. We also do not pool across townships, since doing so would break the within-place comparison between a township and its own farms. See \Cref{app: kansasspatial} for details.}

\begin{figure}[tbp]
\begin{adjustwidth}{-2cm}{-2cm}
\begin{center}
\caption{Comparing risk aversion at the farm and township levels in Kansas \label{fig:histrrakansascomp}}
\subfloat[Mean farm $\widehat{\rho}$ $-$ Township $\widehat{\rho}$]{\includegraphics[width=0.5\textwidth]{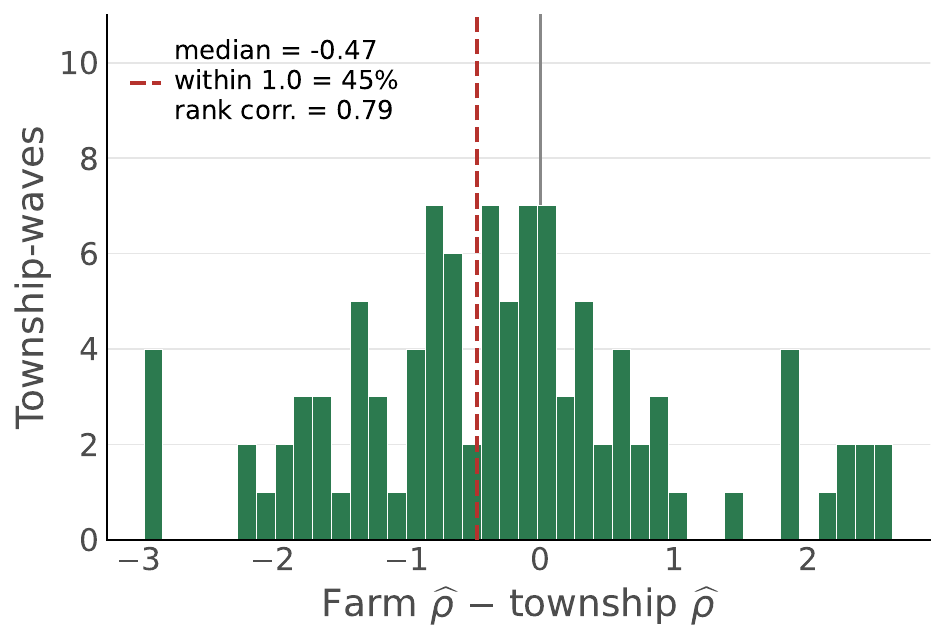}} $\thinspace$
\subfloat[Median farm $\widehat{\rho}$ $-$ Township $\widehat{\rho}$]{\includegraphics[width=0.5\textwidth]{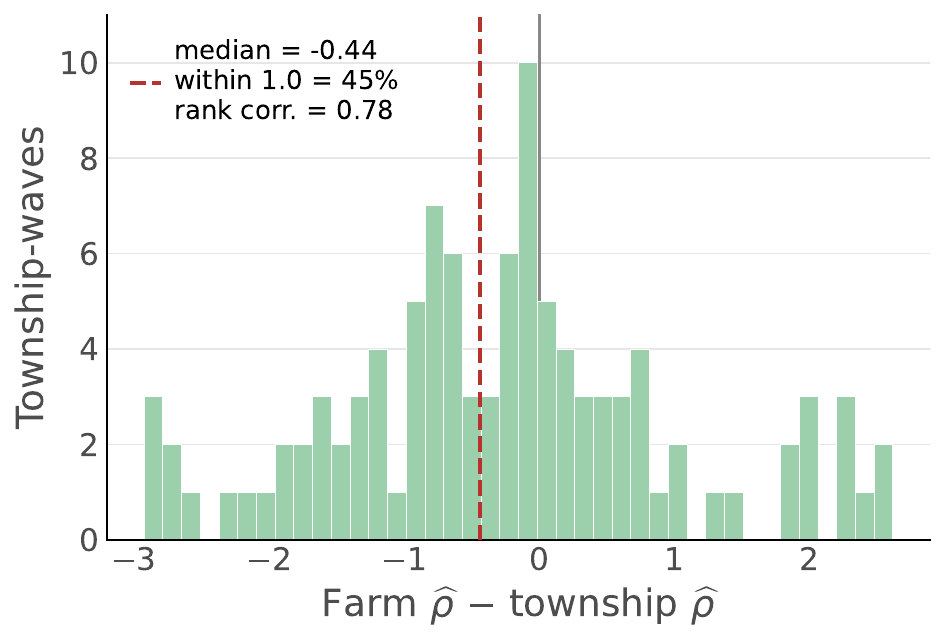}}
\end{center}
\end{adjustwidth}
{\footnotesize \textbf{Notes:} These panels plot the difference between the mean and median of farm $\widehat{\rho}$ in a township-year and that township's $\widehat{\rho}$ in that year. The analysis uses the farm-level data of \cite{sylvester2002} for the 24 Kansas townships in 1895, 1905, 1915, 1920, 1925, and 1930.}
\end{figure}

\Cref{fig:histrrakansascomp} compares the farm- and township-level estimates within the same township-year. For each township-year, we compute the distance between the township-level estimate and the mean farm-level estimate, $Dist_{mean,w,t}
=
\mathrm{Mean}_f(\widehat{\rho}_{f,t})
-
\widehat{\rho}_{w,t},$
and the analogous distance using the median farm-level estimate, $
Dist_{median,w,t}
=
\mathrm{Median}_f(\widehat{\rho}_{f,t})
-
\widehat{\rho}_{w,t}.
$
The results indicate a substantial level of concordance between the aggregate measure of risk preferences and the average individual risk preference. Both distributions are centered close to zero, with dispersion reflecting the noise in estimating risk aversion from a single township-year crop allocation. More importantly, the measures exhibit strong rank agreement: across township-years, the township-level estimate has a rank correlation of $0.79$ and $0.78$ with the mean and median farm-level estimates, respectively, with wave-specific correlations ranging from $0.65$ to $0.93$ (\Cref{fig:disttwscatter}).  This suggests that our county-level $\widehat{\rho}$ measures reflect the average risk preferences of the farmers within them, rather than being artifacts of aggregation. \Cref{fig:histrrakansas} reports the corresponding farm- and township-level distributions of \(\widehat{\rho}\).

\section{Main results}
\label{sec: descriptive results}

In this section we present our risk aversion estimates, and compare them to observable characteristics and contemporaneous measures of risk-taking behavior in other domains.

\begin{figure}[!ht]\begin{center}
\caption{Estimated county median risk aversion, 1889 -- 1929}\label{fig:rhaoverviewmed}
\vspace{0.15cm}

\subfloat[Median risk aversion across waves]{\includegraphics[width=0.49\textwidth]{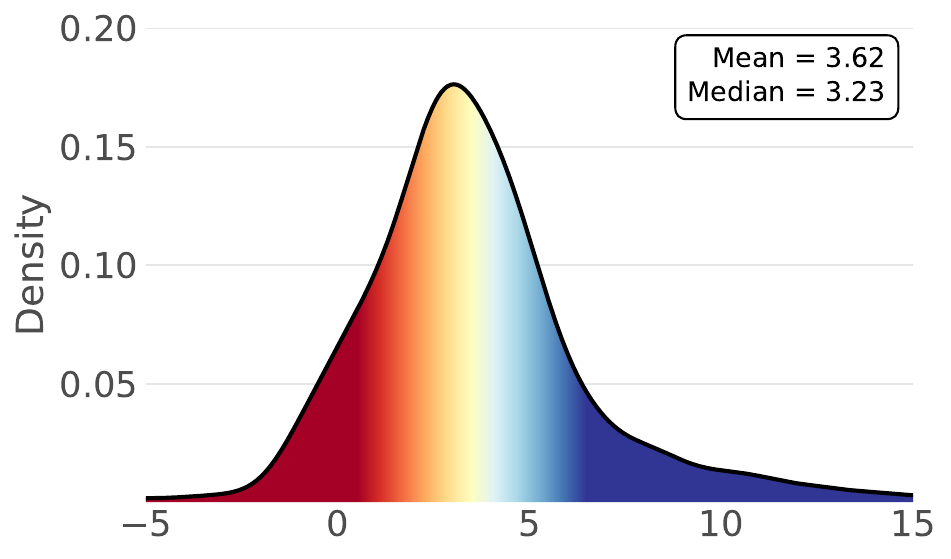}}\hfill
\subfloat[Median risk aversion over time]{\includegraphics[width=0.49\textwidth]{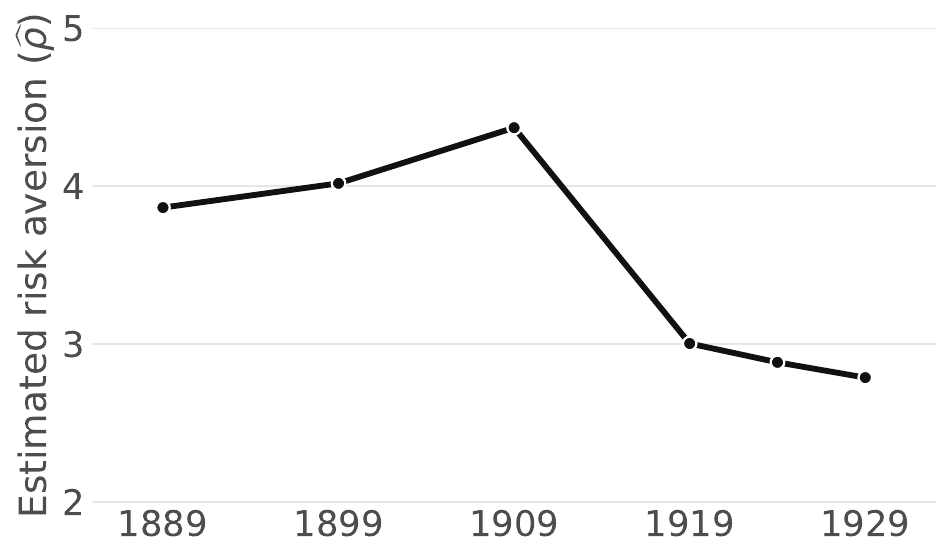}}
\vspace{0.1cm}
\end{center}
{\footnotesize\raggedright \textbf{Notes:} This figure shows the cross-sectional and temporal distribution of the estimated risk aversion coefficient, $\widehat{\rho}$. Panel (a) is the kernel density of each county's median $\widehat{\rho}$ across the waves in which it is observed. Panel (b) plots the cross-county median $\widehat{\rho}$ in each wave.\par}
\end{figure}

\subsection{Cross-sectional, temporal, and spatial distribution of \texorpdfstring{$\widehat{\rho}$}{rho-hat}}
\label{sec: results descriptive}

We first describe the distributional characteristics of estimated risk aversion ($\widehat{\rho}$) in our county-level US data. Panel (a) of \Cref{fig:rhaoverviewmed} plots the kernel density of counties' median $\widehat{\rho}$ across observed waves. Three patterns stand out. First, the distribution is unimodal and centered on moderate risk aversion: the mean of the county medians is $3.62$, and the median of medians is $3.23$. Second, there is substantial dispersion around this center, including a small mass of counties with a median risk-seeking estimate (7.9\% with $\widehat{\rho}<0$), and a long tail of counties with high median risk aversion (23.3\% with $\widehat{\rho}>5$). Third, most estimates (60.7\%) lie between $1$ and $5$, placing them within the range commonly used in macroeconomic models and comparable to modern estimates of risk preferences.

Panel (b) of \Cref{fig:rhaoverviewmed} plots national median risk aversion over time. The national median rises gradually from 3.87 in 1889 to 4.37 1909, before falling sharply to 3.00 in 1919, and 2.79 by 1929. This pattern is broadly consistent with macroeconomic trends over the period, specifically the volatile and deflationary environment at the turn of the century, and the commodity price boom and speculative fever in agriculture around World War I.

\begin{figure}[htbp]\centering
\caption{Estimated risk aversion by wave, 1889 -- 1929}\label{fig:map6}
\vspace{0.15cm}

\includegraphics[width=0.49\textwidth]{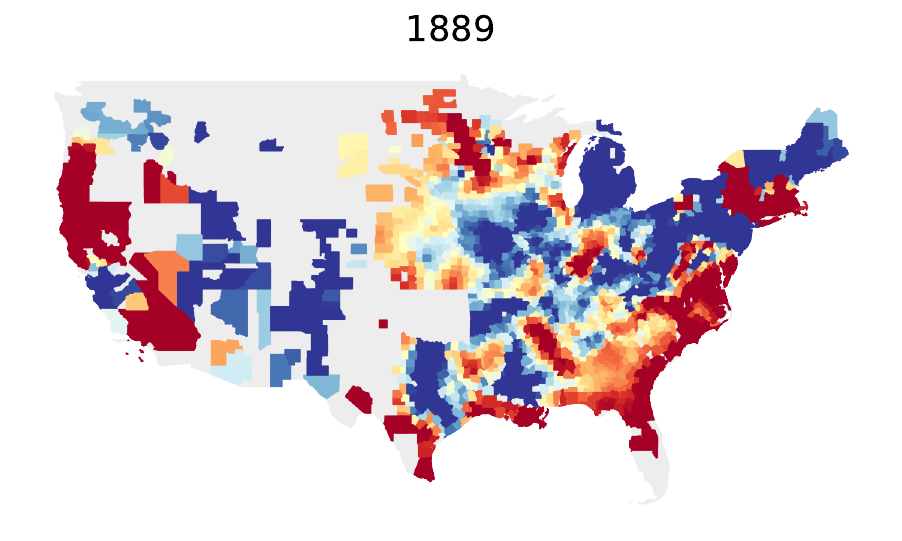}\hfill
\includegraphics[width=0.49\textwidth]{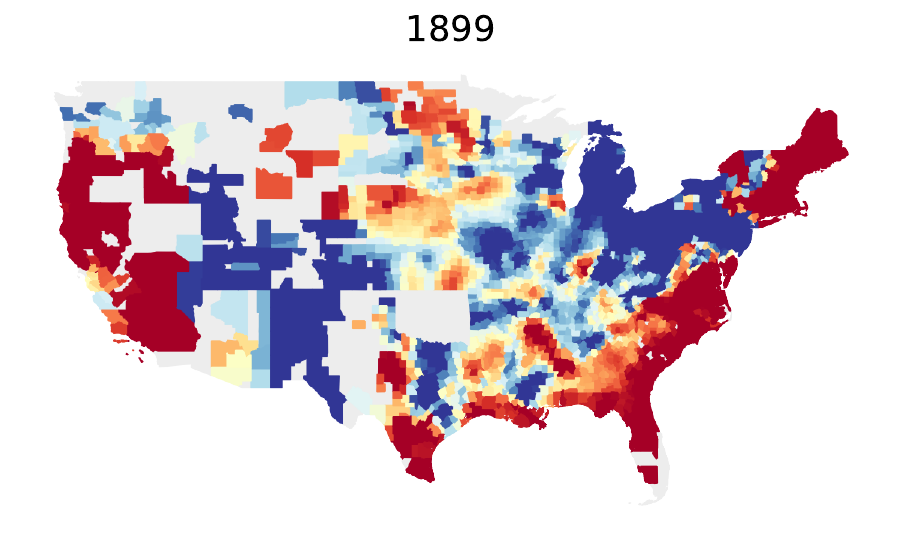}

\includegraphics[width=0.49\textwidth]{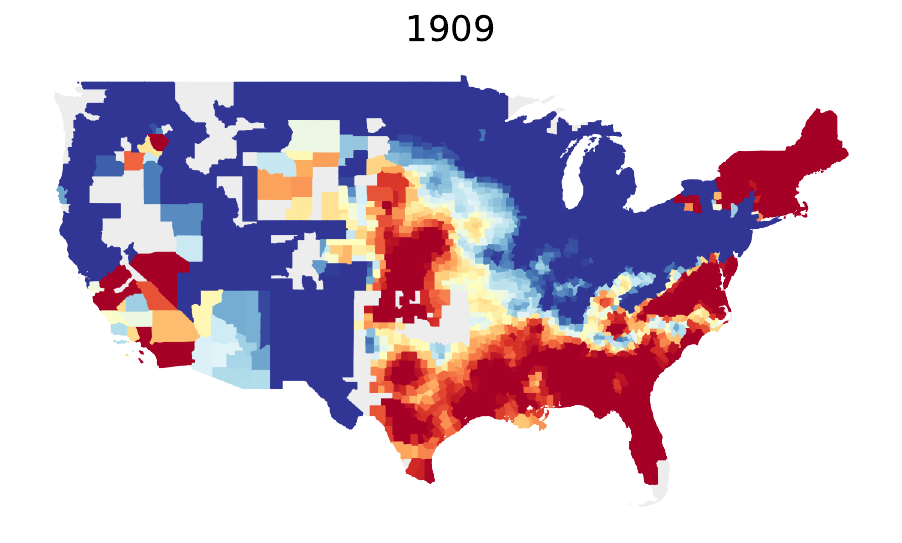}\hfill
\includegraphics[width=0.49\textwidth]{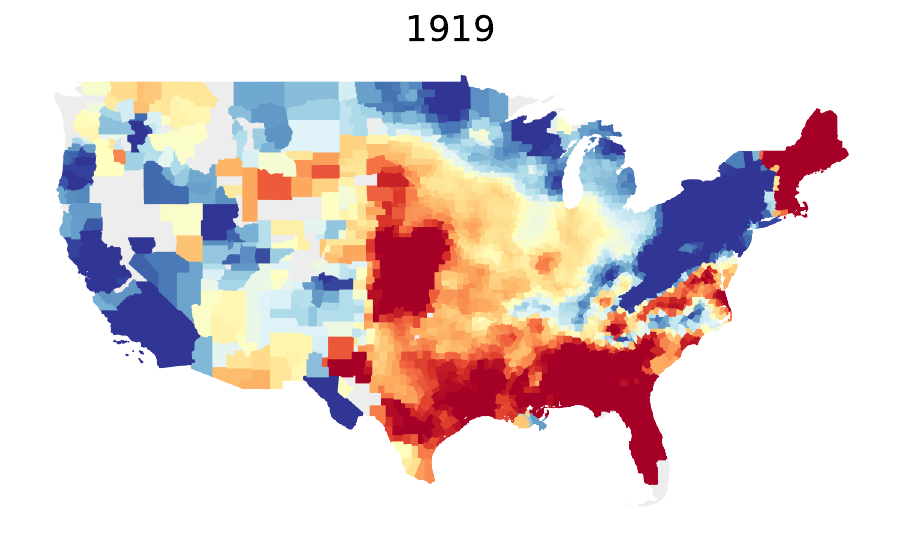}

\includegraphics[width=0.49\textwidth]{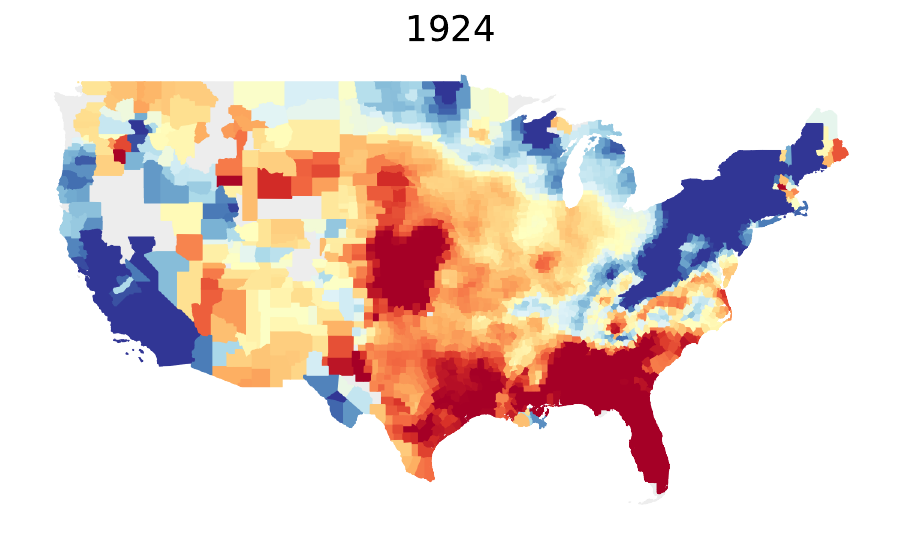}\hfill
\includegraphics[width=0.49\textwidth]{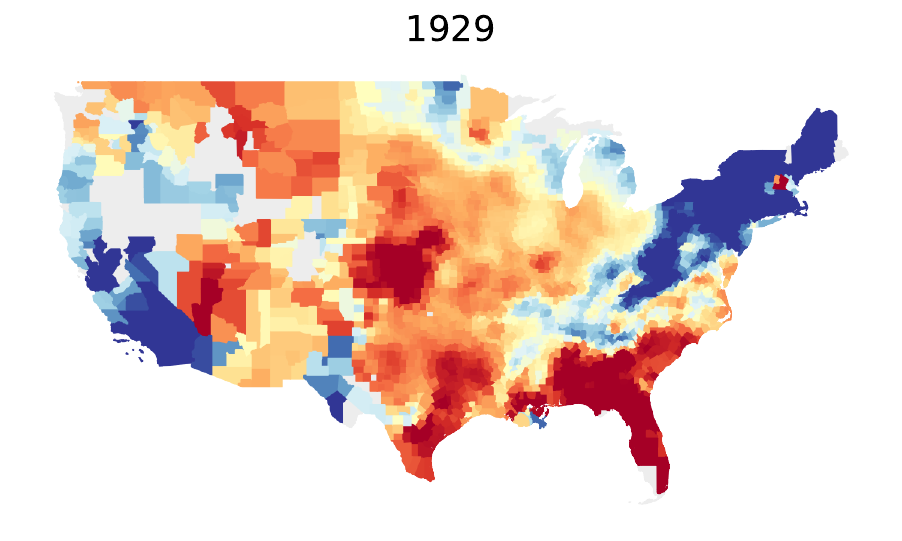}

\vspace{0.2cm}
\includegraphics[width=0.72\textwidth]{rho_cbar.pdf}
\vspace{0.1cm}

{\footnotesize\raggedright \textbf{Notes:} This figure shows the estimated risk aversion coefficient, $\widehat{\rho}$, across counties in different waves. Blue indicates more risk-averse, red more risk-loving, and grey no estimate.\par}

\end{figure}

We next examine spatial variation in risk aversion. \Cref{fig:rc5medmap} maps each county's median $\widehat{\rho}$ across the waves in which it is observed. This reveals some striking spatial patterns . The most risk-loving counties form a broad, contiguous zone across the Deep South and Gulf Coast (Georgia, Alabama, Mississippi, Florida, and the Carolinas) that extends westward through the central and southern Great Plains. The most risk-averse counties cluster in the Northeast and the Mid-Atlantic (New York, New Jersey, and Pennsylvania), together with the Great Lakes states, Appalachian West Virginia, and the Pacific coast. Nevertheless, there is substantial local variation across counties within each region.

These regional patterns are persistent but not fixed over time, as can be seen in \Cref{fig:map6}. A wide arc of the interior --- the Great Lakes, the upper Midwest and northern Plains, and the Mountain West --- grows more risk-averse through 1909 and retreats sharply afterward, consistent with the aggregate temporal pattern in \Cref{fig:rhaoverviewmed}. By contrast, in the first half of the sample period the south and the central plains, which generally exhibit low risk aversion throughout, become more risk-seeking (widening the gap with other regions), before reversing and becoming more risk averse in the 1920s.\footnote{\Cref{app: results} reports additional descriptive results: (1) a full set of distributional summary statistics, both pooled and by wave; (2) maps and distributions for the cross-wave county means; (3) density plots for the distributions in each wave; and (4) maps of the cross-wave change in measured risk aversion by county.}

\subsection{Correlates of risk aversion}
\label{sec: results correlates}

We next examine how risk aversion correlates with observable county characteristics. We regress $\widehat{\rho}$ on a set of county covariates in a pooled specification with state-by-year fixed effects. The covariates span geography and demographics, the economic structure of the local farm sector and indicator variables for the county's cultivation cluster, and two measures of production concentration.\footnote{Demographic covariates (population, rural, white, foreign-born, and female shares) are from the decennial Census of Population. All farm-structure and concentration covariates are from the Census of Agriculture. Elevation and ruggedness are from the NOAA ETOPO 2022 relief model. Distance to railroad is from the historical railroad GIS of \cite{Atack2013}. Distance to the nearest major city is measured to the 50 largest cities by 1900 population.} Both $\widehat{\rho}$ and the covariates are standardized, so each coefficient reports the change in $\widehat{\rho}$, in standard-deviation units, associated with a one-standard-deviation increase in the corresponding variable. \Cref{tab:corrprodstacked} reports the results.

\begin{table}[t]
\centering
\caption{Correlates of risk aversion \label{tab:corrprodstacked}}
\vspace{.1cm}
\resizebox{0.85\textwidth}{!}{%
\begingroup
\setlength\lightrulewidth{\arrayrulewidth}\setlength\cmidrulewidth{\arrayrulewidth}%
\begin{tabular}{c}
\hline
\noalign{\vskip 4pt}
\begin{tabular}[t]{lc}
\multicolumn{2}{l}{\textit{A: Geography and demographics}} \\
\cmidrule(lr){1-2}
Latitude & 0.25*** (0.06) \\
\addlinespace[2pt]
Longitude & -0.11 (0.12) \\
\addlinespace[2pt]
Elevation & 0.01 (0.03) \\
\addlinespace[2pt]
Ruggedness & 0.02 (0.03) \\
\addlinespace[2pt]
Dist.\ to major city & 0.02 (0.03) \\
\addlinespace[2pt]
Dist.\ to railroad & -0.00 (0.01) \\
\addlinespace[2pt]
Log population & -0.03 (0.03) \\
\addlinespace[2pt]
Rural share & -0.01 (0.01) \\
\addlinespace[2pt]
White share & 0.13*** (0.02) \\
\addlinespace[2pt]
Foreign-born share & 0.03 (0.03) \\
\addlinespace[2pt]
Female share & -0.03* (0.01) \\
\addlinespace[14pt]
\multicolumn{2}{l}{\textit{C: Production concentration}} \\
\cmidrule(lr){1-2}
Cropland HHI & -0.23*** (0.03) \\
\addlinespace[2pt]
Crop-value HHI & 0.11*** (0.03) \\
\end{tabular}\hskip 1.8em\begin{tabular}[t]{lc}
\multicolumn{2}{l}{\textit{B: Economic structure}} \\
\cmidrule(lr){1-2}
Log farm value & 0.06* (0.03) \\
\addlinespace[2pt]
Log crop-output value & -0.04 (0.04) \\
\addlinespace[2pt]
Log no.\ farms & 0.01 (0.03) \\
\addlinespace[2pt]
Land-in-farms share & -0.03* (0.02) \\
\addlinespace[2pt]
Improved-land share & 0.14*** (0.03) \\
\addlinespace[2pt]
Log irrigated farms & 0.00 (0.01) \\
\addlinespace[2pt]
Owner-operated share & 0.03 (0.02) \\
\addlinespace[2pt]
Avg.\ farm size & 0.03** (0.01) \\
\addlinespace[2pt]
Farm-size Gini & 0.10*** (0.02) \\
\addlinespace[2pt]
Corn output share & 0.02 (0.02) \\
\addlinespace[2pt]
Core-crop improved share & 0.02 (0.02) \\
\addlinespace[2pt]
Cotton belt & 0.22*** (0.05) \\
\addlinespace[2pt]
Wheat belt & 0.15*** (0.05) \\
\addlinespace[2pt]
Hay belt & 0.14 (0.09) \\
\end{tabular} \\
\midrule
Observations: 15,619 \\
\addlinespace[2pt]
State $\times$ Year fixed effects \\
\noalign{\vskip 4pt}
\hline
\end{tabular}
\endgroup

}
\begin{minipage}{\textwidth}
\smallskip\footnotesize{\textbf{Notes:}  This table shows the coefficients from regressing the standardized relative risk aversion coefficient ($\hat{\rho}$), with higher values indicating greater risk aversion, on a set of standardized covariates, so each coefficient is the change in $\widehat{\rho}$, in standard-deviation units, per one-standard-deviation increase in the covariate. The three panels come from a single pooled regression over the six crop-year waves, 1889--1929. 100 km Conley spatial-HAC errors in parentheses. *** p$<$0.01, ** p$<$0.05, * p$<$0.1.}
\end{minipage}
\end{table}

Among the geographic and demographic covariates, only latitude and racial composition are strongly related to $\widehat{\rho}$: more northern counties within each state are markedly more risk-averse, as are counties with a larger white population. More female-heavy counties are marginally more risk-loving. The remaining covariates --- including longitude, elevation, ruggedness, population, the rural and foreign-born shares, and distance to the nearest city and railroad --- are small and statistically indistinguishable from zero. The latter two null correlations, in particular, indicate that measured risk aversion is not proxied by market access.

Among the economic variables, higher farmland value is weakly associated with higher $\widehat{\rho}$. Counties with a larger improved-land share, larger average farms, and greater dispersion in farm size are more risk-averse, as are counties in the cotton and wheat belts (relative to the held-out corn belt). Risk aversion is essentially uncorrelated with crop-output value, the number of farms, the owner-operated share, and irrigation. It is also uncorrelated with corn's share of core-crop output value (the numeraire crop) and with the share of improved land planted in the core crops. This is reassuring, since it indicates that the estimates are not driven by the local prominence of the numeraire crop or by how much of a county's land the analysis covers.

Turning to the two production concentration measures, we find that risk aversion is negatively correlated with cropland concentration, in line with the model's basic mechanism where more risk-averse counties spread acreage to reduce portfolio variance. We also find that risk aversion is positively correlated with crop-value concentration. This means that more  risk-averse counties devote more land to low-revenue crops, so their income remains concentrated in their high-value crops even as cropland concentration falls. These results are consistent with the empirical pattern of expected returns we estimate. Since crop returns in general are positively correlated with each other (\Cref{fig: returncorr}), the most efficient mechanism for reducing risk is to shift from high-mean high-variance crops to low-mean low-variance crops (\Cref{fig: retmvavg}), rather than to another uncorrelated but risky crop. Note that the two variables explain only about 10\% of the variation in $\widehat{\rho}$, which means they are at best very noisy proxies for risk aversion.\footnote{These patterns relate our measure to \citet{Fiszbein2022}, who shows that a county's agricultural diversity in 1860 (one minus the HHI of its output shares across all agricultural products) had large and persistent effects on long-run development, by affecting the acquisition of new skills and ideas. Observed diversity combines the climate-given menu of feasible products with farmers' allocations across that menu. Fiszbein's object is the menu: his instrument, potential diversity, is constructed from climate-based attainable yields through a profit-maximization model of crop choice, so that determinants of allocation other than climate-driven productivity (which, as he notes, may include risk aversion) are excluded from the identifying variation by construction. Our object is the complement: holding each county's menu of return distributions fixed, we recover the risk preferences that govern allocation within it. The two designs thus decompose the same variation into the breadth of the menu and the position taken on it.}

In the analyses that follow, we use these correlates as controls when studying the relationship between risk aversion and other outcomes. Our baseline specifications always control for the geography and demographic covariates, which are plausibly predetermined. Because the economic covariates may be endogenous to risk aversion, we include them only in expanded specifications. We do not control for the two production-concentration measures, since they are closely related to the crop allocations used to identify $\widehat{\rho}$.

\subsection{Relationship to risk-taking in other domains}
\label{sec: results other risk}

We next validate the risk-aversion measures against contemporaneous indicators of risk-taking. This exercise is necessarily limited: if direct measures of risk preferences existed at scale, there would be little need for our approach. Still, we use data on risk-taking behavior in other domains, allowing us to ask whether counties and individuals classified by our method as more risk-averse also behaved more cautiously in other areas.

\subsubsection{Financial risk-taking: farm mortgages and WWI Liberty Bonds}
\label{sec: model validation}

Based on our framework, measured risk aversion should be closely related to financial risk-taking behavior. We consider two outcomes in this domain: (1) farm mortgages, the primary form of leverage available during this period; and (2) adoption of a novel financial instrument, World War I Liberty Bonds. We regress these outcomes on $\widehat{\rho}$ and report the results in \Cref{tab:downgrouped}. Both the outcomes and $\widehat{\rho}$ are standardized, so coefficients report the change in the outcome, in standard deviation units, associated with a one-SD increase in $\widehat{\rho}$.

\begin{table}[t]
\caption{Risk aversion and financial risk-taking by wave\label{tab:downgrouped}}
\begin{footnotesize}
\begin{center}
\begingroup
\setlength\lightrulewidth{\arrayrulewidth}\setlength\cmidrulewidth{\arrayrulewidth}%
\begin{tabular}{lcccc} \hline
\noalign{\vskip 4pt}
 & \multicolumn{2}{c}{\makecell{Share farms\\mortgaged}} & \multicolumn{2}{c}{\makecell{WWI Liberty\\Bonds}} \\
\cmidrule(lr){2-3} \cmidrule(lr){4-5}
 & (1) & (2) & (3) & (4) \\ \midrule
\multicolumn{1}{c}{$\widehat\rho_{1899}$} & 0.02 & -0.04 &  &  \\
 & (0.05) & (0.05) &  &  \\
\addlinespace[3pt]\cmidrule(lr){1-5}\addlinespace[3pt]
\multicolumn{1}{c}{$\widehat\rho_{1909}$} & 0.02 & -0.02 &  &  \\
 & (0.03) & (0.02) &  &  \\
\addlinespace[3pt]\cmidrule(lr){1-5}\addlinespace[3pt]
\multicolumn{1}{c}{$\widehat\rho_{1919}$} & 0.01 & -0.11** & -0.15** & -0.24*** \\
 & (0.06) & (0.04) & (0.06) & (0.06) \\
\addlinespace[3pt]\cmidrule(lr){1-5}\addlinespace[3pt]
\multicolumn{1}{c}{$\widehat\rho_{1929}$} & 0.02 & -0.15** &  &  \\
 & (0.09) & (0.07) &  &  \\
\midrule
Observations & 2,644 & 2,638 & 1,256 & 1,252 \\
$R^2$ & 0.655 & 0.754 & 0.719 & 0.747 \\
\addlinespace[4pt]
Fixed effects & \multicolumn{2}{c}{\rule[0.5ex]{2.0em}{\arrayrulewidth}\ State\ \rule[0.5ex]{2.0em}{\arrayrulewidth}} & \multicolumn{2}{c}{\rule[0.5ex]{2.0em}{\arrayrulewidth}\ State\ \rule[0.5ex]{2.0em}{\arrayrulewidth}} \\
\addlinespace[4pt]
Geog./Demog.\ controls & X & X & X & X \\
Economic controls &  & X &  & X \\
\noalign{\vskip 4pt}
\hline
\end{tabular}
\endgroup

\end{center}
\end{footnotesize}
\vspace{0.25cm}
\footnotesize{\textbf{Notes:} This table regresses two indicators of financial risk-taking on the risk-aversion estimates. Columns (1)--(2) use the share of a county's owner-operated farms carrying mortgage debt;
columns (3)--(4) use county subscription rates to the fourth World War I Liberty Bond drive. Each outcome is regressed on contemporaneous risk aversion $\widehat{\rho}$ separately by census wave, given by the row. The relative risk aversion coefficient ($\widehat{\rho}$) and the outcomes are standardized. Columns (1) and (3) include geographic and demographic controls; columns (2) and (4) add the full economic-control vector. Controls are specified in \Cref{tab:corrprodstacked}. Because columns (1)--(2) stack four wave-specific regressions, their number of observations and $R^2$ rows report the average across the four regressions; for the full-controls specification the wave-specific values are 1899 ($N{=}2{,}473$, $R^2{=}0.72$), 1909 ($N{=}2{,}678$, $R^2{=}0.72$), 1919 ($N{=}2{,}709$, $R^2{=}0.79$), and 1929 ($N{=}2{,}690$, $R^2{=}0.79$). Columns (3)--(4) are single 1919 cross-sections.  All specifications include state fixed effects. 100 km Conley spatial-HAC standard errors in parentheses. *** p$<$0.01, ** p$<$0.05, * p$<$0.1.}
\end{table}

We begin with farm leverage, measured by the share of a county's farms carrying mortgage debt in the Ag Census in 1899, 1909, 1919, and 1929. Columns (1) and (2) show that more risk-averse counties had fewer mortgaged farms in the war and postwar waves. Conditional on the full economic-control vector, a one-standard-deviation increase in $\widehat{\rho}$ is associated with a $0.11$ standard-deviation lower mortgaged share in 1919 and a $0.15$ standard-deviation lower share in 1929 (column 2). The relationship is also negative in earlier waves, but not statistically significant. This pattern is consistent with more risk-averse counties taking on less leverage during the period of debt-financed land speculation that preceded the postwar collapse in farm prices and the foreclosures that followed \citep{Rajan2015}.

We next turn to financial behavior outside agriculture. To finance the First World War, the federal government ran four public bond drives between 1917 and 1919. Liberty Bonds, which were purchased by at least one third of the adult U.S. population were a highly novel financial instrument, likely the first that many individuals owned outside of a bank account (\cite{Brunet2025}). Although marketed as safe investments, in reality they delivered volatile real returns after the war. We use county-level subscription rates to the fourth (and largest) bond drive, using data from \cite{hilt2020}, who provide subscription rates from Federal Reserve Banks for approximately 48\% of the counties in our data. Since this bond drive took place in late 1918, we link it to measured risk aversion in 1919. 

Columns (3) and (4) of \Cref{tab:downgrouped} show a strong negative relationship between county-level subscription rates in the fourth drive and risk aversion. A one-standard-deviation increase in $\widehat{\rho}_{1919}$ lowers bond participation by $0.15$ to $0.24$ standard deviations. Risk-tolerant households appear to have been quicker to take up an unfamiliar financial product. Because this behavior lies outside the agricultural domain used to estimate
$\widehat{\rho}$, it provides a particularly useful external validation that our measure does indeed capture the underlying preference.

\subsubsection{Risk-sharing: Kansas fraternal organization membership \label{sec: fratorg}}

Next, we turn to the Kansas data, where we have information on membership in fraternal organizations at the individual farmer level. In nineteenth-century America, these organizations served as both social institutions and informal insurance networks: members paid dues into a common fund and, in return, were entitled to benefits in the event of death, disability, or hardship (\cite{Beito2000}). If our estimates capture genuine variation in risk preferences, we would expect more risk-averse farmers to participate more in this form of informal insurance.

To explore this, we draw on digitized registers of deceased members of the four largest fraternal organizations active in Kansas during this period, compiled by \cite{ang2026}: the Independent Order of Odd Fellows (IOOF), its women's auxiliary the Rebekah Lodge (RL), the Ancient Order of United Workmen (AOUW), and the Knights of Pythias (KoP). We identify the farmers in our panel as lodge members if a death-notice matching their name and township is found.

\begin{table}[t]
\caption{Fraternal lodge membership and farm-level risk aversion: Kansas \label{tab: rralodgekansas}}
\begin{footnotesize}
\begin{center}
\begin{tabular}{lcc}
\hline\noalign{\vskip 4pt}
 & \multicolumn{2}{c}{Farm $\widehat{\rho}$} \\
\cmidrule(lr){2-3}
& (1) & (2) \\
\midrule
Lodge member & 0.619*** &  \\
& (0.179) &  \\
\addlinespace[2pt]
Independent Order of Odd Fellows &  & 0.488** \\
&  & (0.229) \\
\addlinespace[2pt]
Rebekah Lodge &  & 0.537* \\
&  & (0.317) \\
\addlinespace[2pt]
Ancient Order of United Workmen &  & 3.462*** \\
&  & (0.980) \\
\addlinespace[2pt]
Knights of Pythias &  & 3.435 \\
&  & (2.413) \\
\midrule
Observations & 8,476 & 8,476 \\
$R^2$ & 0.128 & 0.128 \\
\addlinespace[2pt]
Township FE & X & X \\
\noalign{\vskip 4pt}\hline
\end{tabular}

\end{center}
\end{footnotesize}
\vspace{0.25cm}
{\footnotesize \textbf{Notes:} This table regresses the estimated farmer-level $\widehat{\rho}$ (averaged across a farmer's waves and winsorized at the 1st and 99th percentiles) on fraternal-lodge membership. Both columns include township fixed effects, with standard errors clustered at the township level in parentheses. *** p$<$0.01, ** p$<$0.05, * p$<$0.1.}
\end{table}

\Cref{tab: rralodgekansas} reports regressions of the farmer-level risk-aversion coefficient \(\widehat{\rho}\), averaged across a farmer's observed waves when they appear more than once, and winsorized at the 1st and 99th percentiles, on indicators for fraternal-organization membership. All specifications control for township fixed effects, and standard errors are clustered at the township level.

Column (1) shows that more risk-averse farmers were more likely to participate in fraternal organizations. Column (2) separates membership by organization. The relationship is positive for all four. Point estimates are by far the largest for the AOUW and the KoP, the two orders most oriented toward financial risk-sharing (the AOUW was built around member-financed death benefits and the KoP operated an insurance branch through its endowment rank (\cite{Beito2000}). Of the two, only the AOUW is precisely estimated, as the KoP is identified from fewer matched members. Among the larger, more socially-oriented orders, membership in the IOOF is statistically significant, and in its Rebekah auxiliary marginally significant, indicating that more risk-averse farmers also sorted into the broader mutual-benefit lodges (\cite{Beito2000}, \cite{Kaufman2003}).

\section{Risk aversion and technological adoption}
\label{sec: structural transformation}

As described in \Cref{sec: setting}, the two decades around the First World War brought rapid technological change to American agriculture. The tractor and other powered machines began to displace the draft animals that had long supplied farm power, while commercial fertilizers and other biological innovations became more widely available. This transition was highly uneven across counties, and its determinants remain a topic of debate. 

In this section we ask whether risk preferences help explain which counties led this technological transition. A large literature attributes the uneven diffusion of the tractor to farm scale, credit access, market conditions, and the pace of technical improvement (\cite{olmstead2000}, \cite{OlmsteadRhode2008}, \cite{Rajan2015}). Because tractors were expensive and their returns uncertain, however, risk aversion could also have slowed adoption. A related mechanism appears in the development literature on agricultural technology adoption, which emphasizes how risk preferences, learning, and uninsured risk shape whether and when farmers adopt new technologies and inputs (\cite{Foster1995}, \cite{Liu2013}, \cite{Karlan2014}). However, direct evidence on this channel is scarce as preferences are rarely observed. Pairing risk-aversion estimates with technology-adoption measures from the Ag Census allows us to provide suggestive evidence for it at scale.

\begin{table}[t]
\caption{Risk aversion and technological adoption\label{tab:fwdtd}}
\begin{footnotesize}
\begin{center}
\resizebox{\linewidth}{!}{\begingroup
\setlength\lightrulewidth{\arrayrulewidth}\setlength\cmidrulewidth{\arrayrulewidth}%
\begin{tabular}{lccccccccc} \hline
\noalign{\vskip 4pt}
 & \multicolumn{3}{c}{Tractors / farm} & \multicolumn{3}{c}{Share using fertilizer} & \multicolumn{3}{c}{Draft animals / farm} \\
\cmidrule(lr){2-4} \cmidrule(lr){5-7} \cmidrule(lr){8-10}
 & 1929 & 1924 & 1929 & 1929 & 1924 & 1929 & 1929 & 1924 & 1929 \\
\cmidrule(lr){2-2} \cmidrule(lr){3-3} \cmidrule(lr){4-4} \cmidrule(lr){5-5} \cmidrule(lr){6-6} \cmidrule(lr){7-7} \cmidrule(lr){8-8} \cmidrule(lr){9-9} \cmidrule(lr){10-10}
 & (1) & (2) & (3) & (4) & (5) & (6) & (7) & (8) & (9) \\ \midrule
\multicolumn{1}{c}{$\widehat\rho_{1919}$} & -0.17** &  &  & -0.08* &  &  & 0.03 &  &  \\
 & (0.08) &  &  & (0.05) &  &  & (0.06) &  &  \\
\addlinespace[2pt]
\multicolumn{1}{c}{$\widehat\rho_{1919}$} &  & -0.15** &  &  & -0.13** &  &  & -0.04 &  \\
 &  & (0.07) &  &  & (0.05) &  &  & (0.05) &  \\
\addlinespace[2pt]
\multicolumn{1}{c}{$\widehat\rho_{1924}$} &  &  & -0.20*** &  &  & -0.14** &  &  & -0.03 \\
 &  &  & (0.08) &  &  & (0.06) &  &  & (0.07) \\
\midrule
Observations & 2,709 & 2,709 & 2,649 & 2,709 & 2,709 & 2,649 & 2,638 & 2,638 & 2,649 \\
$R^2$ & 0.713 & 0.692 & 0.754 & 0.777 & 0.772 & 0.785 & 0.728 & 0.775 & 0.748 \\
\addlinespace[4pt]
Fixed effects & \multicolumn{3}{c}{\rule[0.5ex]{3.3em}{\arrayrulewidth}\ State\ \rule[0.5ex]{3.3em}{\arrayrulewidth}} & \multicolumn{3}{c}{\rule[0.5ex]{3.3em}{\arrayrulewidth}\ State\ \rule[0.5ex]{3.3em}{\arrayrulewidth}} & \multicolumn{3}{c}{\rule[0.5ex]{3.3em}{\arrayrulewidth}\ State\ \rule[0.5ex]{3.3em}{\arrayrulewidth}} \\
\addlinespace[4pt]
Geog./Demog.\ controls & X & X & X & X & X & X & X & X & X \\
Economic controls & X & X & X & X & X & X & X & X & X \\
\noalign{\vskip 4pt}
\hline
\end{tabular}
\endgroup
}
\end{center}
\end{footnotesize}
\vspace{0.25cm}
\footnotesize{\textbf{Notes:} This table regresses three technology-adoption outcomes on predetermined risk-aversion measures. Specifically, each column pairs an outcome --- tractors per farm (columns 1--3), share of farms using fertilizer (columns 4--6), and draft animals per farm (columns 7--9) --- measured in the year given by the column heading, with risk aversion from an earlier wave, $\widehat{\rho}_{1919}$ or $\widehat{\rho}_{1924}$, given by the row. The relative risk aversion coefficient ($\widehat{\rho}$) is standardized, with higher values indicating greater risk aversion. Outcomes are standardized as well. Draft animals comprise horses, mules, asses, and burros. All columns are estimated with state fixed effects and controls measured at the risk-aversion wave. Controls are specified in \Cref{tab:corrprodstacked}. 100 km Conley spatial-HAC errors in parentheses. *** p$<$0.01, ** p$<$0.05, * p$<$0.1.}
\end{table}

We begin with three outcomes. The first is tractors per farm, measured in 1924 and 1929, which captures adoption of the new mechanical power technology. The second is the share of farms using commercial fertilizer. Fertilizer was a smaller and more divisible investment than tractors, but its payoff also depends on uncertain soil, weather, and crop-price conditions, making it a canonical risky agricultural input in the development literature. The third is draft animals per farm, the older source of farm power. We include this outcome as a benchmark: if more risk-averse counties were simply poorer, less commercialized, or slower to adopt all inputs, they should also have fewer draft animals. Each outcome is regressed on risk aversion measured in an earlier wave, 1919 or 1924, conditioning on our baseline geographic, demographic, and economic controls and state fixed effects.

\Cref{tab:fwdtd} presents the results. The three outcomes line up with a risk aversion-based interpretation of technology adoption. More risk averse counties operated fewer tractors and used commercial fertilizer on a smaller share of farms across specifications. By contrast, risk aversion is not systematically related to draft animals per farm. This pattern suggests that the estimates are not simply capturing lower input use. Instead, more risk averse counties lagged specifically in the adoption of newer, more uncertain agricultural technologies.

\begin{figure}[t]\centering
\caption{Farm mechanization and placebo technologies in 1929 by risk aversion in 1924}\label{fig:mech2429}
\vspace{0.15cm}

\subfloat[Farm mechanization]{\includegraphics[width=0.49\textwidth]{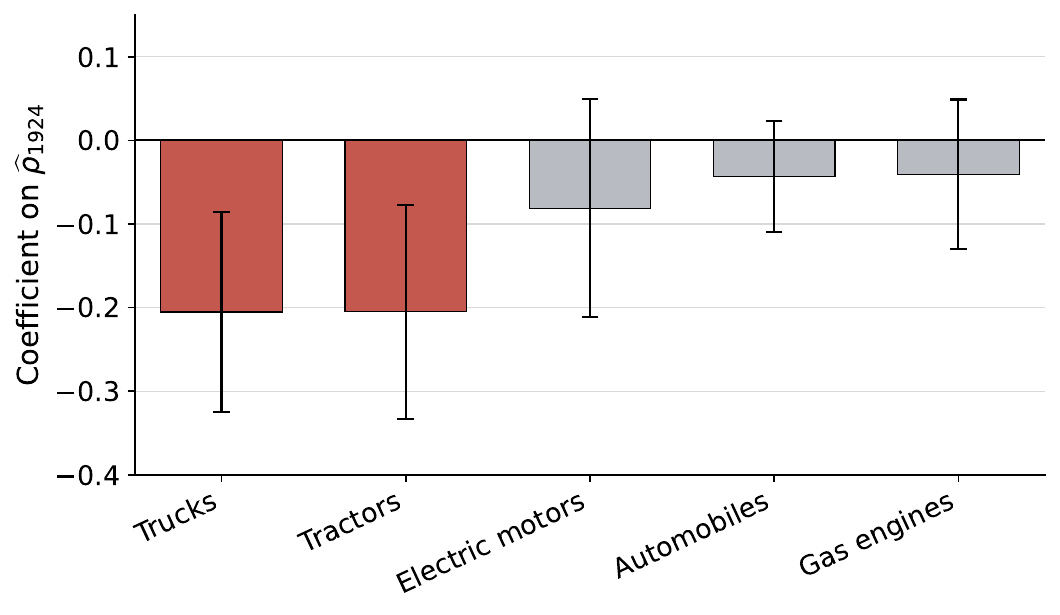}}\hfill
\subfloat[Placebo technologies]{\includegraphics[width=0.49\textwidth]{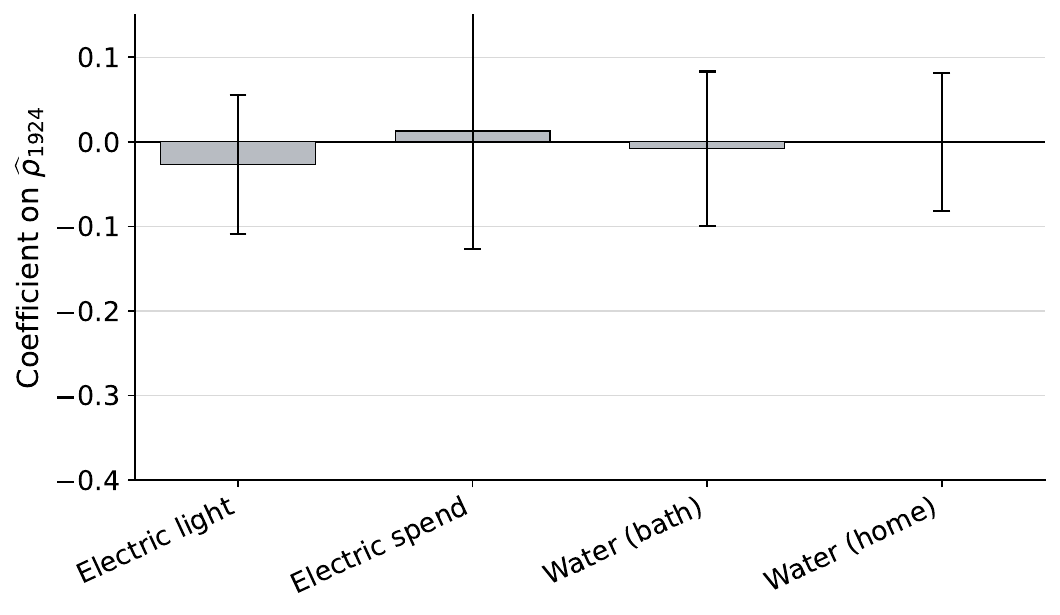}}
\vspace{0.1cm}

{\footnotesize\raggedright \textbf{Notes:} This figure shows the coefficients from regressing several 1929 technology-adoption outcomes on $\widehat{\rho}$ in 1924. Each bar is a separate regression. Panel~(a) reports five
farm-mechanization outcomes, measured per farm: motor trucks, tractors, electric motors, automobiles, and gas engines. Panel~(b) reports four household amenity placebos: the share of farms reporting electric light and power, electric light and power expenditure per farm, the share with water piped to a bathroom, and the share with water piped into the dwelling. $\widehat{\rho}$ is standardized, with higher values indicating greater risk aversion. Outcomes are standardized. Regressions include state fixed effects and controls measured at the risk-aversion wave. Controls are specified in \Cref{tab:corrprodstacked}. 90\% confidence intervals using 100~km Conley spatial-HAC SEs are also plotted.\par}
\end{figure}

We then examine whether this pattern extends beyond tractors and fertilizer to other forms of powered farm equipment which are measured in the 1929 Ag Census: gas engines, automobiles, electric motors, and trucks. For each technology, we regress the number of units per farm in a county in 1929 on risk aversion measured in 1924. Panel~(a) of \Cref{fig:mech2429} shows that more risk-averse counties also lagged in the adoption of trucks, another important farm technology which, like tractors, involved a large and lumpy investment. Risk averse counties also appear to lag in adopting other farm technology (electric motors, automobiles, gas engines), but the results are noisy and not statistically significant. 

Panel~(b) of \Cref{fig:mech2429} reports a placebo check using two household technologies less directly tied to farmers' own risky production choices: electric lighting (the share of farms reporting it and expenditure per farm) and running water (the share of farms with water piped into a bathroom or into the dwelling). Availability of these technologies was typically shaped by infrastructure decisions made by state or national authorities, utilities, and other actors outside the farm. Adoption of these technologies has a much smaller and statistically insignificant relationship with risk aversion. This supports the interpretation that our estimates capture attitudes toward risky productive investment, rather than a general propensity to live in technologically advanced places.

We conclude that our measure of risk aversion helps explain which counties led and which lagged in the adoption of new agricultural technologies, and particularly the shift from animal to mechanical power. With this, we also speak to the long-standing question in development economics of why farmers in the developing world are often slow to adopt beneficial technologies. Specifically, we provide evidence for a mechanism the literature has theorized but seldom been able to observe: under uncertain returns and uninsured downside risk, more risk-averse producers hold back from profitable but risky investments.

\section{Conclusions}
\label{sec: conclusion}

There can be little doubt about the risk preferences of George Ellery Hale, the man who built the Hooker telescope. Beginning in 1906, Hale spent eleven years making a half-million-dollar gamble on its construction, starting the project with less than 10\% of the necessary funding, and persisting even after its never-before-cast 100-inch optical glass disk arrived from France so badly flawed that his optician urged him to abandon it. Eventually his gamble paid off. On the night of November 1, 1917, the first image of a distant star came into focus in the telescope's mirror. The Hooker was the largest telescope in the world until 1949. Edwin Hubble used it to prove that there exist galaxies outside the Milky Way, and redshift data he collected from it with Milton Humason helped reveal that the universe is expanding.

Astronomers have a beautiful name for the first observation recorded from a new telescope. They call it ``first light.'' The first light from our behavioral telescope is certainly exciting to see: a first systematic measure of the risk preferences of people separated from us by over a century. The method we used to obtain this can be aimed at other times and places. One can imagine that it will be used to measure risk preferences deeper into history, and in the present in developing countries, where behavioral data is often difficult to obtain.

The most exciting part of this work, we believe, is not the specific results, or even the particular method we develop for recovering risk preferences. It is the broader idea behind the method, that sophisticated structural methods can be used as a form of \textit{behavioral cliometry} to reconstruct economic preferences in the deep past. This idea goes beyond risk preferences. Versions of our method for recovering time preferences and other behavioral primitives are possible. Perhaps one day, with enough such data, we will be able to trace not just the physical evolution of the economy, but also the behavioral evolution that accompanied and shaped it.

\vspace{1cm}

\begin{multicols}{2}
\begin{singlespacing}
\footnotesize
\bibliographystyle{apalike}
\bibliography{referencesarxiv}

\begin{thebibliography}{}

\bibitem[Acemoglu and Zilibotti, 1997]{AcemogluZilibotti1997}
Acemoglu, D. and Zilibotti, F. (1997).
\newblock Was {Prometheus} {Unbound} by {Chance}? {Risk}, {Diversification},
  and {Growth}.
\newblock {\em Journal of Political Economy}, 105(4):709--751.

\bibitem[Allen and Atkin, 2022]{AllenAtkin2022}
Allen, T. and Atkin, D. (2022).
\newblock Volatility and the {Gains} {From} {Trade}.
\newblock {\em Econometrica}, 90(5):2053--2092.

\bibitem[Anderson et~al., 1977]{Anderson1977}
Anderson, J.~R., Dillon, J.~L., and Hardaker, J.~B. (1977).
\newblock {\em Agricultural {Decision} {Analysis}}.
\newblock The Iowa State University Press.

\bibitem[Andrikogiannopoulou and Papakonstantinou,
  2020]{AndrikogiannopoulouPapakonstantinou2020}
Andrikogiannopoulou, A. and Papakonstantinou, F. (2020).
\newblock History-{Dependent} {Risk} {Preferences}: {Evidence} from
  {Individual} {Choices} and {Implications} for the {Disposition} {Effect}.
\newblock {\em The Review of Financial Studies}, 33(8):3674--3718.

\bibitem[Ang and Chinoy, 2026]{ang2026}
Ang, D. and Chinoy, S. (2026).
\newblock World {{War I}} and the {{Rise}} of the {{Ku Klux Klan}}.

\bibitem[Ankli and Olmstead, 1981]{AnkliOlmstead1981}
Ankli, R.~E. and Olmstead, A.~L. (1981).
\newblock The {Adoption} of the {Gasoline} {Tractor} in {California}.
\newblock {\em Agricultural History}, 55(3):213--230.

\bibitem[Antle, 1987]{Antle1987}
Antle, J.~M. (1987).
\newblock Econometric {Estimation} of {Producers}' {Risk} {Attitudes}.
\newblock {\em American Journal of Agricultural Economics}.

\bibitem[Araujo and Possebom, 2026]{araujoPotatoPotahtoFAOGAEZ2026a}
Araujo, R. and Possebom, V. (2026).
\newblock Potato {Potahto} in the {FAO}-{GAEZ} {Productivity} {Measures}?
  {Nonclassical} {Measurement} {Error} with {Multiple} {Proxies}.
\newblock arXiv:2502.12141 [econ.GN].

\bibitem[Atack, 2013]{Atack2013}
Atack, J. (2013).
\newblock On the {Use} of {Geographic} {Information} {Systems} in {Economic}
  {History}: {The} {American} {Transportation} {Revolution} {Revisited}.
\newblock {\em The Journal of Economic History}, 73(2):313--338.

\bibitem[Banerjee and Duflo, 2005]{BanerjeeDuflo2005}
Banerjee, A.~V. and Duflo, E. (2005).
\newblock Growth {Theory} through the {Lens} of {Development} {Economics}.
\newblock In Aghion, P. and Durlauf, S.~N., editors, {\em Handbook of
  {Economic} {Growth}, {Volume} {1A}}, pages 473--552. North-Holland
  (Elsevier), Amsterdam.

\bibitem[Bar-Shira et~al., 1997]{BarShira1997}
Bar-Shira, Z., Just, R., and Zilberman, D. (1997).
\newblock Estimation of farmers' risk attitude: an econometrie approach.
\newblock {\em Agricultural Economics}, 17(2-3):95--264.

\bibitem[Barseghyan et~al., 2013]{BarseghyanMolinari2013}
Barseghyan, L., Molinari, F., O'Donoghue, T., and Teitelbaum, J.~C. (2013).
\newblock The {Nature} of {Risk} {Preferences}: {Evidence} from {Insurance}
  {Choices}.
\newblock {\em American Economic Review}, 103(6):2499--2529.

\bibitem[Barseghyan et~al., 2016]{Barseghyan2016}
Barseghyan, L., Molinari, F., and Teitelbaum, J.~C. (2016).
\newblock Inference under stability of risk preferences.
\newblock {\em Quantitative Economics}, 7(2):367--409.

\bibitem[Barseghyan et~al., 2021]{BarseghyanMolinari2021}
Barseghyan, L., Molinari, F., and Thirkettle, M. (2021).
\newblock Discrete {Choice} under {Risk} with {Limited} {Consideration}.
\newblock {\em American Economic Review}, 111(6):1972--2006.

\bibitem[Becker et~al., 2020]{Becker2020}
Becker, A., Enke, B., and Falk, A. (2020).
\newblock Ancient {Origins} of the {Global} {Variation} in {Economic}
  {Preferences}.
\newblock {\em AEA Papers and Proceedings}, (110).

\bibitem[Beito, 2000]{Beito2000}
Beito, D.~T. (2000).
\newblock {\em From Mutual Aid to the Welfare State: Fraternal Societies and
  Social Services, 1890-1967}.
\newblock University of North Carolina Press, Chapel Hill.

\bibitem[Black, 1972]{Black1972}
Black, F. (1972).
\newblock Capital {Market} {Equilibrium} with {Restricted} {Borrowing}.
\newblock {\em The Journal of Business}, 45(3):444--455.

\bibitem[Bombardini and Trebbi, 2012]{BombardiniTrebbi2012}
Bombardini, M. and Trebbi, F. (2012).
\newblock Risk {Aversion} and {Expected} {Utility} {Theory}: {An} {Experiment}
  with {Large} and {Small} {Stakes}.
\newblock {\em Journal of the European Economic Association}, 10(6):1348--1399.

\bibitem[Bouchouicha and Vieider, 2019]{Bouchouicha2019}
Bouchouicha, R. and Vieider, F.~M. (2019).
\newblock Growth, entrepreneurship, and risk-tolerance: a risk-income paradox.
\newblock {\em Journal of Economic Growth}, 24(3):257--282.

\bibitem[Brunet et~al., 2025]{Brunet2025}
Brunet, G., Hilt, E., and Jaremski, M. (2025).
\newblock `{Invest}!': {Liberty} {Bonds} and {Stock} {Ownership} over the
  {Twentieth} {Century}.
\newblock {\em Working paper}.

\bibitem[Bucciol and Miniaci, 2011]{Bucciol2011}
Bucciol, A. and Miniaci, R. (2011).
\newblock Household {Portfolios} and {Implicit} {Risk} {Preference}.
\newblock {\em Review of Economics and Statistics}, 93(4):1235--1250.

\bibitem[Cain et~al., 2018]{cain2018}
Cain, L.~P., Fishback, P.~V., and Rhode, P.~W., editors (2018).
\newblock {\em The {{Oxford Handbook}} of {{American Economic History Volume}}
  1}.
\newblock Oxford University Press, 1 edition.

\bibitem[Calvet et~al., 2026]{CalvetCampbellGomesSodini2026}
Calvet, L.~E., Campbell, J.~Y., Gomes, F., and Sodini, P. (2026).
\newblock The {Cross}-{Section} of {Household} {Preferences}.
\newblock {\em The Journal of Finance}.

\bibitem[Calvet et~al., 2007]{CalvetCampbellSodini2007}
Calvet, L.~E., Campbell, J.~Y., and Sodini, P. (2007).
\newblock Down or {Out}: {Assessing} the {Welfare} {Costs} of {Household}
  {Investment} {Mistakes}.
\newblock {\em Journal of Political Economy}, 115(5):707--747.

\bibitem[Chakraborty et~al., 2016]{ChakrabortyThompsonYehoue2016}
Chakraborty, S., Thompson, J.~C., and Yehoue, E.~B. (2016).
\newblock The culture of entrepreneurship.
\newblock {\em Journal of Economic Theory}, 163:288--317.

\bibitem[Chiappori and Paiella, 2011]{Chiappori2011}
Chiappori, P.-A. and Paiella, M. (2011).
\newblock Relative {Risk} {Aversion} {Is} {Constant}: {Evidence} {From} {Panel}
  {Data}.
\newblock {\em Journal of the European Economic Association}, 9(6):1021--1052.

\bibitem[Chiappori et~al., 2019]{ChiapporiSalanie2019}
Chiappori, P.-A., Salanie, B., Salanie, F., and Gandhi, A. (2019).
\newblock From {Aggregate} {Betting} {Data} to {Individual} {Risk}
  {Preferences}.
\newblock {\em Econometrica}, 87(1):1--36.

\bibitem[Chiappori et~al., 2014]{Chiappori2014}
Chiappori, P.-A., Samphantharak, K., Schulhofer-Wohl, S., and Townsend, R.~M.
  (2014).
\newblock Heterogeneity and risk sharing in village economies.
\newblock {\em Quantitative Economics}, 5(1):1--27.

\bibitem[Cicchetti and Dubin, 1994]{CicchettiDubin1994}
Cicchetti, C.~J. and Dubin, J.~A. (1994).
\newblock A {Microeconometric} {Analysis} of {Risk} {Aversion} and the
  {Decision} to {Self}-{Insure}.
\newblock {\em Journal of Political Economy}, 102(1):169--186.

\bibitem[Clarke, 1991]{Clarke1991}
Clarke, S. (1991).
\newblock New {Deal} {Regulation} and the {Revolution} in {American} {Farm}
  {Productivity}: {A} {Case} {Study} of the {Diffusion} of the {Tractor} in the
  {Corn} {Belt}, 1920-1940.
\newblock {\em The Journal of Economic History}, 51(1):101--123.

\bibitem[Cohen and Einav, 2007]{cohenEstimatingRiskPreferences2007}
Cohen, A. and Einav, L. (2007).
\newblock Estimating {Risk} {Preferences} from {Deductible} {Choice}.
\newblock {\em American Economic Review}, 97(3):745--788.

\bibitem[{Committee on Finance, United States Senate}, 1893]{aldrich1893}
{Committee on Finance, United States Senate} (1893).
\newblock {\em Wholesale {Prices}, {Wages}, and {Transportation}: {Report} by
  {Mr}. {Aldrich}, from the {Committee} on {Finance}}.
\newblock U.S. Government Printing Office.
\newblock Google-Books-ID: dvxKAAAAYAAJ.

\bibitem[Cooley et~al., 1977]{cooley1977}
Cooley, T.~F., DeCanio, S.~J., and Matthews, M.~S. (1977).
\newblock An {Agricultural} {Time} {Series}-{Cross} {Section} {Data} {Set}.
\newblock {\em NBER Working Paper Series}.

\bibitem[Dercon, 1996]{Dercon1996}
Dercon, S. (1996).
\newblock Risk, {Crop} {Choice}, and {Savings}:{Evidence} from {Tanzania}.
\newblock {\em Economic Development and Cultural Change}, 44(3):484--700.

\bibitem[Doepke and Klasing, 2026]{DoepkeKlasing2026}
Doepke, M. and Klasing, M. (2026).
\newblock Preparing {Kids} for {Capitalism}: {The} {Effect} of {German}
  {Reunification} on the {Intergenerational} {Transmission} of {Preferences}.
\newblock Working {Paper}, London School of Economics and University of
  Groningen.

\bibitem[Doepke and Zilibotti, 2014]{Doepke2014}
Doepke, M. and Zilibotti, F. (2014).
\newblock Culture, {Entrepreneurship}, and {Growth}.
\newblock In Aghion, P. and Durlauf, S.~N., editors, {\em Handbook of
  {Economic} {Growth}, {Volume} {2A}}, pages 1--48. Elsevier, Amsterdam.

\bibitem[Doepke and Zilibotti, 2017]{Doepke2017}
Doepke, M. and Zilibotti, F. (2017).
\newblock Parenting {With} {Style}: {Altruism} and {Paternalism} in
  {Intergenerational} {Preference} {Transmission}.
\newblock {\em Econometrica}, 85(5):1331--1371.

\bibitem[Dohmen et~al., 2012]{Dohmen2012}
Dohmen, T., Falk, A., Huffman, D., and Sunde, U. (2012).
\newblock The intergenerational transmission of risk and trust attitudes.
\newblock {\em Review of Economic Studies}, 79(2):645--677.

\bibitem[Eckert et~al., 2020]{Eckert2020b}
Eckert, F., Gvirtz, A., Liang, J., and Peters, M. (2020).
\newblock A {Method} to {Construct} {Geographical} {Crosswalks} with an
  {Application} to {US} {Counties} since 1790.
\newblock {\em NBER Working Paper Series}.

\bibitem[Egan et~al., 2026]{EganMacKayYang2025}
Egan, M., MacKay, A., and Yang, H. (2026).
\newblock What {Drives} {Variation} in {Investor} {Portfolios}? {Estimating}
  the {Roles} of {Beliefs} and {Risk} {Preferences}.
\newblock Working {Paper}, Harvard University, University of Virginia, and
  London Business School.

\bibitem[Fafchamps, 1992]{Fafchamps1992a}
Fafchamps, M. (1992).
\newblock Cash {Crop} {Production}, {Food} {Price} {Volatility}, and {Rural}
  {Market} {Integration} in the {Third} {World}.
\newblock {\em American Journal of Agricultural Economics}, 74(1):90--99.

\bibitem[Fiszbein, 2022]{Fiszbein2022}
Fiszbein, M. (2022).
\newblock Agricultural {Diversity}, {Structural} {Change}, and {Long}-{Run}
  {Development}: {Evidence} from the {United} {States}.
\newblock {\em American Economic Journal: Macroeconomics}, 14(2):1--43.

\bibitem[Florentsen et~al., 2019]{FlorentsenEtAl2019}
Florentsen, B., Nielsson, U., Raahauge, P., and Rangvid, J. (2019).
\newblock The {Aggregate} {Cost} of {Equity} {Underdiversification}.
\newblock {\em The Financial Review}, 54(4):833--856.

\bibitem[Foster and Rosenzweig, 1995]{Foster1995}
Foster, A.~D. and Rosenzweig, M.~R. (1995).
\newblock Learning by {Doing} and {Learning} from {Others}: {Human} {Capital}
  and {Technical} {Change} in {Agriculture}.
\newblock {\em Journal of Political Economy}, 103(6):1176--1209.

\bibitem[Frazzini and Pedersen, 2014]{FrazziniPedersen2014}
Frazzini, A. and Pedersen, L.~H. (2014).
\newblock Betting against beta.
\newblock {\em Journal of Financial Economics}, 111(1):1--25.

\bibitem[Frieden, 1997]{Frieden1997}
Frieden, J.~A. (1997).
\newblock Monetary {Populism} in {Nineteenth}-{Century} {America}: {An} {Open}
  {Economy} {Interpretation}.
\newblock {\em The Journal of Economic History}, 57(2):367--395.

\bibitem[Fullenkamp et~al., 2003]{FullenkampTenorioBattalio2003}
Fullenkamp, C., Tenorio, R., and Battalio, R. (2003).
\newblock Assessing {Individual} {Risk} {Attitudes} {Using} {Field} {Data}.
\newblock {\em The Review of Economics and Statistics}, 85(1):218--226.

\bibitem[Galor and Michalopoulos, 2012]{Galor2012}
Galor, O. and Michalopoulos, S. (2012).
\newblock Evolution and the {Growth} {Process}: {Natural} selection of
  {Entrepreneurial} {Traits}.
\newblock {\em Journal of Economic Theory}, 147(2):759--780.

\bibitem[Gibbons et~al., 1989]{GibbonsRossShanken1989}
Gibbons, M.~R., Ross, S.~A., and Shanken, J. (1989).
\newblock A {Test} of the {Efficiency} of a {Given} {Portfolio}.
\newblock {\em Econometrica}, 57(5):1121--1152.

\bibitem[Gordon, 2016]{gordonRiseFallAmerican2016}
Gordon, R.~J. (2016).
\newblock {\em The {Rise} and {Fall} of {American} {Growth}: {The} {U}.{S}.
  {Standard} of {Living} since the {Civil} {War}}.
\newblock Princeton University Press, Princeton Oxford.

\bibitem[Greif, 1994]{Greif1994}
Greif, A. (1994).
\newblock Cultural {Beliefs} and the {Organization} of {Society}: {A}
  {Historical} and {Theoretical} {Reflection} on {Collectivist} and
  {Individualist} {Societies}.
\newblock {\em Journal of Political Economy}, 102(5):912--950.

\bibitem[Gross, 2018]{Gross2018}
Gross, D.~P. (2018).
\newblock Scale versus scope in the diffusion of new technology: evidence from
  the farm tractor.
\newblock {\em The RAND Journal of Economics}, 49(2):427--452.

\bibitem[Haines et~al., 2018]{Haines2018}
Haines, M., Fishback, P., and Rhode, P. (2018).
\newblock United {States} {Agriculture} {Data}, 1840 - 2012.
\newblock https://www.icpsr.umich.edu/web/ICPSR/studies/35206.

\bibitem[Handel, 2013]{Handel2013}
Handel, B.~R. (2013).
\newblock Adverse {Selection} and {Inertia} in {Health} {Insurance} {Markets}:
  {When} {Nudging} {Hurts}.
\newblock {\em American Economic Review}, 103(7):2643--2682.

\bibitem[Henrich et~al., 2010]{HenrichHeineNorenzayan2010}
Henrich, J., Heine, S.~J., and Norenzayan, A. (2010).
\newblock The weirdest people in the world?
\newblock {\em Behavioral and Brain Sciences}, 33(2-3):61--83.

\bibitem[Hilt and Rahn, 2020]{hilt2020}
Hilt, E. and Rahn, W. (2020).
\newblock Financial {Asset} {Ownership} and {Political} {Partisanship}:
  {Liberty} {Bonds} and {Republican} {Electoral} {Success} in the 1920s.
\newblock {\em The Journal of Economic History}, 80(3):746--781.

\bibitem[Karlan et~al., 2014]{Karlan2014}
Karlan, D., Osei, R., Osei-Akoto, I., and Udry, C. (2014).
\newblock Agricultural {Decisions} after {Relaxing} {Credit} and {Risk}
  {Constraints}.
\newblock {\em The Quarterly Journal of Economics}, 129(2):597--652.

\bibitem[Katz and Margo, 2014]{Katz2014}
Katz, L. and Margo, R. (2014).
\newblock Technical {Change} and the {Relative} {Demand} for {Skilled} {Labor}:
  {The} {United} {States} in {Historical} {Perspective}.
\newblock In {\em Human {Capital} in {History}}, pages 15--57. University of
  Chicago Press and NBER, Cambridge, MA.

\bibitem[Kaufman, 2002]{Kaufman2003}
Kaufman, J. (2002).
\newblock {\em For the Common Good? American Civic Life and the Golden Age of
  Fraternity}.
\newblock Oxford University Press, New York.

\bibitem[Kim and Margo, 2004]{Kim2004}
Kim, S. and Margo, R. (2004).
\newblock Historical {Perspectives} on {U}.{S}. {Economic} {Geography}.
\newblock In {\em Handbook of {Regional} and {Urban} {Economics}}, volume~4.
  Elsevier North Holland, Amsterdam.

\bibitem[Klasing, 2014]{Klasing2014}
Klasing, M.~J. (2014).
\newblock Cultural change, risk-taking behavior and implications for economic
  development.
\newblock {\em Journal of Development Economics}, 110:158--169.

\bibitem[Knight, 1921]{Knight1921}
Knight, F.~H. (1921).
\newblock {\em Risk, {Uncertainty} and {Profit}}.
\newblock Number~31 in Hart, {Schaffner} \& {Marx} {Prize} {Essays}. Houghton
  Mifflin Company, Boston and New York.

\bibitem[Koundouri et~al., 2009]{KoundouriLaukkanen2009}
Koundouri, P., Laukkanen, M., Myyra, S., and Nauges, C. (2009).
\newblock The effects of {EU} agricultural policy changes on farmers' risk
  attitudes.
\newblock {\em European Review of Agricultural Economics}, 36(1):53--77.

\bibitem[Kurosaki and Fafchamps, 2002]{Kurosaki2002}
Kurosaki, T. and Fafchamps, M. (2002).
\newblock Insurance market efficiency and crop choices in {Pakistan}.
\newblock {\em Journal of Development Economics}, 67(2):419--453.

\bibitem[Landes, 1949]{Landes1949}
Landes, D.~S. (1949).
\newblock French {Entrepreneurship} and {Industrial} {Growth} in the
  {Nineteenth} {Century}.
\newblock {\em The Journal of Economic History}, 9(1):45--61.

\bibitem[Lange et~al., 2009]{Lange2009}
Lange, F., Olmstead, A.~L., and Rhode, P.~W. (2009).
\newblock The {Impact} of the {Boll} {Weevil}, 1892--1932.
\newblock {\em Journal of Economic History}, 69(3):685--718.

\bibitem[Levine, 2005]{Levine2005}
Levine, R. (2005).
\newblock Finance and {Growth}: {Theory} and {Evidence}.
\newblock In Aghion, P. and Durlauf, S.~N., editors, {\em Handbook of
  {Economic} {Growth}, {Volume} {1A}}, pages 865--934. North-Holland
  (Elsevier), Amsterdam.

\bibitem[Liu, 2013]{Liu2013}
Liu, E.~M. (2013).
\newblock Time to {Change} {What} to {Sow}: {Risk} {Preferences} and
  {Technology} {Adoption} {Decisions} of {Cotton} {Farmers} in {China}.
\newblock {\em Review of Economics and Statistics}, 95(4):1386--1403.

\bibitem[Manuelli and Seshadri, 2014]{ManuelliSeshadri2014}
Manuelli, R.~E. and Seshadri, A. (2014).
\newblock Frictionless {Technology} {Diffusion}: {The} {Case} of {Tractors}.
\newblock {\em American Economic Review}, 104(4):1368--1391.

\bibitem[Martini and Silberberg, 2006]{MartiniSilberberg2006}
Martini, D.~D. and Silberberg, E. (2006).
\newblock The {Diffusion} of {Tractor} {Technology}.
\newblock {\em The Journal of Economic History}, 66(2):354--389.

\bibitem[McCloskey, 1976]{McCloskey1976}
McCloskey, D.~N. (1976).
\newblock English {Open} {Fields} as {Behavior} {Towards} {Risk}.
\newblock {\em Research in Economic History}, 1:124--170.

\bibitem[Moschini and Hennessy, 1999]{Moschini1999}
Moschini, G. and Hennessy, D.~A. (1999).
\newblock Uncertainty, risk aversion and risk management for agricultural
  producers.
\newblock In {\em Handbook of {Agricultural} {Economics}}, volume~1A, pages
  87--153. North-Holland.

\bibitem[North, 1991]{North1991}
North, D.~C. (1991).
\newblock Institutions.
\newblock {\em Journal of Economic Perspectives}, 5(1):97--112.

\bibitem[Olmstead and Rhode, 2000]{olmstead2000}
Olmstead, A.~L. and Rhode, P.~W. (2000).
\newblock The diffusion of the tractor in {American} agriculture: 1910-60.
\newblock {\em NBER Working Paper Series}.

\bibitem[Olmstead and Rhode, 2001]{OlmsteadRhode2001}
Olmstead, A.~L. and Rhode, P.~W. (2001).
\newblock Reshaping the {Landscape}: {The} {Impact} and {Diffusion} of the
  {Tractor} in {American} {Agriculture}, 1910-1960.
\newblock {\em The Journal of Economic History}, 61(3):663--698.

\bibitem[Olmstead and Rhode, 2008]{OlmsteadRhode2008}
Olmstead, A.~L. and Rhode, P.~W. (2008).
\newblock {\em Creating {Abundance}: {Biological} {Innovation} and {American}
  {Agricultural} {Development}}.
\newblock Cambridge University Press, Cambridge.

\bibitem[Olmstead and Rhode, 2018]{Olmstead2018}
Olmstead, A.~L. and Rhode, P.~W. (2018).
\newblock Agriculture in {American} {Economic} {History}.
\newblock In {\em The {Oxford} {Handbook} of {American} {Economic} {History}},
  volume~1, pages 159--182.

\bibitem[Paravisini et~al., 2017]{paravisiniRiskAversionWealth2017}
Paravisini, D., Rappoport, V., and Ravina, E. (2017).
\newblock Risk {Aversion} and {Wealth}: {Evidence} from {Person}-to-{Person}
  {Lending} {Portfolios}.
\newblock {\em Management Science}, 63(2):279--297.

\bibitem[Rajan and Ramcharan, 2015]{Rajan2015}
Rajan, R. and Ramcharan, R. (2015).
\newblock The {Anatomy} of a {Credit} {Crisis}: {The} {Boom} and {Bust} in
  {Farm} {Land} {Prices} in the {United} {States} in the 1920s.
\newblock {\em American Economic Review}, 105(4):1439--1477.

\bibitem[Rhode, 2024]{Rhode2024}
Rhode, P.~W. (2024).
\newblock How {Suitable} are {FAO}-{GAEZ} {Crop} {Suitability} {Indices} for
  {Historical} {Analysis}?

\bibitem[Rosenberg, 1996]{Rosenberg1996}
Rosenberg, N. (1996).
\newblock Uncertainty and {Technological} {Change}.
\newblock In {\em Technology and {Growth}: {Conference} {Proceedings}}, pages
  91--110. Federal Reserve Bank of Boston, Boston.

\bibitem[Rosenzweig and Binswanger, 1993]{Rosenzweig1993}
Rosenzweig, M.~R. and Binswanger, H.~P. (1993).
\newblock Wealth, {Weather} {Risk} and the {Composition} and {Profitability} of
  {Agricultural} {Investments}.
\newblock {\em The Economic Journal}, 103(416):56--78.

\bibitem[Schumpeter, 1934]{Schumpeter1934}
Schumpeter, J.~A. (1934).
\newblock {\em The {Theory} of {Economic} {Development}: {An} {Inquiry} into
  {Profits}, {Capital}, {Credit}, {Interest}, and the {Business} {Cycle}}.
\newblock Harvard University Press, Cambridge, MA.

\bibitem[Slivinski et~al., 2019]{slivinski2019}
Slivinski, L.~C., Compo, G.~P., Whitaker, J.~S., Sardeshmukh, P.~D., Giese,
  B.~S., McColl, C., Allan, R., Yin, X., Vose, R., Titchner, H., Kennedy, J.,
  Spencer, L.~J., Ashcroft, L., Br{\"o}nnimann, S., Brunet, M., Camuffo, D.,
  Cornes, R., Cram, T.~A., Crouthamel, R., Dom{\'{\i}}nguez-Castro, F.,
  Freeman, J.~E., Gergis, J., Hawkins, E., Jones, P.~D., Jourdain, S., Kaplan,
  A., Kubota, H., Blancq, F.~L., Lee, T.-C., Lorrey, A., Luterbacher, J.,
  Maugeri, M., Mock, C.~J., Moore, G.~K., Przybylak, R., Pudmenzky, C., Reason,
  C., Slonosky, V.~C., Smith, C.~A., Tinz, B., Trewin, B., Valente, M.~A.,
  Wang, X.~L., Wilkinson, C., Wood, K., and Wyszy{\'n}ski, P. (2019).
\newblock Towards a more reliable historical reanalysis: {Improvements} for
  version 3 of the {Twentieth} {Century} {Reanalysis} system.
\newblock {\em Quarterly Journal of the Royal Meteorological Society},
  145(724):2876--2908.

\bibitem[Stulz and Williamson, 2003]{StulzWilliamson2003}
Stulz, R.~M. and Williamson, R. (2003).
\newblock Culture, openness, and finance.
\newblock {\em Journal of Financial Economics}, 70(3):313--349.

\bibitem[Sylvester et~al., 2002]{sylvester2002}
Sylvester, K.~M., Leonard, S.~H., Gutmann, M.~P., and Cunfer, G. (2002).
\newblock Demography and {{Environment}} in {{Grassland Settlement}}: {{Using
  Linked Longitudinal}} and {{Cross-Sectional Data}} to {{Explore Household}}
  and {{Agricultural Systems}}.
\newblock {\em History and Computing}, 14(1-2):31--60.

\bibitem[{US Bureau Of Labor Statistics}, 1927]{bls1927}
{US Bureau Of Labor Statistics} (1927).
\newblock Wholesale {Prices} 1890 to 1926.
\newblock Technical Report Bulletin No. 440, Bureau of Labor Statistics.

\bibitem[{US Census Bureau}, 1975]{uscensusbureau1975}
{US Census Bureau} (1975).
\newblock Historical {Statistics} of the {United} {States}, {Colonial} {Times}
  to 1970.
\newblock Technical report.
\newblock Section: Government.

\bibitem[{U.S. Department of Agriculture}, 1936]{usda1936}
{U.S. Department of Agriculture} (1936).
\newblock Agricultural statistics, 1936.

\bibitem[von Gaudecker, 2015]{vonGaudecker2015}
von Gaudecker, H.-M. (2015).
\newblock How {Does} {Household} {Portfolio} {Diversification} {Vary} with
  {Financial} {Literacy} and {Financial} {Advice}?
\newblock {\em The Journal of Finance}, 70(2):489--507.

\bibitem[Whatley, 1985]{Whatley1985}
Whatley, W.~C. (1985).
\newblock A {History} of {Mechanization} in the {Cotton} {South}: {The}
  {Institutional} {Hypothesis}.
\newblock {\em The Quarterly Journal of Economics}, 100(4):1191--1215.

\end{thebibliography}
\end{singlespacing}
\end{multicols}

\pagebreak

\appendix

\renewcommand{\thesection}{\Alph{section}}
\renewcommand{\thesubsection}{\thesection.\arabic{subsection}}

\counterwithin*{equation}{section}
\renewcommand\theequation{\thesection\arabic{equation}}

\newpage
\section*{Online Appendix}

\section{Model derivation}
\label{app: irs}

To solve the farmer's problem in \Cref{eq: farmerproblem}, we difference the first-order conditions against a numeraire crop (crop $1$). Because the data record observed per-acre yield $\tilde q_i = q_i(l_i\bar L)^{\alpha_i-1}$ rather than the productivity object $q_i$, we write the problem in terms of observed per-acre revenue $p_i\tilde q_i$. Consumption is total observed revenue,
\[ c = \sum_{j=1}^N p_jq_j(l_j\bar L)^{\alpha_j} = \sum_{j=1}^N p_j\tilde q_j(l_j\bar L). \]
Taking first-order conditions for all planted crops and subtracting the numeraire crop's first order condition yields
\[ \mathbb E\!\left[U'(c)\left(\alpha_i p_i\tilde q_i - \alpha_1 p_1\tilde q_1\right)\right] = 0 \qquad \forall\, i\in\{2,\dots,N\}\mid l_i>0. \]
These conditions equate the expected marginal revenue of an extra unit of land share in crop $i$, weighted by its returns-to-scale parameter, to that of the numeraire crop. We linearize each condition with a first-order Taylor expansion of the bracketed term around expected consumption $\mathbb E(c)$:\footnote{This makes the recovered risk aversion a local estimate around expected consumption and relies on a linear approximation of the farmer's choices.}
\[
U'(c)(\alpha_i p_i\tilde q_i - \alpha_1 p_1\tilde q_1) \approx U'(\mathbb E (c))(\alpha_i p_i\tilde q_i - \alpha_1 p_1\tilde q_1) + U''(\mathbb E (c))(\alpha_i p_i\tilde q_i - \alpha_1 p_1\tilde q_1)(c-\mathbb E (c)).
\]
Because land shares, total land, and the scale parameters are all known at planting,
\[ c - \mathbb E(c) = \sum_{j=1}^N\bigl(p_j\tilde q_j - \mathbb E(p_j\tilde q_j)\bigr)(l_j\bar L). \]
Multiplying through by $(\alpha_i p_i\tilde q_i-\alpha_1 p_1\tilde q_1)$, taking expectations, and setting the result to zero to match the first-order condition gives, after using the definition of covariance and the fact that $\alpha_i$ and $\alpha_1$ are non-stochastic,
\[ 0 = U'(\mathbb E (c))\bigl(\alpha_i\mathbb E(p_i\tilde q_i)-\alpha_1\mathbb E(p_1\tilde q_1)\bigr) + U''(\mathbb E (c))\sum_{j=1}^N \bigl[\alpha_i\,\mathrm{Cov}(p_j\tilde q_j,p_i\tilde q_i) - \alpha_1\,\mathrm{Cov}(p_j\tilde q_j,p_1\tilde q_1)\bigr](l_j\bar L). \]
Rearranging, and multiplying and dividing the right-hand side by $\mathbb E(c)=\sum_j\mathbb E(p_j\tilde q_j)(l_j\bar L)$, delivers \Cref{eq: main model}. The returns-to-scale parameter therefore enters linearly, as a marginal-product wedge $\alpha_i$ on observed revenue, with planted acres $(l_j\bar L)$ as the consumption weights. Further, the numeraire crop's own mean revenue $\alpha_1\mathbb E(p_1\tilde q_1)$ and its covariances 
\vspace{-.2cm}

$\alpha_1\sum_j\mathrm{Cov}(p_j\tilde q_j,p_1\tilde q_1)(l_j\bar L)$ enter additively, reflecting its risk.

\section{Data appendix}

\subsection{Core-crop coverage}

\begin{figure}[H]
\centering
\caption{Core-crop coverage of improved farmland, by wave \label{fig: cropcoveragewave}}
\vspace{0.15cm}

\includegraphics[width=.72\textwidth]{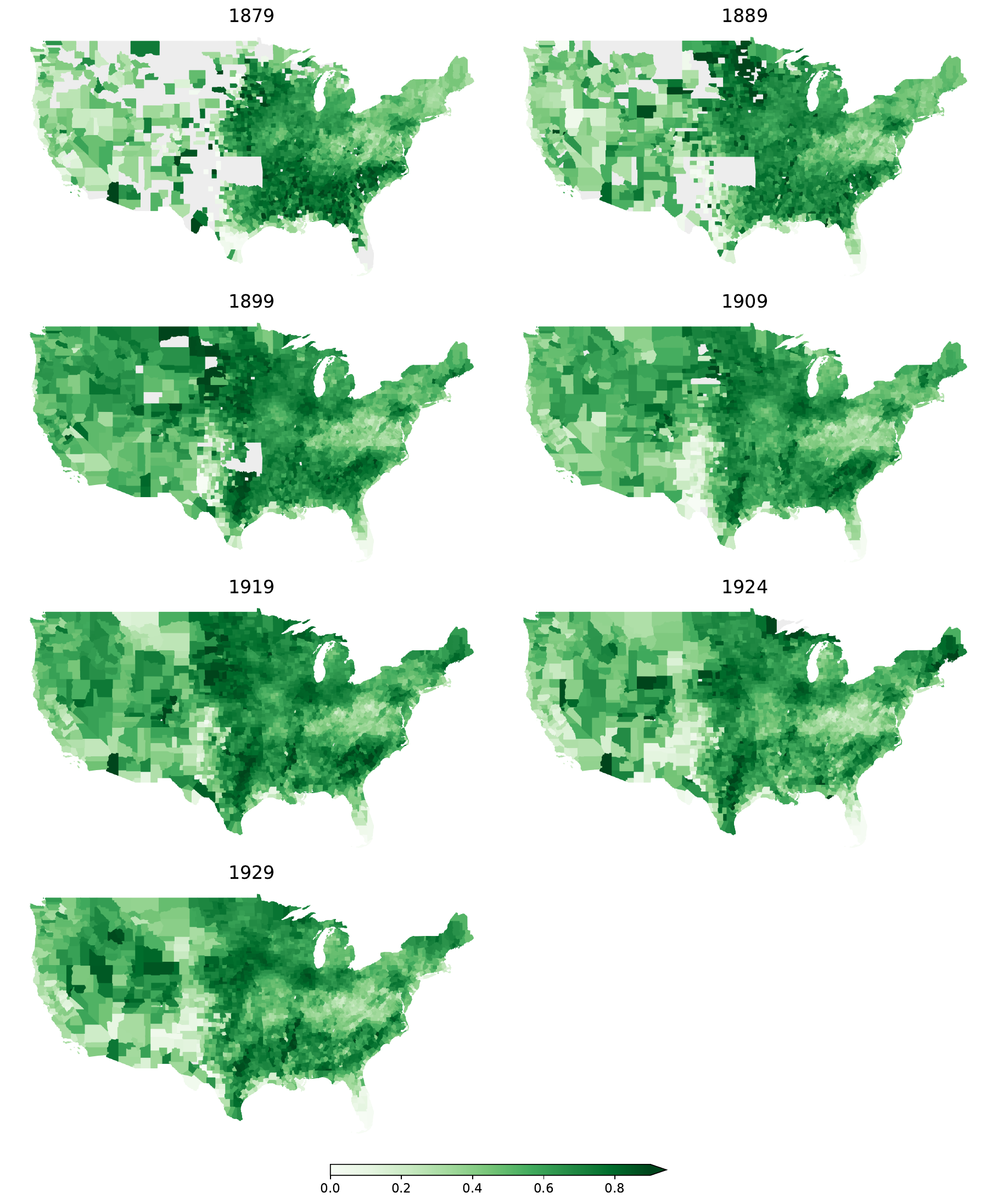}
\vspace{0.1cm}

{\footnotesize\raggedright \textbf{Notes:} This figure maps the core-crop coverage (the share of a county's improved farmland planted in the seven core crops) in each county across agricultural census years.\par}
\end{figure}

\subsection{Cropland shares}
\label{App: cropshares}

\begin{table}[H]
\caption{Summary statistics of share of land planted in each crop in 1880--1930}
\begin{center}
\small
\begin{tabular}{l cc cc cc}
\hline \hline
 & \multicolumn{2}{c}{1880} & \multicolumn{2}{c}{1890} & \multicolumn{2}{c}{1900} \\
 & Mean & Median & Mean & Median & Mean & Median \\ \hline
\addlinespace
Share of barley & 0.014 & 0.000 & 0.011 & 0.000 & 0.011 & 0.000 \\
Share of corn & 0.388 & 0.419 & 0.344 & 0.366 & 0.360 & 0.387 \\
Share of cotton & 0.103 & 0.000 & 0.114 & 0.000 & 0.102 & 0.000 \\
Share of hay & 0.198 & 0.110 & 0.279 & 0.211 & 0.284 & 0.191 \\
Share of oats & 0.096 & 0.086 & 0.114 & 0.105 & 0.083 & 0.052 \\
Share of rye & 0.011 & 0.002 & 0.010 & 0.002 & 0.008 & 0.001 \\
Share of wheat & 0.190 & 0.157 & 0.128 & 0.069 & 0.152 & 0.089 \\
\hline
Share of farmland w/ selected crops & 0.577 & 0.600 & 0.573 & 0.606 & 0.605 & 0.630 \\
\addlinespace[8pt]
\hline
 & \multicolumn{2}{c}{1910} & \multicolumn{2}{c}{1920} & \multicolumn{2}{c}{1930} \\
 & Mean & Median & Mean & Median & Mean & Median \\ \hline
\addlinespace
Share of barley & 0.017 & 0.000 & 0.014 & 0.001 & 0.024 & 0.001 \\
Share of corn & 0.362 & 0.389 & 0.318 & 0.325 & 0.314 & 0.329 \\
Share of cotton & 0.130 & 0.000 & 0.121 & 0.000 & 0.153 & 0.000 \\
Share of hay & 0.282 & 0.209 & 0.273 & 0.198 & 0.302 & 0.225 \\
Share of oats & 0.093 & 0.058 & 0.091 & 0.056 & 0.079 & 0.029 \\
Share of rye & 0.009 & 0.001 & 0.017 & 0.003 & 0.008 & 0.001 \\
Share of wheat & 0.108 & 0.047 & 0.166 & 0.109 & 0.121 & 0.036 \\
\hline
Share of farmland w/ selected crops & 0.559 & 0.579 & 0.598 & 0.623 & 0.552 & 0.569 \\
\hline \hline
\end{tabular}

\label{tab: acressumstats}
\end{center}
\vspace{0.5cm}
{\footnotesize\raggedright \textbf{Notes:} This table reports the mean and median share of land allocated to each of the selected crops across all counties in 1880, 1890, 1900, 1910, 1920, and 1930. Shares are calculated relative to the total acreage planted with the full set of crops considered. The table also reports the mean and median share of improved farmland devoted to these crops in each year.}
\end{table}

\subsection{Construction of cultivation clusters}
\label{app: cultclusters}

We construct cultivation clusters from county crop-share vectors in the Ag Census. For each county $c$ and census wave $t$, let
\[
\mathbf{l}_{c,t}=\left(l_{1,c,t},\ldots,l_{N,c,t}\right)
\]
denote the vector of land shares across the seven core crops, where $l_{i,c,t}$ is the share of core-crop acreage planted with crop $i$. We apply $k$-means clustering to these crop-share vectors separately in each census wave.

\begin{figure}[H]\centering
\caption{Choosing the number of cultivation clusters \label{fig: kselect}}
\includegraphics[width=\textwidth]{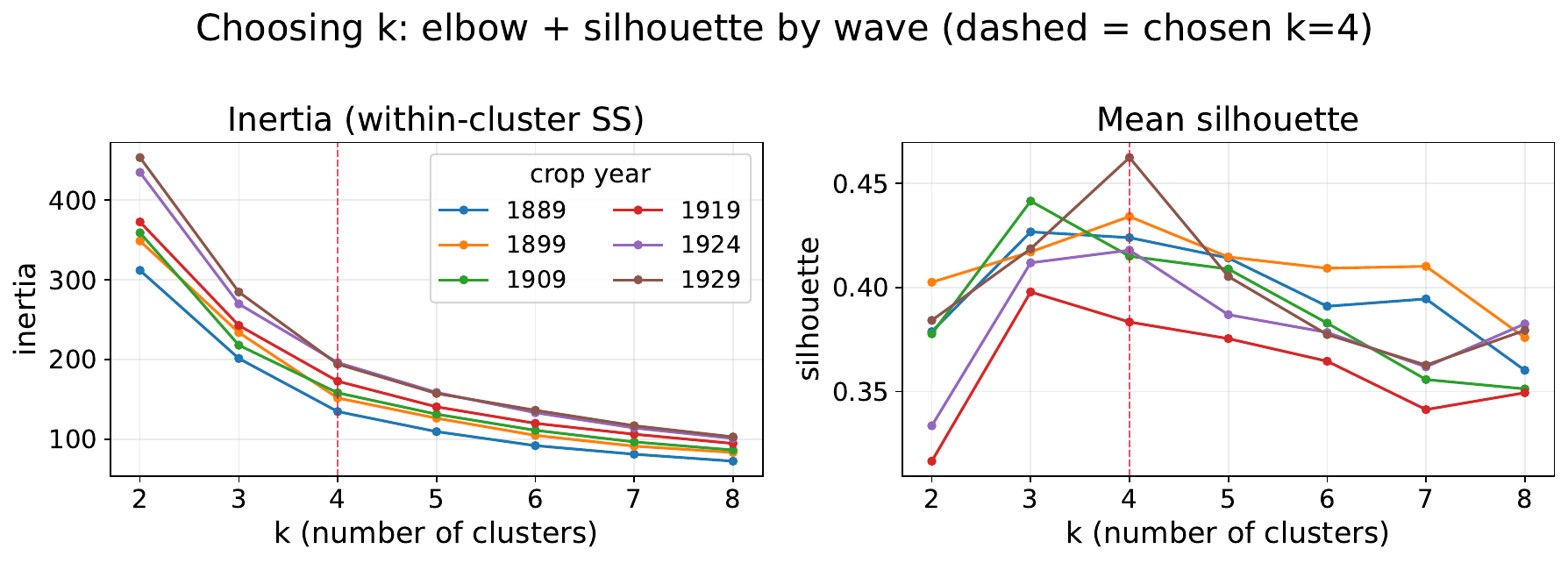}
\begin{minipage}{\textwidth}\smallskip{\footnotesize\raggedright \textbf{Notes:} This figure shows the within-cluster inertia (elbow) and mean silhouette for a different number of clusters $k$, by agricultural census year.}\end{minipage}
\end{figure}

We choose $k=4$ broad clusters based on three criteria: internal fit (the mean silhouette peaks at $k=3$--$4$ with a clear inertia elbow as shown in \Cref{fig: kselect}), geographic coherence (the clusters are largely spatially contiguous), and feasibility (bounded by need to estimate each crop's returns to scale parameter within a cluster). 

\begin{figure}[H]\centering
\caption{The four cultivation clusters by county and census year}\label{fig: clusters4}
\vspace{.1cm}
\includegraphics[width=0.85\textwidth]{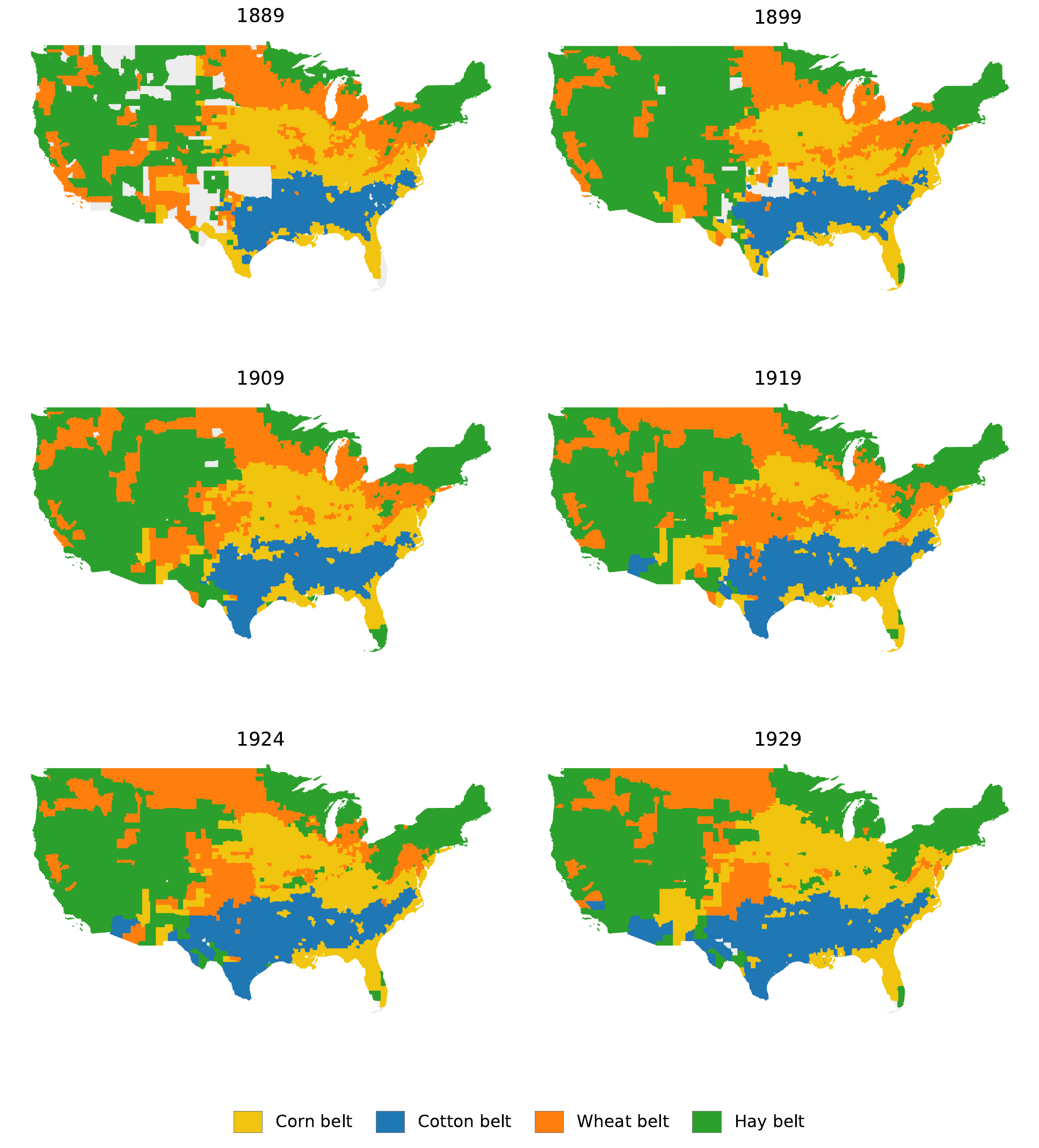}
\vspace{0.1cm}

{\footnotesize\raggedright \textbf{Notes:} This figure maps the four cultivation clusters obtained by the $k$-means clustering procedure across agricultural census years.\par}
\end{figure}

\begin{figure}[H]\centering
\caption{Average crop mix by wave across cultivation clusters \label{fig: clustercentroids}}
\includegraphics[width=0.85\textwidth]{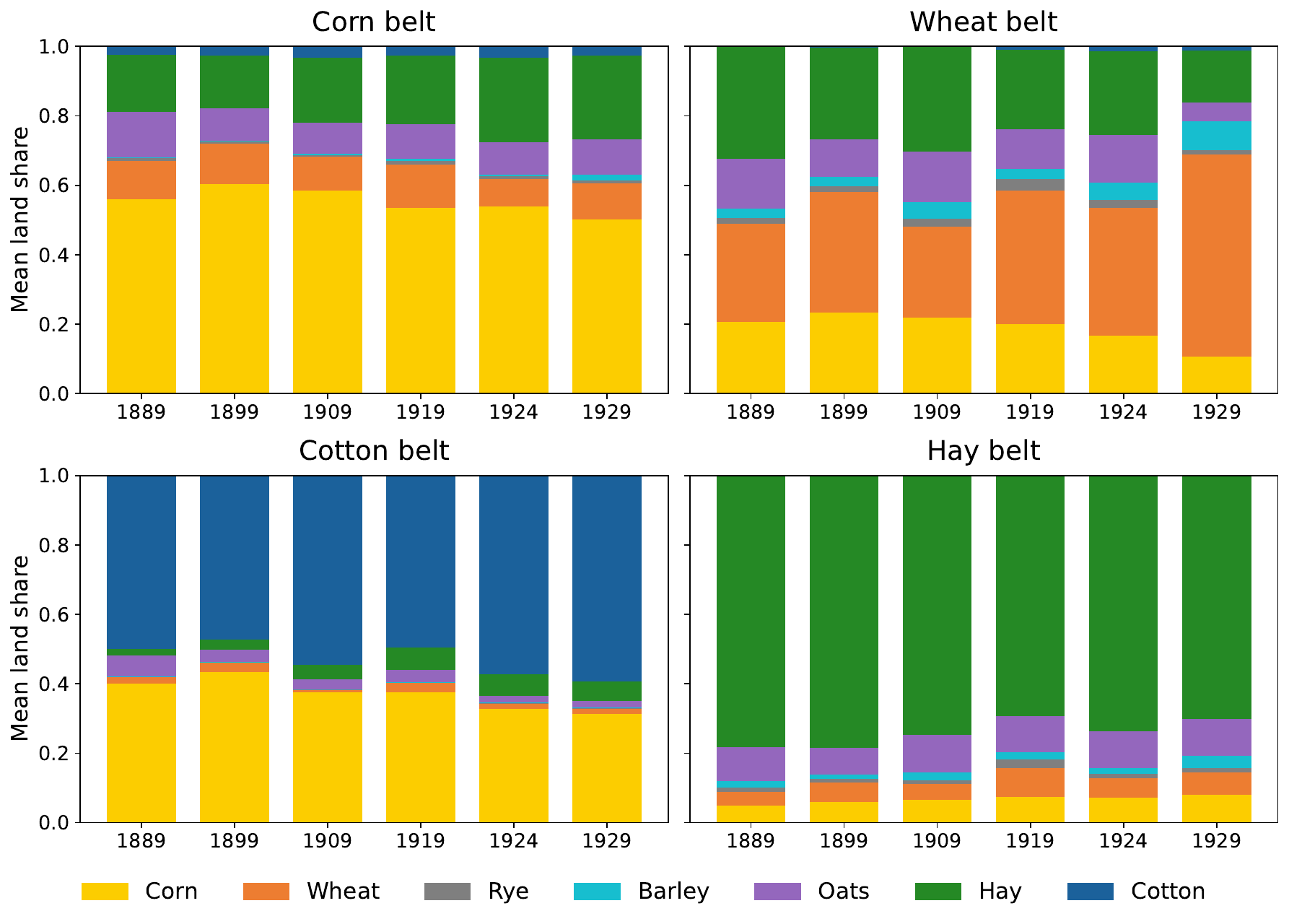}
\begin{minipage}{\textwidth}\smallskip\footnotesize{\textbf{Notes}: These figures show the mean land shares of our core set of crops in each cultivation cluster across agricultural census years.}\end{minipage}
\end{figure}

\subsection{Additional information on price data}
\label{App: prices}

We use three procedures to combine the price datasets described in \Cref{tab:pricedata}. First, when the \cite{cooley1977} series is missing a state-year price for one of our core crops but at least two other crops in the dataset have both state- and national-level prices that year, we compute the median ratio of state- to national-level prices across those available crops and impute the missing state-level price by multiplying this ratio by the national price of the missing crop. This imputation procedure draws on the crops for which both a state-level price in \cite{cooley1977} and a national-level price from one of the other sources are available. This encompasses the seven core crops together with additional crops (rice, tobacco, potatoes, sweet potatoes, buckwheat, flaxseed, and hops). 

Second, when a state's \cite{cooley1977} series does not span the full period, we extend it using national-level price changes: where state-level coverage ends before 1930, we carry the last observed state-level price forward by applying the national-level percentage change in each subsequent year; and where coverage begins after the start of the period, we carry the earliest observed state-level price backward by removing the national-level percentage change in each preceding year. Third, in the cases where a crop has no state-level data in a state, we assign the national price directly.

\begin{center}

\begin{table}[!ht]

\caption{Data sources used to construct price time-series \label{tab:pricedata}}

\centering
\begin{footnotesize}

\begin{tabular}{p{45mm}p{40mm}p{60mm}H} 
		\hline \hline

\rowcolor{lightgray}\multicolumn{4}{c}{\textbf{State-level prices (1880--1930)}} \\ \addlinespace
\textbf{Data source} & \textbf{Years \& crops} & \textbf{Description} & \textbf{More info} \\  \addlinespace
\hline

Agricultural Time Series-Cross Section Data Set (\cite{cooley1977}) & 1880--1930; barley, corn, cotton, hay, oats, rye, and wheat &  Agricultural Time Series-Cross Section (ATICS) dataset compiled using bulletins and circulars of the United States Department of Agriculture. & Appendix xx \\  \addlinespace

\rowcolor{lightgray}\multicolumn{4}{c}{\textbf{Auxiliary national-level data used to construct state-level prices}} \\ \addlinespace
\textbf{Data source} & \textbf{Years \& crops} & \textbf{Description} & \textbf{More info} \\  \addlinespace
\hline

Wholesale Prices 1890 to 1926 (\cite{bls1927}) & 1890--1920; barley, corn, oats, and rye & Bulletin of the Bureau of Labor Statistics (BLS) with a record of nation-wide wholesale prices. & Appendix xx \\ \addlinespace

Wholesale Prices, Wages, and Transportation (\cite{aldrich1893}) & 1850--1893; barley, corn, cotton, oats, rye, and wheat & Report prepared for the Committee of Finance of the United States Senate with wholesale crop prices. Used to fill pre-1890 gaps where BLS data is unavailable. & Appendix xx \\ \addlinespace

Historical Statistics of the United States, Colonial Times to 1970 (\cite{uscensusbureau1975}) & 1880--1930; hay &  Compendium providing a statistical history of the U.S. on a variety of topics, including crop prices & Appendix xx \\ \addlinespace

USDA Agricultural Statistics Yearbook, 1936 (\cite{usda1936}) & 1927--1930; barley, corn, cotton, oats, rye, and wheat
 & Annual yearbook of US agricultural statistics. Used to extend the BLS national price series from 1927 through 1930 via year-over-year growth rates. & Appendix xx \\ \addlinespace

\hline \hline

\end{tabular}
\end{footnotesize}
\end{table}
\end{center}
\vspace{-1cm}

\begin{table}[H]
\caption{Summary statistics of crop price time-series in each decade in 1880--1930, in dollars of 1913}
\centering
\resizebox{\linewidth}{!}{%
\begin{tabular}{lcc|cc|cc|cc|cc}
\hline \hline
 & \multicolumn{2}{c}{1880--1889} &  \multicolumn{2}{c}{1890--1899} & \multicolumn{2}{c}{1900--1909} & \multicolumn{2}{c}{1910--1919} & \multicolumn{2}{c}{1920--1929}  \\
 & Mean & Median & Mean & Median & Mean & Median & Mean & Median & Mean & Median \\ \hline

Price of barley (bushels) & 0.777 & 0.772 & 0.602 & 0.600 & 0.697 & 0.681 & 0.828 & 0.794 & 0.474 & 0.461 \\
Price of corn (bushels) & 0.607 & 0.605 & 0.533 & 0.535 & 0.675 & 0.677 & 0.833 & 0.809 & 0.495 & 0.502 \\
Price of cotton (400lb bales) & 42.029 & 41.015 & 37.874 & 35.578 & 48.999 & 47.098 & 59.884 & 58.591 & 51.422 & 49.175 \\
Price of hay (tons) & 10.102 & 9.814 & 11.276 & 11.057 & 12.182 & 11.711 & 13.373 & 13.142 & 8.233 & 7.943 \\
Price of oats (bushels) & 0.451 & 0.440 & 0.411 & 0.407 & 0.505 & 0.495 & 0.517 & 0.513 & 0.319 & 0.326 \\
Price of rye (bushels) & 0.784 & 0.769 & 0.717 & 0.713 & 0.823 & 0.802 & 0.997 & 0.945 & 0.617 & 0.585 \\
Price of wheat (bushels) & 1.040 & 1.019 & 0.887 & 0.855 & 0.991 & 0.969 & 1.217 & 1.165 & 0.777 & 0.722 \\

\hline \hline
\end{tabular}%
}
{\footnotesize\raggedright \textbf{Notes:} This table presents the mean and median of the prices of different crops across states in different decades. See text for the details about the construction of these price variables. All prices are deflated to 1913 dollars using the Consumer Price Index (Federal Reserve Bank of Minneapolis).}

\label{tab:pricestats}
\end{table}

\begin{figure}[H]
\label{fig: price timeseries}
\begin{adjustwidth}{-2cm}{-2cm}
\begin{center}
\caption{Time series of crop prices \label{fig:pricetseries}}

\subfloat[Barley]{\includegraphics[width=0.37\textwidth]{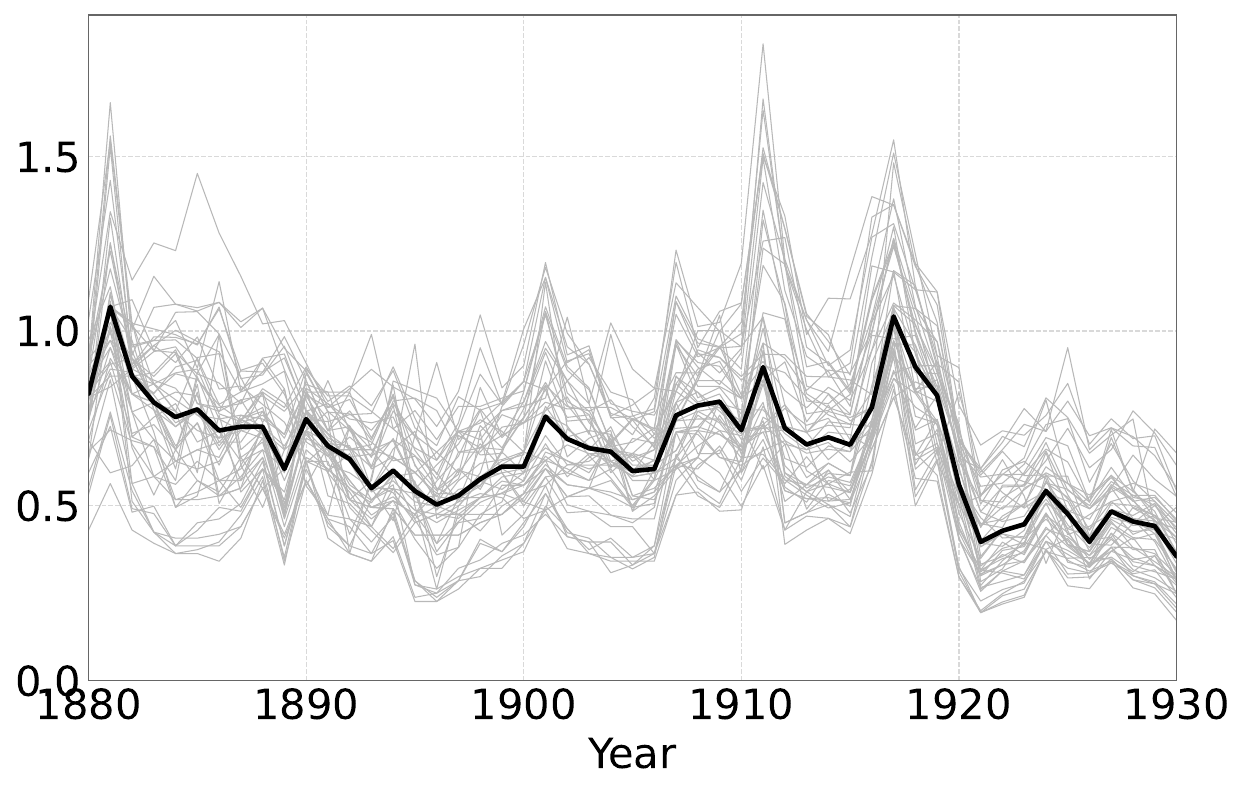}} 
\subfloat[Corn]{\includegraphics[width=0.37\textwidth]{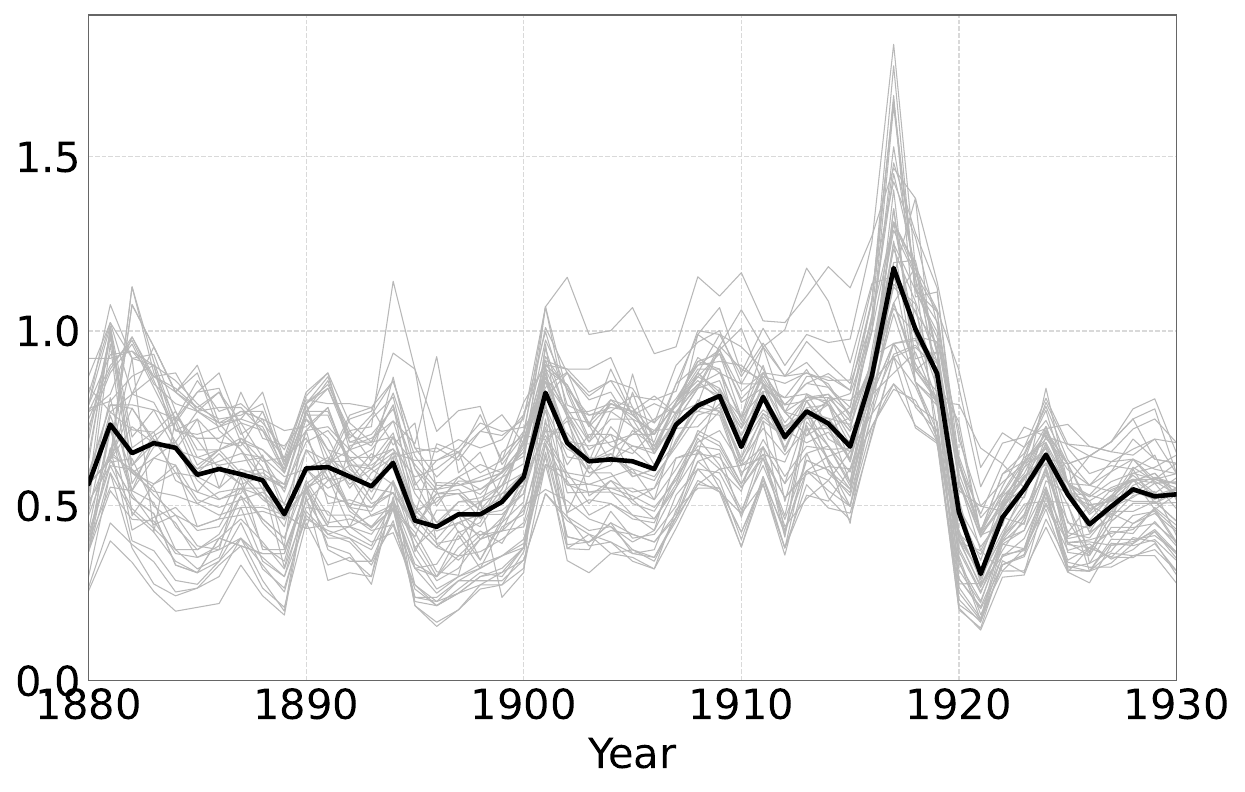}} \\
\subfloat[Cotton]{\includegraphics[width=0.37\textwidth]{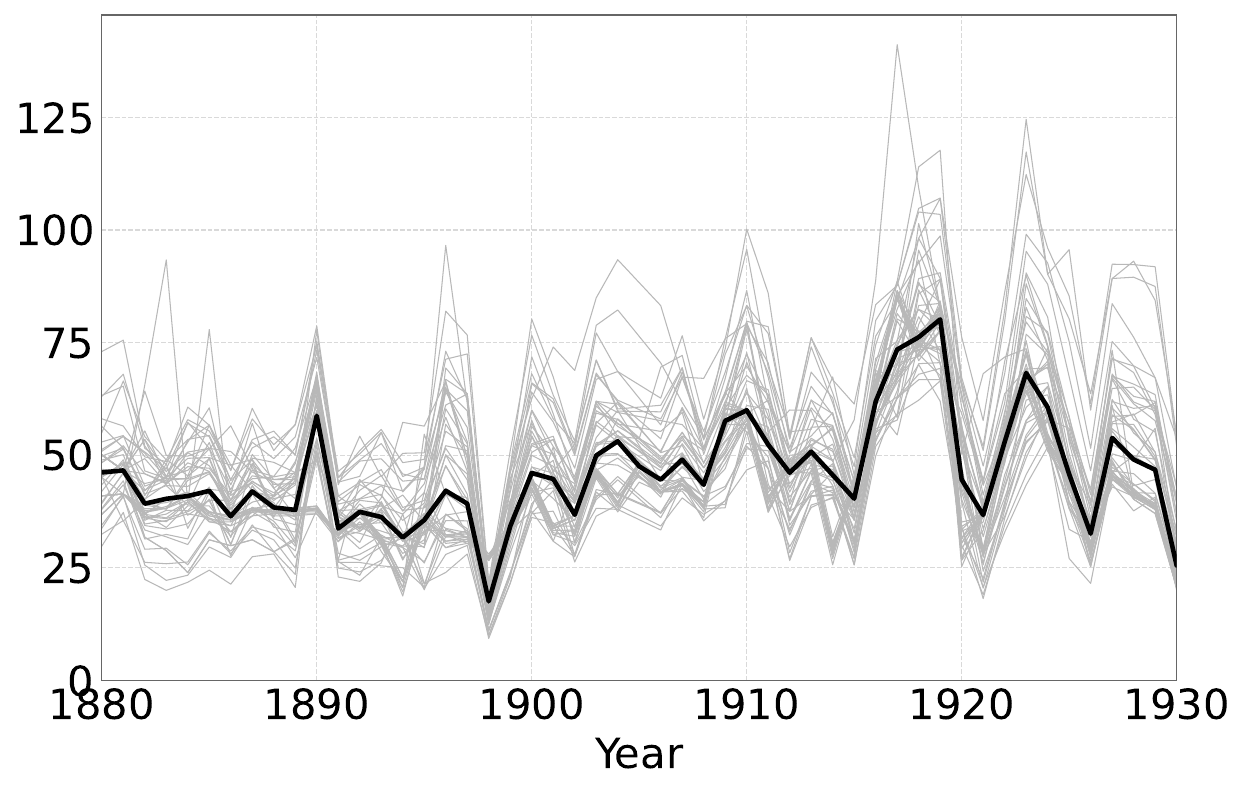}} 
\subfloat[Hay]{\includegraphics[width=0.37\textwidth]{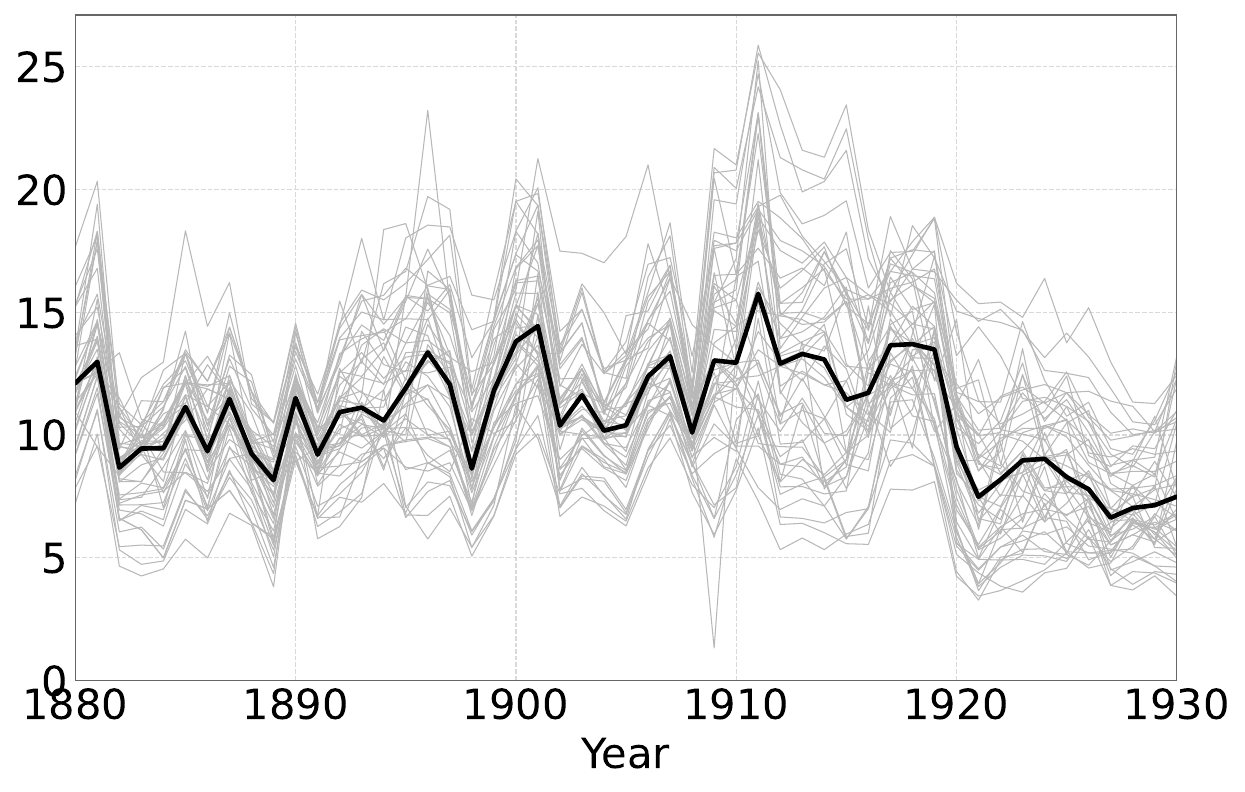}} \\
\subfloat[Oats]{\includegraphics[width=0.37\textwidth]{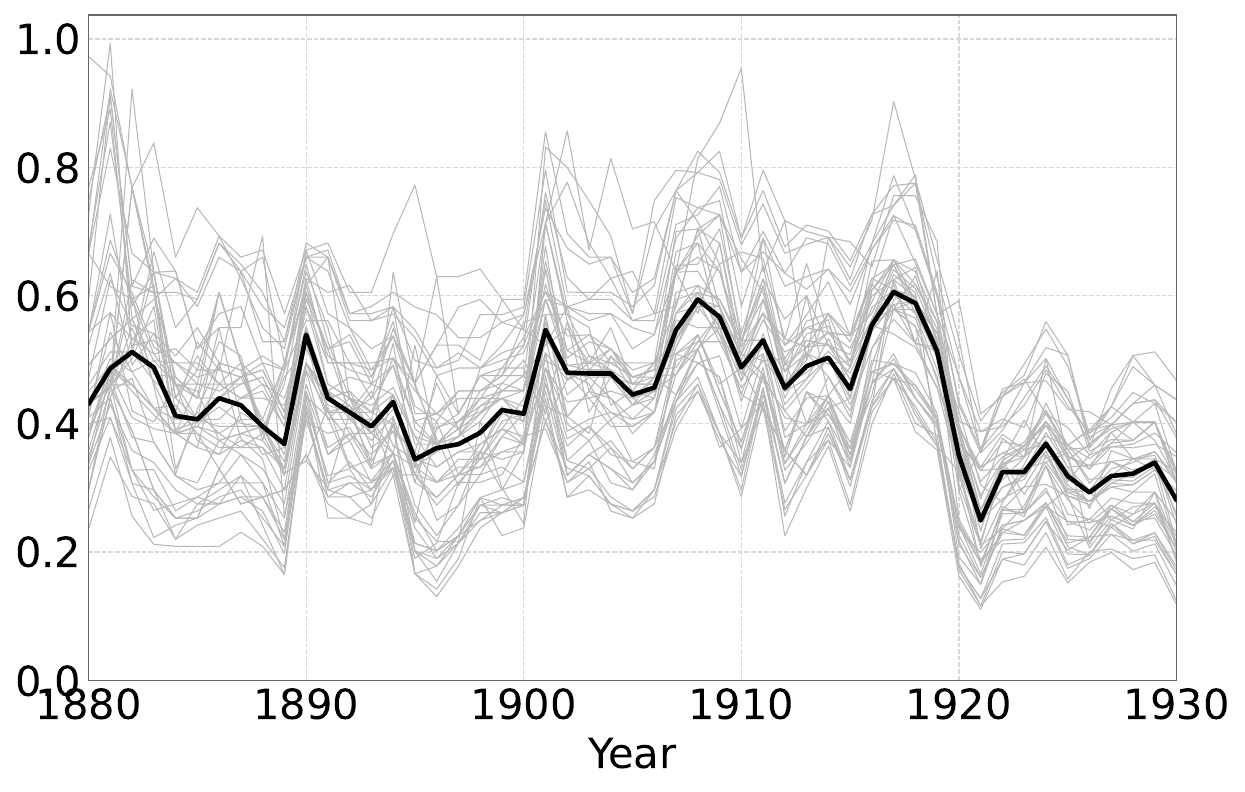}} 
\subfloat[Rye]{\includegraphics[width=0.37\textwidth]{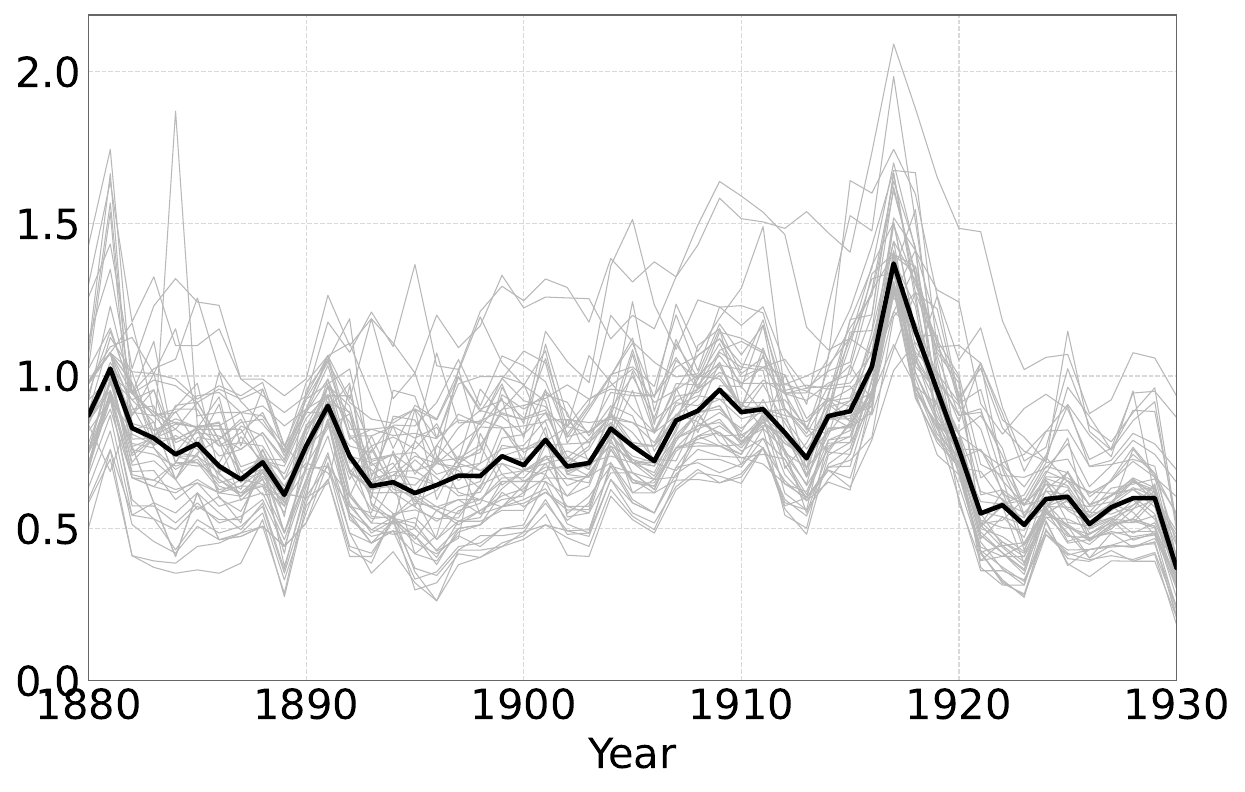}} \\
\subfloat[Wheat]{\includegraphics[width=0.37\textwidth]{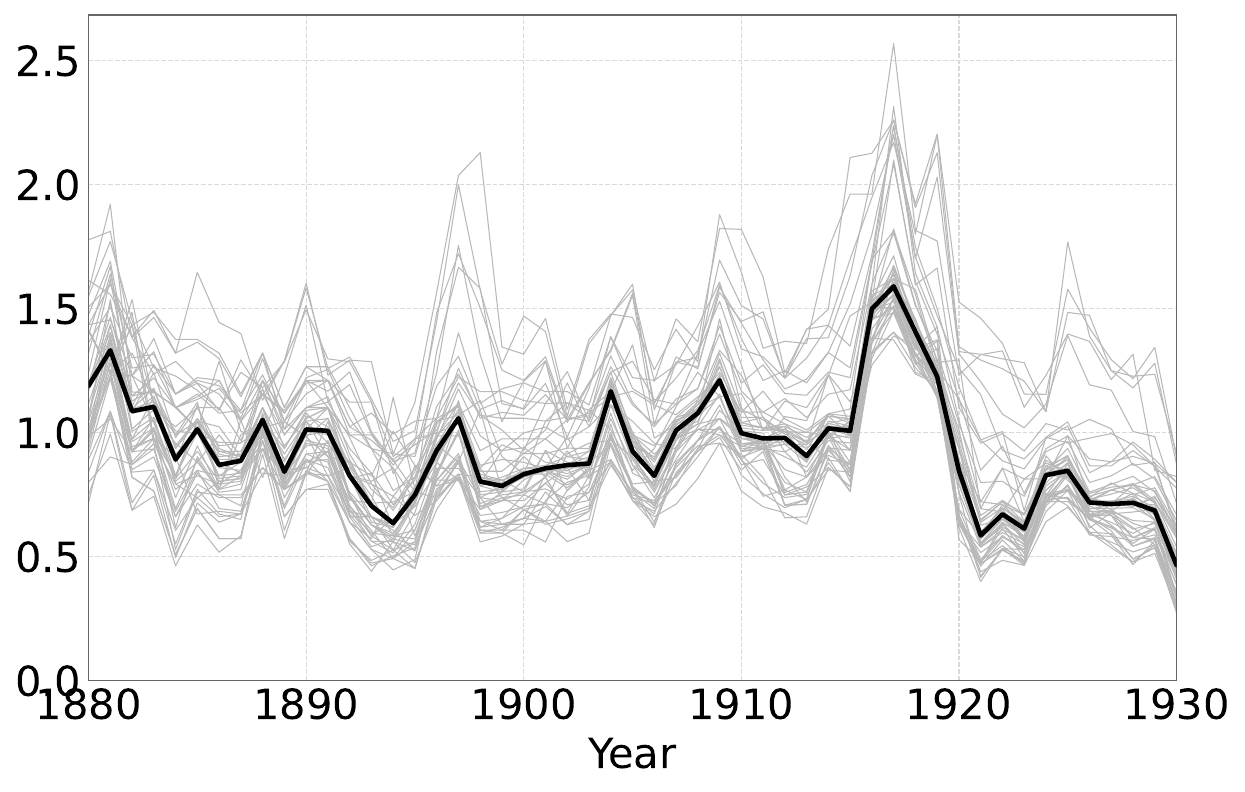}}
\end{center}
\end{adjustwidth}
\vspace{0.2cm}
{\footnotesize\raggedright \textbf{Notes:} These figures plot the time series of prices for each crop by state (grey lines), together with the national average across states (black line).}
\end{figure}

\subsection{Additional information on Kansas farm-level dataset}
\label{app: kansas data}

\subsubsection{Descriptive information}

\begin{figure}[H]
\begin{center}
\caption{Location of townships in the Kansas farm-level dataset \label{fig:mapkansastownships}}
{\includegraphics[width=0.75\textwidth]{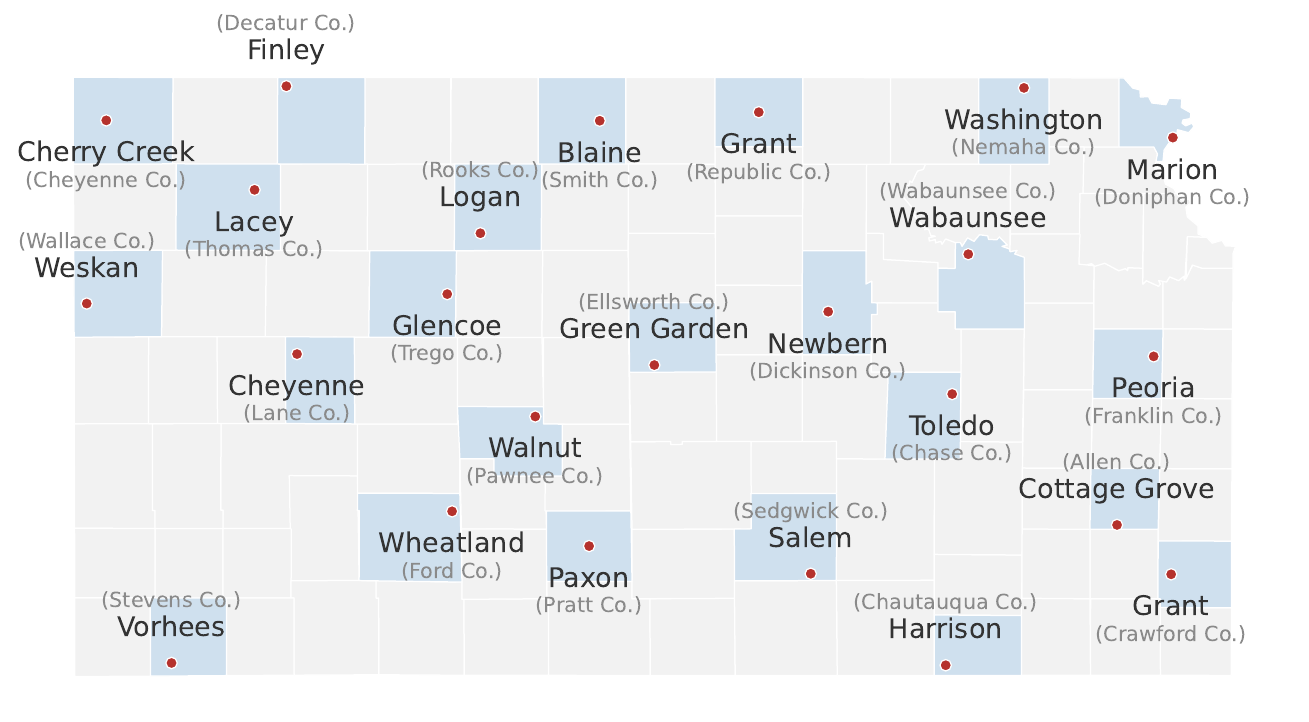}}

{\footnotesize\raggedright \textbf{Notes:} This map shows the names and locations of the 24 townships and corresponding counties included in the Kansas farm-level dataset.\par}
\vspace{-0.25cm}
\end{center}
\end{figure}

\begin{table}[H]
\caption{Summary statistics of cropland shares in the Kansas farm-level dataset \label{tab: sumstatskansas}}
\begin{center}
\resizebox{\textwidth}{!}{\begin{tabular}{l cc cc cc cc cc cc }
\hline \hline
 & \multicolumn{2}{c}{1895} & \multicolumn{2}{c}{1905} & \multicolumn{2}{c}{1915} & \multicolumn{2}{c}{1920} & \multicolumn{2}{c}{1925} & \multicolumn{2}{c}{1930} \\
 & Mean & Median & Mean & Median & Mean & Median & Mean & Median & Mean & Median & Mean & Median \\ \hline
Share of corn & 0.528 & 0.533 & 0.386 & 0.375 & 0.266 & 0.231 & 0.223 & 0.170 & 0.301 & 0.254 & 0.313 & 0.279 \\
Share of wheat & 0.024 & 0.000 & 0.010 & 0.000 & 0.004 & 0.000 & 0.041 & 0.000 & 0.000 & 0.000 & 0.013 & 0.000 \\
Share of oats & 0.130 & 0.091 & 0.064 & 0.000 & 0.079 & 0.000 & 0.097 & 0.052 & 0.074 & 0.000 & 0.068 & 0.000 \\
Share of hay & 0.278 & 0.238 & 0.508 & 0.512 & 0.632 & 0.649 & 0.608 & 0.631 & 0.602 & 0.631 & 0.590 & 0.600 \\
Share of rye & 0.016 & 0.000 & 0.003 & 0.000 & 0.005 & 0.000 & 0.006 & 0.000 & 0.001 & 0.000 & 0.003 & 0.000 \\
Share of barley & 0.024 & 0.000 & 0.029 & 0.000 & 0.013 & 0.000 & 0.026 & 0.000 & 0.022 & 0.000 & 0.013 & 0.000 \\
\hline \hline
\end{tabular}
}
\end{center}
{\footnotesize \textbf{Notes:} This table reports the mean and median share of land allocated to each of the six selected crops across all farms in 1895, 1905, 1915, 1920, 1925, and 1930. Shares are calculated relative to the total acreage planted with the six crops considered.}
\end{table}

\subsubsection{Farmer panel construction \label{app: kansasmatch}}

The \cite{sylvester2002} data record each farmer's last name and given names in every wave, but carry no identifier linking a farm across waves. We construct the farmer panel by matching names across adjacent census waves within township. We first standardize names: we capitalize last names and given names, remove punctuation, expand common given-name abbreviations (e.g., ``Wm'' to ``William'', ``Jno'' to ``John''), and extract middle initials and suffixes. Candidate links are pairs of farms in adjacent waves of the same township whose standardized last names share the same Soundex phonetic encoding. Each candidate pair is assigned the similarity score
\[
\text{score} \;=\; 7 \times JW_{\text{given}} \;+\; 3 \times JW_{\text{last name}},
\]
where $JW$ denotes the Jaro--Winkler similarity of the standardized names, plus $1.5$ if the two records' middle initials agree, minus $2$ if they conflict, and minus $5$ if their suffixes (Sr., Jr., II, III) conflict. We accept candidate pairs scoring at least $7$ (of a maximum of $11.5$) and resolve competing matches by keeping the highest-scoring link for each farm on either side.

\Cref{tab: kansasmatch} reports, for each adjacent pair of waves, the number of farm records and the number and share of links, which ranges between 33 and 49 percent across pairs.  Of the 14{,}560 farm records in the six waves, the procedure identifies 9{,}437 unique farmers, 2{,}773 of whom are observed in two or more waves.

\begin{table}[H]
\centering\small
\caption{Match counts in the farmer panel \label{tab: kansasmatch}}
\vspace{.1cm}
\begin{tabular}{lcccc}
\toprule
Waves & Records ($t$) & Records ($t'$) & Linked & \% of $t$ linked \\
\midrule
1895--1905 & 2,240 & 2,458 & 811 & 36.2 \\
1905--1915 & 2,458 & 2,561 & 815 & 33.2 \\
1915--1920 & 2,561 & 2,499 & 1,124 & 43.9 \\
1920--1925 & 2,499 & 2,486 & 1,217 & 48.7 \\
1925--1930 & 2,486 & 2,316 & 1,156 & 46.5 \\
\bottomrule
\end{tabular}

\begin{minipage}{0.9\textwidth}
\smallskip\footnotesize{\textbf{Notes:} This table reports, for each adjacent pair of census waves $(t, t')$, the number of farm records with usable names in each wave and the number and share of links between them.}
\end{minipage}
\end{table}

\subsubsection{Matching farmers to fraternal organizations \label{app: kansasmatchfrat}}

We use the same name standardization and Jaro--Winkler weights to link farmers to the fraternal death-notice registers, omitting the middle-initial and suffix terms. A candidate match is penalized by 3 points if the recorded death year precedes the farmer's first census appearance, and the same acceptance threshold of 7 applies. Of the 53{,}823 death notices, 10{,}995 unique persons are located in the 24 sample counties, and 583 are matched to a farmer in our data.

It is worth noting, however, that death-notice records understate true lifetime membership, since a member must have died during a window in which the relevant organization's records survived.

\section{Methodology appendix}

\subsection{Yield prediction and imputation}

\subsubsection{Climate feature construction}
\label{app: climate features}

\begin{table}[H]
\begin{center}
\small
\renewcommand{\arraystretch}{1.15}
\caption{Climate-feature construction \label{tab: climatefeatures}}
\begin{tabularx}{\textwidth}{@{} l X r @{}}
\toprule
\multicolumn{3}{@{}l}{\textit{(a) Base statistics (10), computed per window}}\\
\midrule
Temperature (6) & mean, maximum, minimum, std.\ dev.; frost days ($T_{\min}<0^\circ$C), hot days ($T_{\max}>30^\circ$C) & \\
Precipitation (4) & total, maximum daily; wet days ($>1$\,mm), longest dry spell & \\
\addlinespace
\multicolumn{3}{@{}l}{\textit{(b) Temporal windows (24-month lag horizon, most recent $\to$ oldest)}}\\
\midrule
\textbf{Frequency} & \textbf{Lag window} & \textbf{\# windows}\\
\midrule
Dekadal (3/month) & Dec$(Y\!-\!1)\to$ Jan$(Y\!-\!2)$ (24 months) & 72\\
Monthly           & Dec$(Y\!-\!1)\to$ Jan$(Y\!-\!2)$ (same 24-month span) & 24\\
Seasonal (3$\times$month)    & offset one month: drops Dec$(Y\!-\!1)$, adds Dec$(Y\!-\!3)$ & 8\\
\addlinespace
                  & \textit{Windows per statistic} & \textbf{104}\\
\midrule
\multicolumn{2}{@{}l}{\textbf{Total features} $=10\ \text{statistics}\times104\ \text{windows}$} & \textbf{1{,}040}\\
\bottomrule
\end{tabularx}
\end{center}
{\footnotesize\raggedright \textbf{Notes:} This table describes the construction of the climate features. $Y$ is the census year (e.g. 1900), with data for the cultivation year $Y-1$ (e.g. 1899). The seasonal (meteorological) windows are offset by one month so that the prior-prior winter is fully enclosed.}
\end{table}

\subsubsection{Census-wave yield prediction \label{app: yield prediction}}

The goal of the prediction exercise is to use climate information in each county and census year to predict yields above and beyond the fixed-effect baseline given by county, year, and state-year effects. For each crop and census wave pair, we use the $1{,}040$ county-level climate variables to estimate six prediction models: LASSO, ridge, elastic net, gradient boosting, random forest, and quantile gradient boosting.

\begin{figure}[H]\centering
\caption{Out-of-fold predictive performance of the yield-prediction model}\label{fig: oofr2}
\vspace{.15cm}
\includegraphics[width=0.78\textwidth]{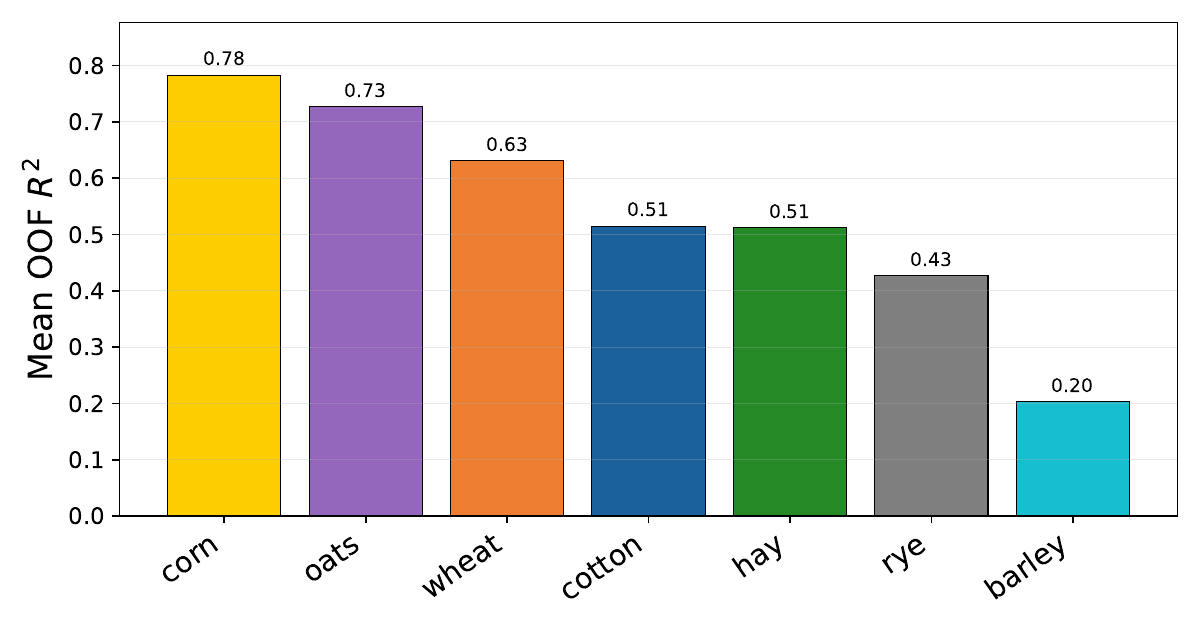}

{\footnotesize\raggedright \textbf{Notes:} This figure reports the mean out-of-fold $R^2$ of the stacked yield predictions by crop, averaged over the six census-wave pairs. \Cref{tab: stackerr2} reports the out-of-fold $R^2$ by crop and wave pair.\par}
\end{figure}

We evaluate each method using county-aware five-fold cross-validation, re-fitting the fixed-effect residualization within each training fold to avoid leakage from the held-out observations. We then stack the out-of-fold predictions separately for each crop and census wave pair, using non-negative weights that sum to one.\footnote{The cross-validation stage also included a seventh candidate, a spatial $k$-nearest-neighbor learner that predicts a county's residual from those of nearby counties using county coordinates rather than the climate features. Because it is not a climate response, it is excluded from the ensemble used for intercensal imputation, and the six climate methods' weights are renormalized; its average stacking weight is $0.046$. The out-of-fold $R^2$ in \Cref{fig: oofr2} and \Cref{tab: stackerr2} is computed for the full cross-validated stack, while \Cref{tab: methodweights} reports the renormalized weights of the deployed six-method ensemble.} The weights are chosen to minimize the out-of-fold prediction error, so methods that perform better for a given crop and period receive larger weights. The non-negativity and unit-sum constraints keep the ensemble within the convex hull of its component predictions and reduce extrapolation when the model is applied to intercensal years. \Cref{app: yieldpredweights} reports the resulting weights. \Cref{fig: oofr2} reports the average out-of-fold $R^2$ of the stacked predictions for each crop, while  \Cref{tab: stackerr2} reports the out-of-fold $R^2$ by crop and census wave pair.

\begin{table}[H]
\centering
\caption{Stacked out-of-fold $R^2$ by crop and wave pair \label{tab: stackerr2}}
\vspace{.1cm}
\small
\begin{tabular}{l r r r r r r r r}
\toprule
crop & avg $n$ & 1879--89 & 1889--99 & 1899--09 & 1909--19 & 1919--24 & 1924--29 & \textbf{avg} \\
\midrule
corn   & 2{,}673 & 0.813 & 0.783 & 0.782 & 0.776 & 0.759 & 0.781 & \textbf{0.782} \\
oats   & 2{,}623 & 0.782 & 0.770 & 0.772 & 0.671 & 0.716 & 0.651 & \textbf{0.727} \\
wheat  & 2{,}422 & 0.685 & 0.601 & 0.621 & 0.619 & 0.631 & 0.625 & \textbf{0.630} \\
cotton &  894  & 0.577 & 0.417 & 0.415 & 0.673 & 0.554 & 0.452 & \textbf{0.515} \\
hay    & 2{,}680 & 0.389 & 0.361 & 0.516 & 0.612 & 0.575 & 0.615 & \textbf{0.512} \\
rye    & 2{,}078 & 0.501 & 0.445 & 0.343 & 0.366 & 0.476 & 0.428 & \textbf{0.426} \\
barley & 1{,}644 & $-0.040$ & 0.134 & 0.260 & 0.173 & 0.325 & 0.328 & \textbf{0.203} \\
\bottomrule
\end{tabular}
\begin{minipage}{\textwidth}
\smallskip{\footnotesize\raggedright \textbf{Notes:} This table shows the out-of-fold $R^2$ of the stacked yield predictions by crop and wave pair. ``Avg $n$'' is the average number of counties with a positive yield observation across the seven census waves considered. The reported $R^2$ is for the full cross-validated stack.}
\end{minipage}
\end{table}

\Cref{tab: climateuplift} reports the predictive gain from adding climate features to the fixed-effect baseline. Specifically, it shows the difference in out-of-fold $R^2$ between the climate-augmented model and the county, year, and state-year fixed-effect baseline. The gain is largest for cotton, whose fixed-effect baseline is relatively weak. In contrast, the gain is lower for rye, for which county fixed effects already capture much of the observed variation.

\begin{table}[H]\centering\small
\caption{Climate-feature uplift over the fixed-effect baseline, by crop and wave pair \label{tab: climateuplift}}
\begin{tabular}{l r r r r r r r}
\toprule
crop & 1879--89 & 1889--99 & 1899--09 & 1909--19 & 1919--24 & 1924--29 & \textbf{avg} \\
\midrule
corn   & 0.047 & 0.077 & 0.064 & 0.087 & 0.087 & 0.069 & \textbf{0.072} \\
oats   & 0.055 & 0.040 & 0.048 & 0.130 & 0.086 & 0.113 & \textbf{0.079} \\
wheat  & 0.100 & 0.108 & 0.078 & 0.096 & 0.161 & 0.102 & \textbf{0.108} \\
cotton & 0.377 & 0.277 & 0.129 & 0.513 & 0.342 & 0.300 & \textbf{0.323} \\
hay    & 0.088 & 0.055 & 0.077 & 0.043 & 0.202 & 0.105 & \textbf{0.095} \\
rye    & 0.032 & 0.015 & 0.020 & 0.041 & 0.038 & 0.027 & \textbf{0.029} \\
barley & 0.065 & 0.063 & 0.092 & 0.058 & 0.113 & 0.133 & \textbf{0.087} \\
\bottomrule
\end{tabular}
\begin{minipage}{\textwidth}\smallskip{\footnotesize\raggedright \textbf{Notes:} This table shows the difference in out-of-fold $R^2$ between the climate-augmented model and the county, year, and state-year fixed-effect baseline, by crop and census-wave pair.}\end{minipage}
\end{table}

\subsubsection{Stacking weights \label{app: yieldpredweights}}

\Cref{tab: methodweights} reports the average stacking weights by crop and prediction method. Ridge regression receives the largest average weight, followed by gradient boosting and quantile gradient boosting.

\begin{table}[H]\centering\small
\caption{Average stacking weight by crop and method \label{tab: methodweights}}
\setlength{\tabcolsep}{6pt}
\begin{tabular}{l r r r r r r}
\toprule
crop & lasso & ridge & elastic net & GBM & RF & qGBM \\
\midrule
corn   & 0.000 & 0.710 & 0.000 & 0.278 & 0.000 & 0.011 \\
oats   & 0.000 & 0.680 & 0.000 & 0.270 & 0.044 & 0.006 \\
wheat  & 0.000 & 0.678 & 0.000 & 0.303 & 0.005 & 0.014 \\
cotton & 0.000 & 0.765 & 0.000 & 0.102 & 0.000 & 0.133 \\
hay    & 0.000 & 0.590 & 0.000 & 0.172 & 0.056 & 0.182 \\
rye    & 0.000 & 0.276 & 0.000 & 0.053 & 0.137 & 0.535 \\
barley & 0.000 & 0.383 & 0.000 & 0.154 & 0.054 & 0.409 \\
\midrule
Avg & 0.000 & 0.583 & 0.000 & 0.190 & 0.042 & 0.184 \\
\bottomrule
\end{tabular}
\begin{minipage}{\textwidth}\smallskip{\footnotesize\raggedright \textbf{Notes:} This table shows the stacking weights used for each crop, averaged across the six wave pairs. The ``Avg'' row averages each method's weight across the seven core crops. Lasso and elastic net receive negligible weight for these crops because ridge absorbs the linear-prediction signal.}\end{minipage}
\end{table}

\subsubsection{Intercensal yield imputation}
\label{app: yieldpredictionfallbackadd}

The baseline imputation for each crop interpolates each county's fixed-effect yield level between two census waves and adds the stacked climate-predicted residual. However, because the climate signal can be less reliable for crops grown sparsely within a cultivation cluster, we assess out-of-sample fit, and in cells where the within-cluster out-of-fold $R^2$ is negative, we omit the climate prediction.

Let $k$ denote the cultivation cluster of county $c$, and let $R^2_{i,k,p}$ denote the within-cluster out-of-fold $R^2$ for crop $i$ in wave pair $p$. We impute
\[
\widehat{\tilde q}_{i,c,\tau} =
\begin{cases}
(1-f_\tau)\widehat{\text{FE}}_{i,c,p_s} + f_\tau\widehat{\text{FE}}_{i,c,p_e} + \hat\varepsilon^{\text{stack}}_{i,c,\tau}
& \text{if } R^2_{i,k,p}\geq 0,\\
(1-f_\tau)\widehat{\text{FE}}_{i,c,p_s} + f_\tau\widehat{\text{FE}}_{i,c,p_e}
& \text{if } R^2_{i,k,p}< 0.
\end{cases}
\]
Thus, when the within-cluster fit is adequate, we use the climate prediction. When it is poor, we drop the unreliable local climate residual, leaving only the interpolated fixed-effect level.

\begin{figure}[H]
\centering
\caption{Imputed national-average per-acre yields by crop, 1878--1929 \label{fig: imputedpanels}}
\vspace{.1cm}
\includegraphics[width=\textwidth]{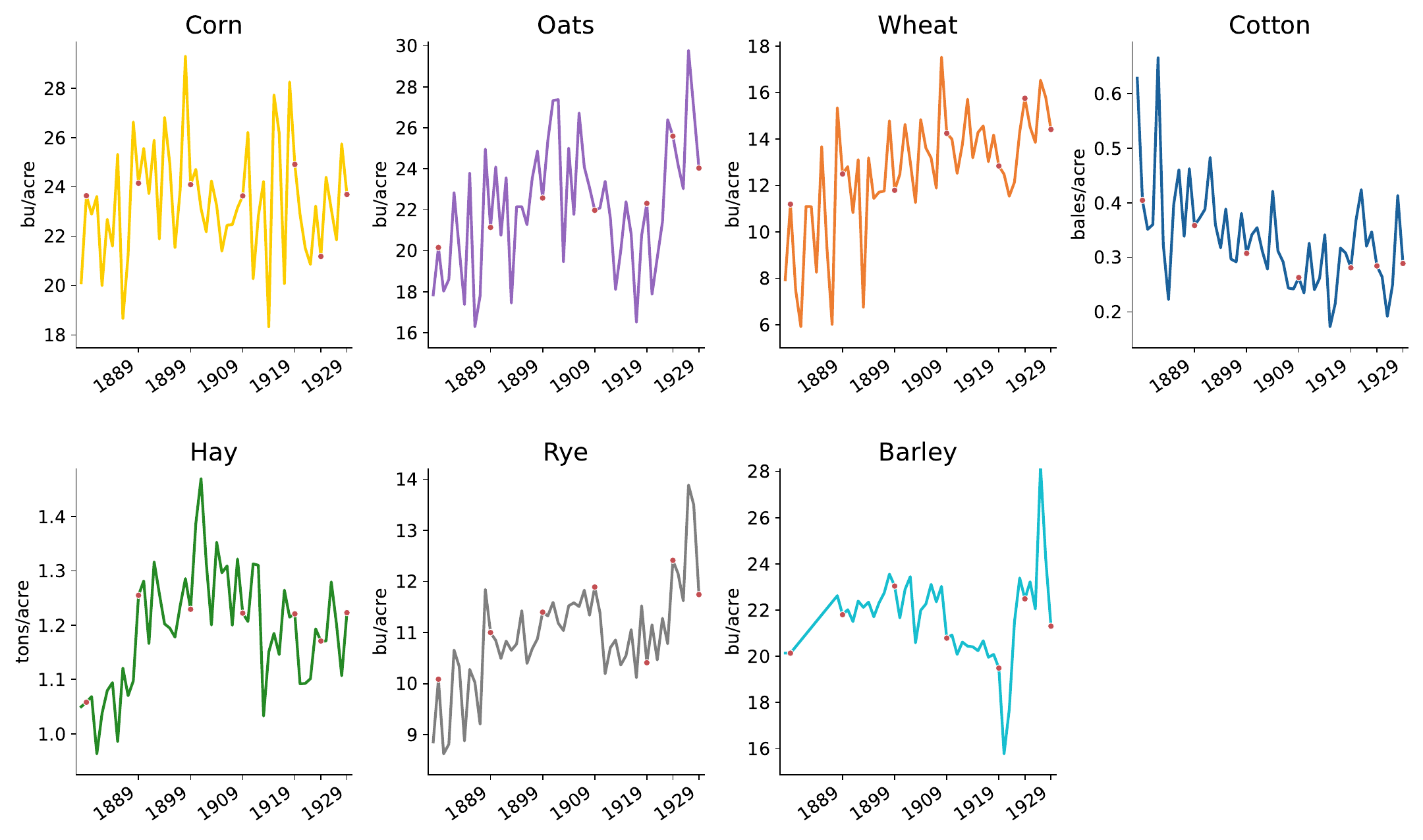}
\begin{minipage}{\textwidth}
\smallskip{\footnotesize\raggedright \textbf{Notes:} This figure shows the average imputed per-acre yield across counties, by crop and crop year. Red points mark the census years.}

\end{minipage}
\end{figure}

\subsection{Returns to scale parameters}
\label{app: RTS}

A crop--cluster--wave cell is estimated locally when at least 30 counties in the cluster report positive acreage and output for the crop in both waves of the pair; cells below this floor instead use the national estimate $\hat\alpha_{i,t}$, obtained from the same long-difference specification estimated on all counties nationally. Of the 168 crop--cluster--wave cells, 11 (6.5\%) use the national fallback.

\begin{figure}[H]\centering
\caption{Returns-to-scale parameter by crop}\label{fig: alphacropseries}
\vspace{.1cm}
\includegraphics[width=\textwidth]{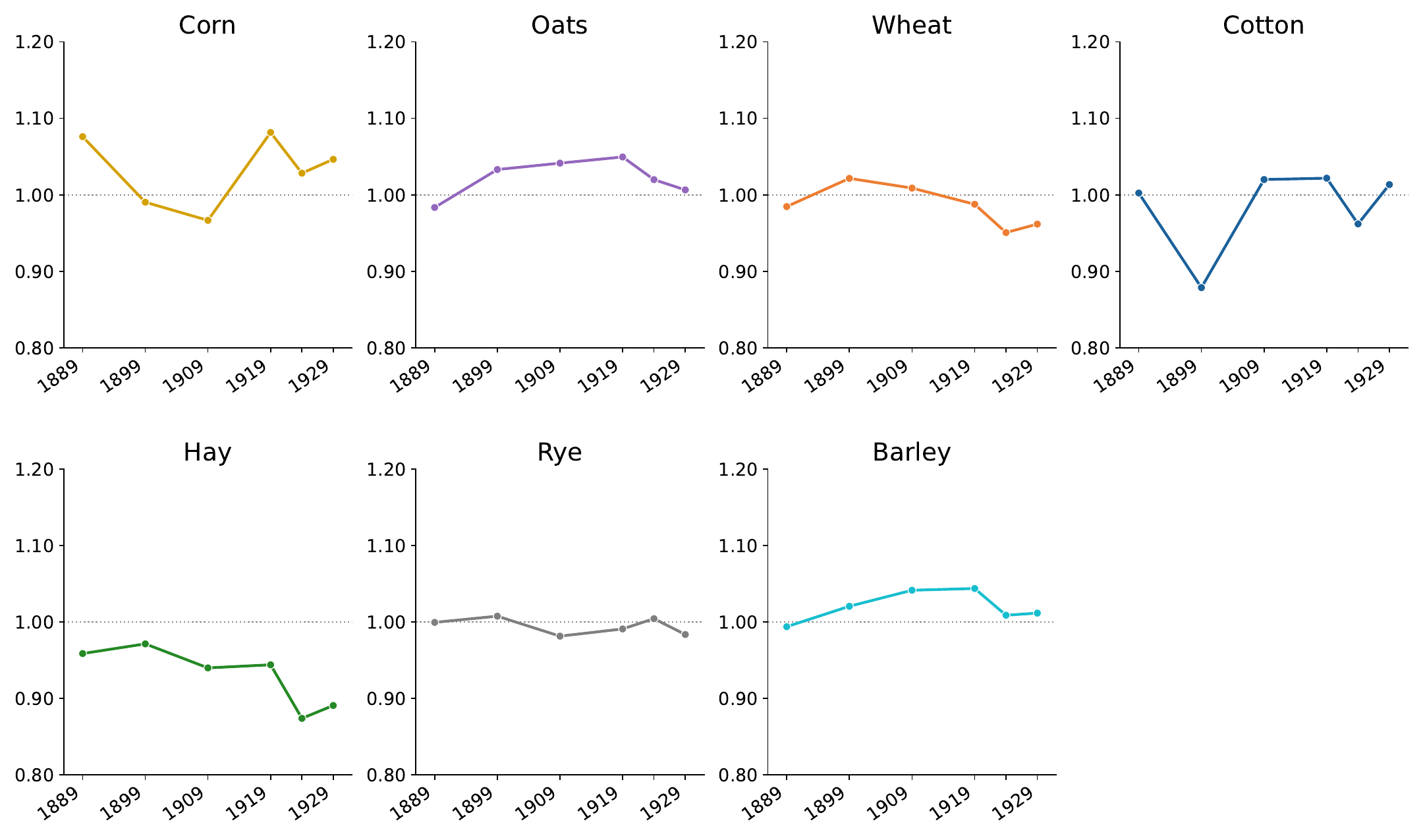}

{\footnotesize\raggedright \textbf{Notes:} This figure shows the national average estimated returns-to-scale parameter $\widehat\alpha$ by crop and
census wave. The dotted line
marks constant returns to scale ($\alpha=1$).\par}
\end{figure}

\begin{figure}[H]
\centering
\caption{Estimated returns-to-scale parameters by cultivation cluster}
\label{fig: alphacluster}
\vspace{.1cm}
\includegraphics[width=\textwidth]{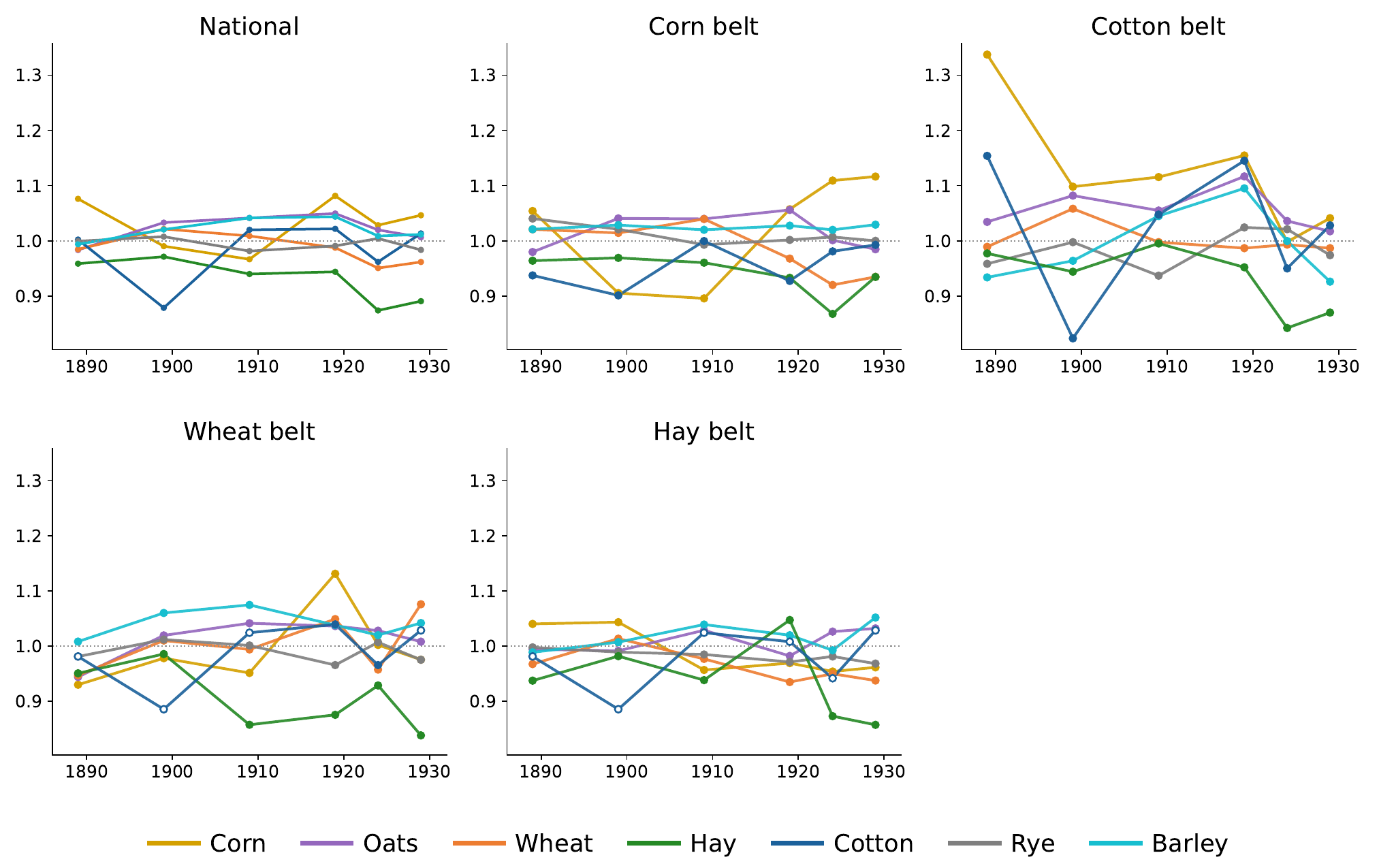}
\begin{minipage}{\textwidth}
\smallskip{\footnotesize\raggedright \textbf{Notes:} These figures show the estimated returns-to-scale parameter $\hat\alpha$ by crop and crop year nationally and in each cultivation cluster. The national panel is the cross-county mean. All panels share a common, tightened $\hat\alpha$ axis centered on the constant-returns value $\alpha=1$ (dotted); filled markers denote locally estimated (non-fallback) cells and hollow markers the cells that fall back to the national estimate.}
\end{minipage}
\end{figure}

\subsection{Model estimation}
\label{app: rho}

\subsubsection{Estimation procedure \label{app: modelest}}

We estimate the coefficient of relative risk aversion by \emph{spatio-temporal pooled OLS}: a no-intercept regression of the first-order conditions, fit on a local pool of nearby counties and adjacent census waves whose observations are weighted by a Gaussian distance kernel.

Recall that for each crop $i$ grown in county $c$ in
census year $t$, lying in cultivation belt $k(c,t)$, we form
\begin{equation}
    Y_{i,c,t}=\hat\alpha_{i,k(c,t),t}\times\hat\mu(i,c,t)-\hat\alpha_{\text{corn},k(c,t),t}\times\hat\mu(\text{corn},c,t)
\end{equation}

and
\begin{equation}
    X_{i,c,t}= 
\frac{\hat\alpha_{i,k(c,t),t}\sum_j (l_{j,c,t}\bar L_{c,t})\hat\Sigma(ji,c,t)-\hat\alpha_{\text{corn},k(c,t),t}\sum_j (l_{j,c,t}\bar L_{c,t})\hat\Sigma(j,\text{corn},c,t)}
{\sum_j (l_{j,c,t}\bar L_{c,t})\hat\mu(j,c,t)},
\end{equation}

where $l_{j,c,t}\bar L_{c,t}$ is the acreage planted with crop $j$, and
$\hat\mu$, $\hat\Sigma$, and $\hat\alpha$ are the expected-return moments and
returns-to-scale parameters. The model's first-order conditions imply the linear relationship
$Y_{i,c,t}=\rho_{c,t}\,X_{i,c,t}+\varepsilon_{i,c,t}$.\footnote{Note that because the moments are computed from observed per-acre revenues, they embed each window year's own acreage rather than the census-year allocation at which the first-order conditions are evaluated. Because intercensal acreage is unobserved, accounting for this requires an assumption of its own. We implement one such correction --- de-scaling each window year's revenue by its log-linearly interpolated acreage and re-evaluating the moments at the census-year allocation --- which yields estimates with a rank correlation of $0.99$ with the baseline and leaves the results substantively unchanged. We therefore retain the simpler baseline construction.}

Because a single county-year supplies at most six such crop-level conditions, we do not estimate the slope cell by cell. Instead, we pool the crop-level observations across space and time with a separable Gaussian distance kernel.

\subsubsection{Spatial kernel}
In space, let $d_{c,j}$ be the great-circle distance between the centroids of the (1900-harmonized) counties $c$ and $j$, and let $h_c$ be the distance from $c$ to its sixth-nearest county. The spatial weight is the adaptive-bandwidth Gaussian
\[
K^{s}_{c,j}=\exp\!\Big(-\tfrac12\big(d_{c,j}/h_c\big)^{2}\Big)\,
            \mathbf 1\{d_{c,j}\le h_c\},
\]
truncated at the sixth neighbor: county $c$ itself ($d_{c,c}=0$, weight one) and its six
nearest neighbors receive positive, distance-decaying weight, and all other counties
receive zero. The adaptive bandwidth $h_c$ makes the pool tighter where counties are dense
and wider where they are sparse.

\subsubsection{Choosing the spatial bandwidth}
We select the six-neighbor bandwidth by cross-validation. The bandwidth trades off bias against variance: too tight a pool leaves sampling noise too high, while too wide a pool averages counties with genuinely different preferences. For each candidate $K$, we predict every county-wave's unpooled $\widehat{\rho}$ from a $K$-nearest-neighbor moment pool of its neighbors, excluding the county itself, and score by the rank correlation between the held-out predictions and the held-out values. We use two criteria: a \emph{same-wave} criterion, which predicts a county's wave-$t$ value from its neighbors' wave-$t$ moments, and a stricter \emph{wave-held-out} criterion, which predicts it from the neighbors' moments in the other waves only.

\begin{figure}[H]\centering
\caption{Cross-validating the spatial bandwidth}\label{fig: cvbandwidth}
\vspace{.1cm}
\includegraphics[width=0.74\textwidth]{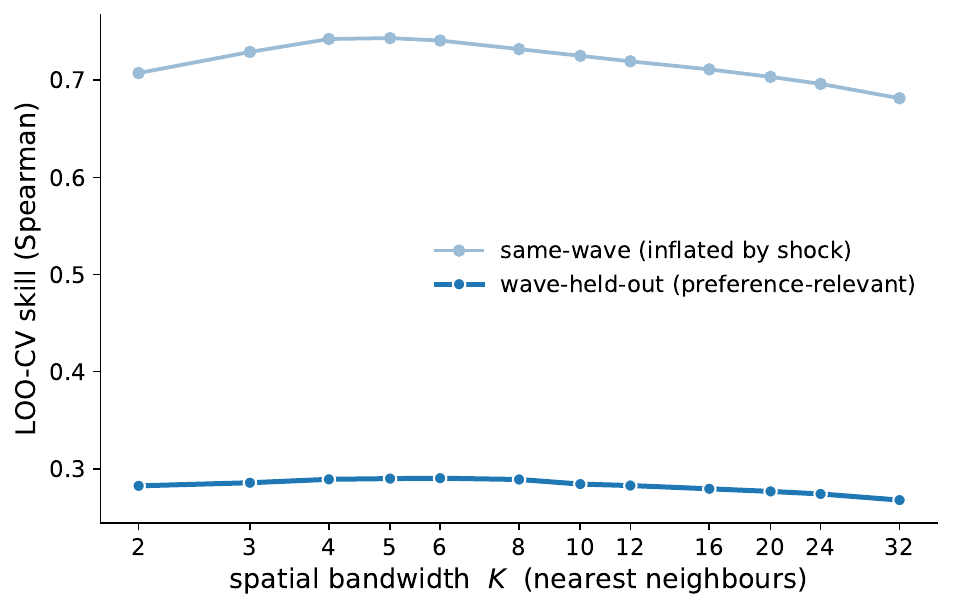}

{\footnotesize\raggedright \textbf{Notes:} Spatial leave-one-out CV skill (Spearman rank correlation between held-out predictions and held-out unpooled $\widehat{\rho}$) against the number of nearest neighbors $K$, pooled across the six census waves. The same-wave criterion predicts each county from its neighbors' same-wave moments; the wave-held-out criterion uses the neighbors' moments from other waves only, removing the shared within-wave regional shock. The gap between the curves is that shock. \par}
\end{figure}

\Cref{fig: cvbandwidth} reports both criteria. The same-wave skill peaks at $K=5$ (a median pool radius of about $51$ km) and the wave-held-out skill at $K=4$ (about $47$ km); both curves are nearly flat over $K=4$--$8$, with $K=6$ (the midpoint of the flat band) within $0.002$ of either optimum, and decline steadily at larger bandwidths. The $\sim$50 km scale coincides with the $0.7^{\circ}$ cell of the underlying climate reanalysis ($55$--$78$ km across). The estimates are in any case insensitive to the choice within the cross-validated range: re-estimating the pooled measure with $K=4$ or $K=8$ moves the median $\widehat{\rho}$ by less than $0.02$ on the identical sample, and a per-county adaptive-bandwidth variant, in which each county selects its own $K$ by local cross-validation, averages $K\approx 6$ and leaves the median unchanged.

\subsubsection{Temporal kernel}
In time, we use a rolling three-wave window consisting of the target wave $t$ together with
the immediately preceding and following census waves. Letting $\Delta_{t,t'}$ denote the
crop-year gap between waves $t$ and $t'$, and $h_t$ the largest gap in the window, the
temporal weight is
\[
K^{t}_{t,t'}=
\begin{cases}
1, & t'=t,\\[2pt]
\exp\!\big(-\tfrac12\big(\Delta_{t,t'}/h_t\big)^{2}\big),
   & t'\ \text{the preceding or following wave.}
\end{cases}
\]
The census years are $1889,1899,1909,1919,1924,1929$, so adjacent gaps are ten years
through 1919 and five years thereafter.

\subsubsection{Pooled estimator}

$\hat\rho_{c,t}$ is the OLS slope of $Y$ on $X$ on the pooled sample that stacks every crop-level observation from these counties and waves,
each weighted by the product $K^{s}_{c,j}K^{t}_{t,t'}$ of its spatial and temporal kernel
weights. The kernel supplies \emph{distance} weights: observations are down-weighted
by their spatial and temporal distance from the target county-year.

\subsection{Kansas farm-level risk aversion}
\label{app: kansas methodology}

\subsubsection{Expected returns \label{app: kansasexpr}}

We construct the expected-return and covariance moments used in the farm- and township-level first-order conditions using the same procedure as in the main analysis, adapted to the Kansas waves. Let \(w\) index townships, \(c(w)\) the county containing township \(w\), \(i\) crops, and \(t\) Kansas survey waves. For each township-year, we assign crop \(i\)'s expected per-acre revenue and covariance moments from the corresponding county-level return history:
\[
\mu_{i,w,t}
=
E_t\left[p_{i,KS,\tau}\widehat{\tilde q}_{i,c(w),\tau}\right],
\qquad
\Sigma_{ij,w,t}
=
\operatorname{Cov}_t\left(
p_{i,KS,\tau}\widehat{\tilde q}_{i,c(w),\tau},
p_{j,KS,\tau}\widehat{\tilde q}_{j,c(w),\tau}
\right),
\]
where \(\widehat{\tilde q}_{i,c(w),\tau}\) is the imputed county-level per-acre yield for crop \(i\), \(p_{i,KS,\tau}\) is the Kansas state price for crop \(i\), and the expectation and covariance are computed over the relevant ten-year window.

\subsubsection{Returns to scale parameters \label{app: kansasalpha}}

The returns-to-scale parameters are also borrowed from the county-level analysis. For each Kansas township-year, we assign the township to its cultivation cluster and use the corresponding \(\alpha_{i,g(w,t),p(t)}\), where \(g(w,t)\) denotes the cultivation cluster of township \(w\) in year \(t\) and \(p(t)\) denotes the nearest available census wave in the main analysis. In particular, 1895 uses the 1889 estimates; 1905 uses the 1899 estimates; 1915 uses the 1909 estimates; 1920 uses the 1919 estimates; 1925 uses the 1924 estimates; and 1930 uses the 1929 estimates.

\subsubsection{Estimation procedure and spatial kernel \label{app: kansasspatial}}

We recover farm- and township-level risk aversion with the same pooled first-order-condition estimator as in the main analysis. The township-level coefficient is formed from the aggregate crop allocation of all farms in township \(w\) in year \(t\).  By contrast, each farm's first-order-condition moments are pooled with those of its \(k=30\) nearest neighbors in the within-township census enumeration order.\footnote{This number of neighbors strikes a balance between the size of each township and the need for enough moments to estimate farm-level risk aversion reliably. The comparison of farm- and township-level risk-aversion coefficients is not sensitive to the pooling bandwidth.} The enumeration order is a useful proxy for geographic proximity because the census was conducted door to door within township.

Let \(d_{ff'}\) denote the absolute distance between farms \(f\) and \(f'\) in the enumeration order, and let \(h_f\) be the distance from farm \(f\) to its \(k\)-th nearest neighbor. We assign weights
\[
\omega_{ff'}
=
\exp\!\left(-\frac{1}{2}\big( d_{ff'}/h_f\big)^2\right)
\mathbf{1}\{d_{ff'}\le h_f\}.
\]
Thus, farm \(f\) itself receives weight one, nearby farms receive positive distance-decaying weight, and farms outside the \(k\)-nearest-neighbor window receive zero weight. The farm-level estimate \(\widehat{\rho}_{f,t}\) is the slope from a weighted first-order-condition regression that stacks the crop-level observations of farms in this local neighborhood, with each neighbor farm's crop-level moments weighted by \(\omega_{ff'}\).

Unlike the county estimator, this pooling is purely spatial and within a single wave: farms cannot be tracked across waves, so there is no temporal kernel. We also do not pool across townships, since that would break the comparison between a township and its own farms.

\begin{table}[H]
\centering\small
\caption{Kansas farm-level sample size and farm-level risk aversion estimates by wave \label{tab: kansassamplesize}}
\vspace{.1cm}
\begin{tabular}{lcc}
\hline\noalign{\vskip 4pt}
Wave & Farm records & Farms with $\widehat{\rho}$ \\
\midrule
1895 & 2,283 & 1,833 \\
1905 & 2,482 & 2,232 \\
1915 & 2,585 & 2,386 \\
1920 & 2,505 & 2,354 \\
1925 & 2,497 & 2,253 \\
1930 & 2,326 & 2,148 \\
\midrule
Total & 14,678 & 13,206 \\
\noalign{\vskip 4pt}\hline
\end{tabular}

\begin{minipage}{0.72\textwidth}
\smallskip\footnotesize{\textbf{Notes:} Number of farm records in the \cite{sylvester2002} Kansas data by wave, and the number of farms for which a farm-level risk-aversion coefficient $\widehat{\rho}$ is estimated (farms with positive core-crop acreage and a valid first-order-condition slope). All 24 sample townships have a township-level $\widehat{\rho}$ in every wave.}
\end{minipage}
\end{table}

\section{Method validation appendix}

\subsection{Revenue panel and expected returns}
\label{app: returns}

\subsubsection{Mean--variance positions by census wave}

\begin{figure}[H]
\begin{center}
\caption{Expected crop returns: mean--variance positions by census wave (1889, 1899, 1909)}\label{retmvwavesA}
\includegraphics[width=0.6\textwidth]{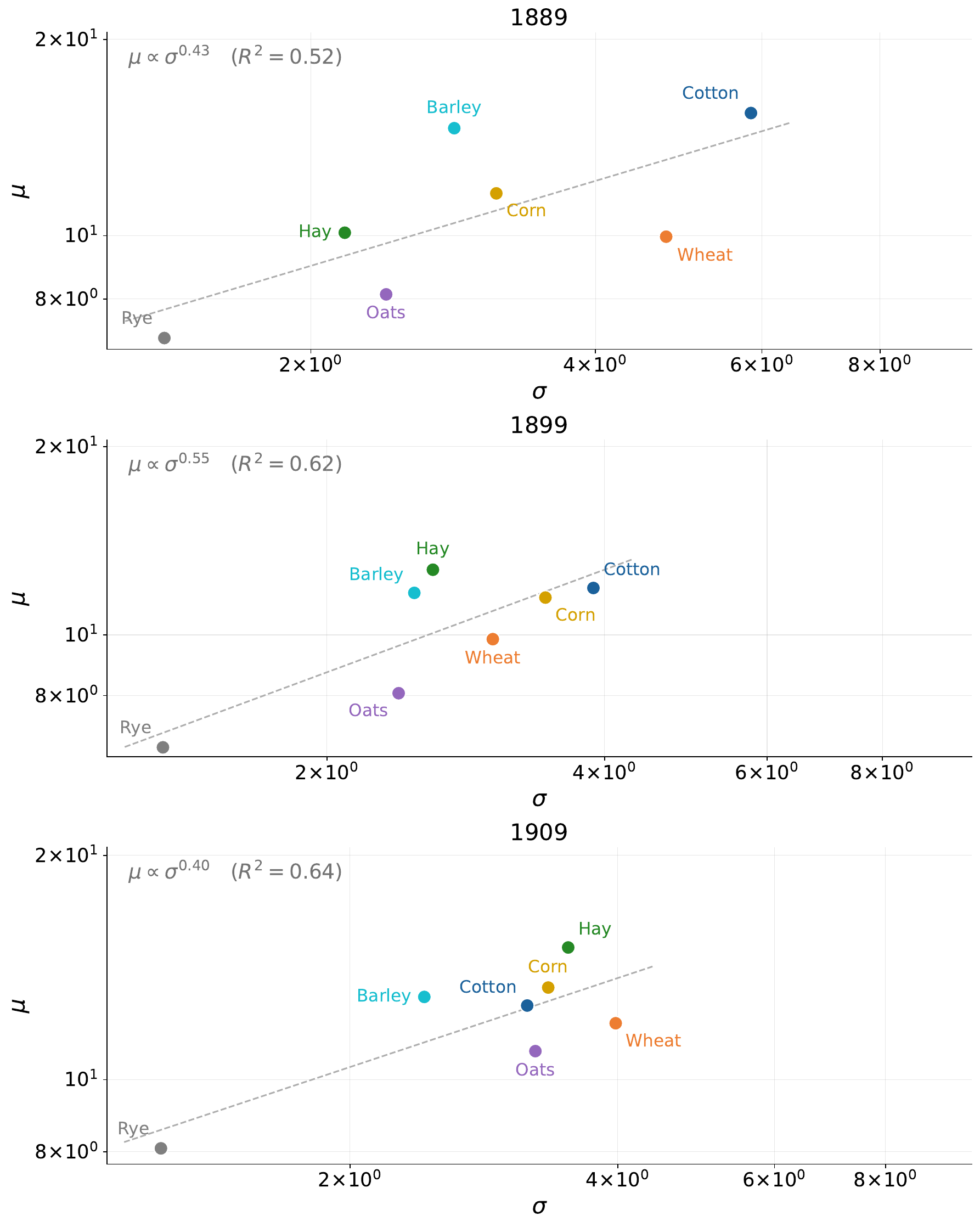}
\end{center}
{\footnotesize\raggedright \textbf{Notes:} These figures show the  mean--variance positions across crops in each census wave. Each dot is one crop at its expected per-acre return $\mu$ against the standard deviation $\sigma$ of that return, both in real 1913 \$/acre. Both axes use a common log--log scale, and
the dashed line is the power-law fit.}
\end{figure}

\begin{figure}[H]
\begin{center}
\caption{Expected crop returns: mean--variance positions by census wave (1919, 1924, 1929)}\label{retmvwavesB}
\includegraphics[width=0.6\textwidth]{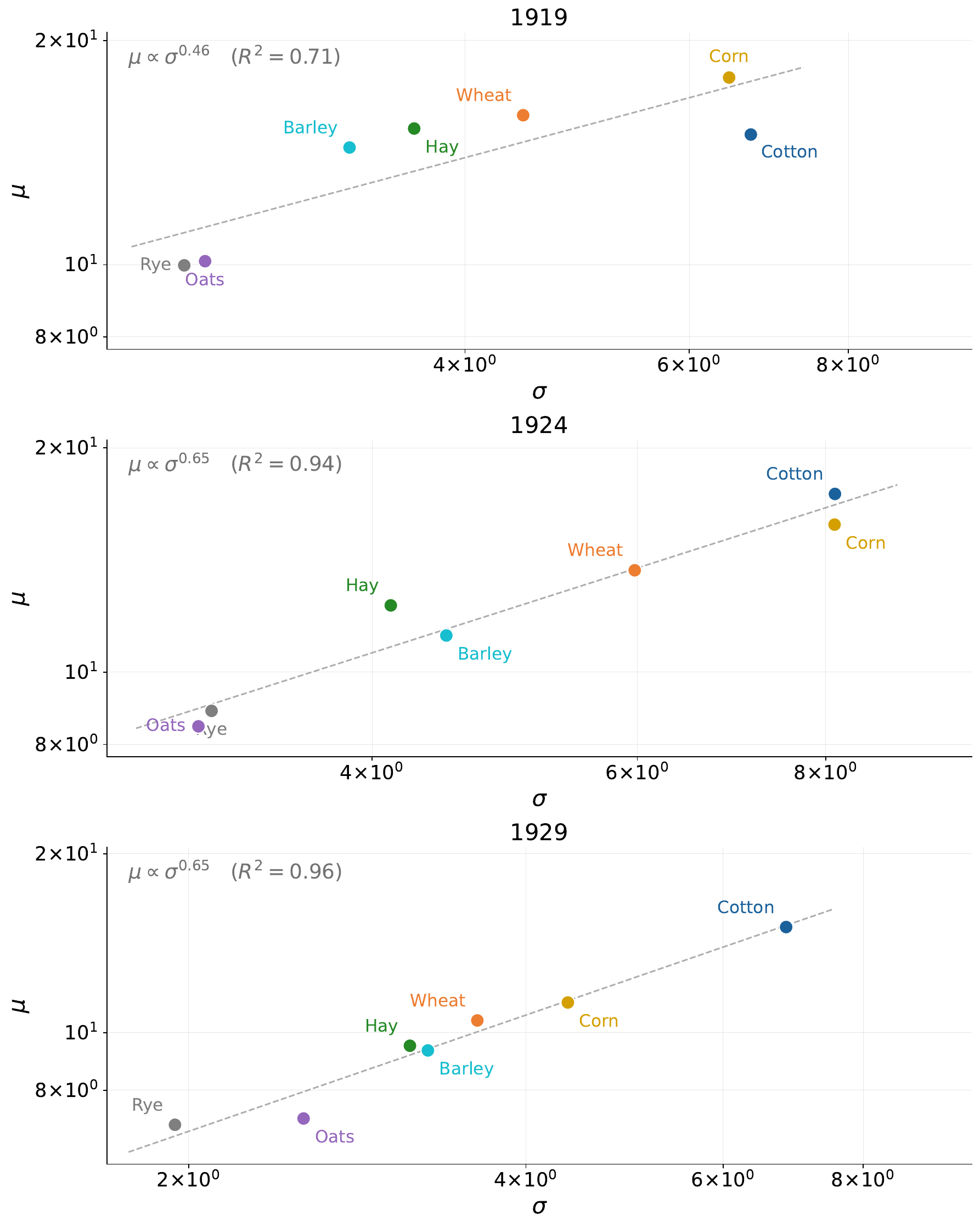}
\end{center}
{\footnotesize\raggedright \textbf{Notes:} These figures show the  mean--variance positions across crops in each census wave. Each dot is one crop at its expected per-acre return $\mu$ against the standard deviation $\sigma$ of that return, both in real 1913 \$/acre. Both axes use a common log--log scale, and
the dashed line is the power-law fit.}
\end{figure}

\subsubsection{Correlation in returns across crops \label{app: corrcrop}}

\begin{figure}[H]
\centering
\caption{Cross-crop correlation of returns \label{fig: returncorr}}
\includegraphics[width=0.72\textwidth]{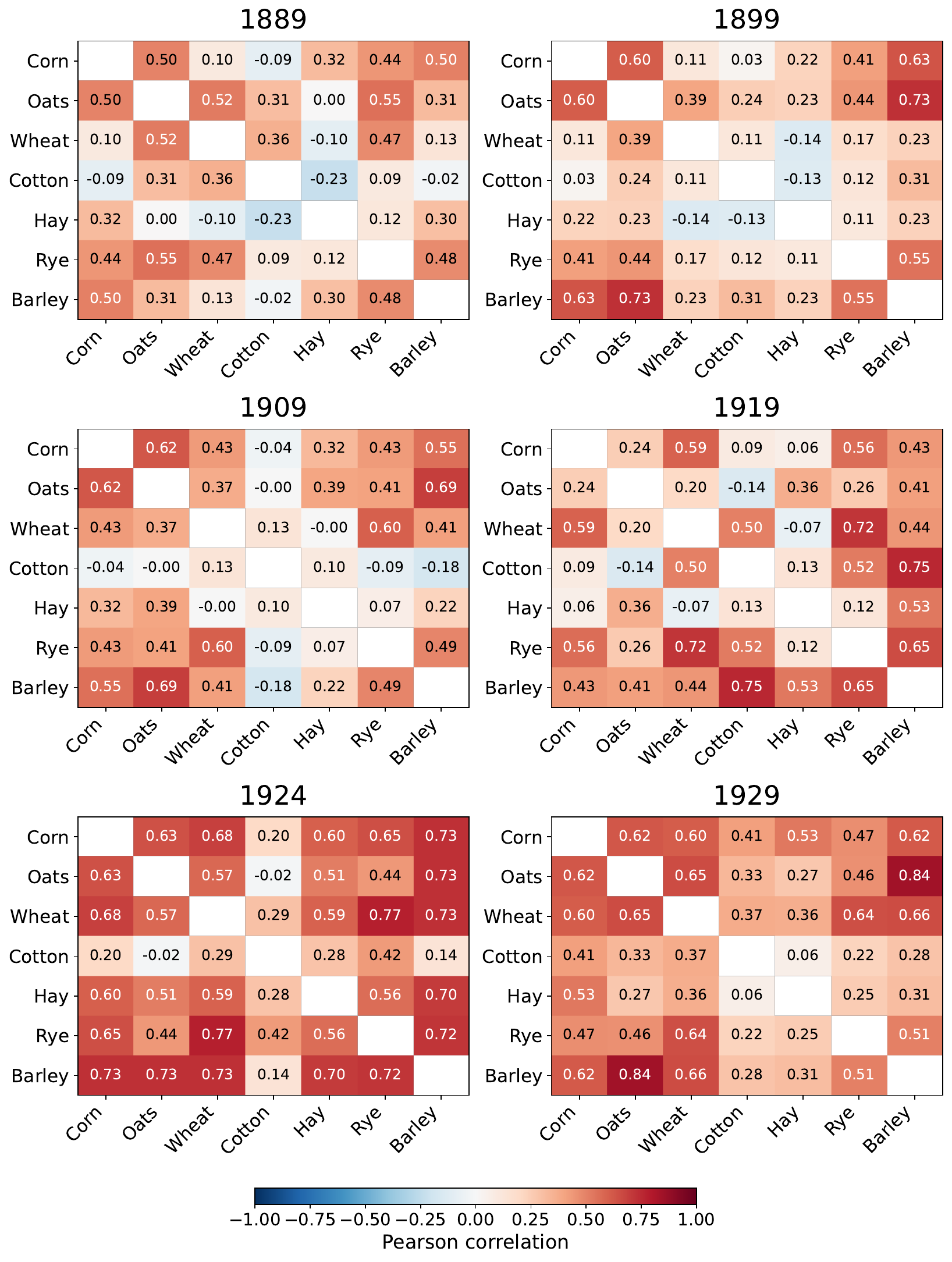}
\begin{minipage}{\textwidth}\smallskip{\footnotesize\raggedright \textbf{Notes:} These figures show the cross-county average correlation between crop returns by census year.}\end{minipage}
\end{figure}

Crop returns are substantially correlated within a county: the average pairwise correlation ranges from $0.24$ to $0.52$ across waves, as shown in \Cref{fig: returncorr}. This reflects the common weather and price shocks that move a county's crops together.

\subsection{Construction of the Sharpe-efficiency ratio \label{app: phiconstruction}}

For each county and census wave we form the ratio $\varphi$ from the county's own ten-year revenue window. Because the maximum-Sharpe portfolio is obtained by optimizing over the covariance matrix, we impose stricter data requirements here than in the estimation of $\widehat{\rho}$: a crop enters the menu only if it is observed in at least eight of the ten window years, only years in which all included crops are observed are used, and we require at least six such years and the presence of the corn numeraire. 

To the resulting covariance matrices we add a Tikhonov adjustment equal to five percent of the average variance along the diagonal, which prevents near-singular matrices from inflating the estimated frontier in the short windows the bootstrap generates. We then draw the returns-to-scale parameters from their estimated sampling distributions, de-scale the window revenues at a common reference acreage, resample the window years with replacement, estimate the maximum-Sharpe portfolio on the in-bag years, and score both it and the realized allocation on the held-out years. We repeat this 35 times per county-wave and report the median $\varphi$. These requirements leave 15,665 county-waves, slightly fewer than the sample for which we estimate $\widehat{\rho}$.

\subsubsection{Sharpe efficiency by census wave \label{app: phiwaves}}

\begin{figure}[H]\centering
\caption{Out-of-bag Sharpe efficiency of county crop allocations, by census wave}\label{fig: phiwaves}
\vspace{.1cm}
\includegraphics[width=\textwidth]{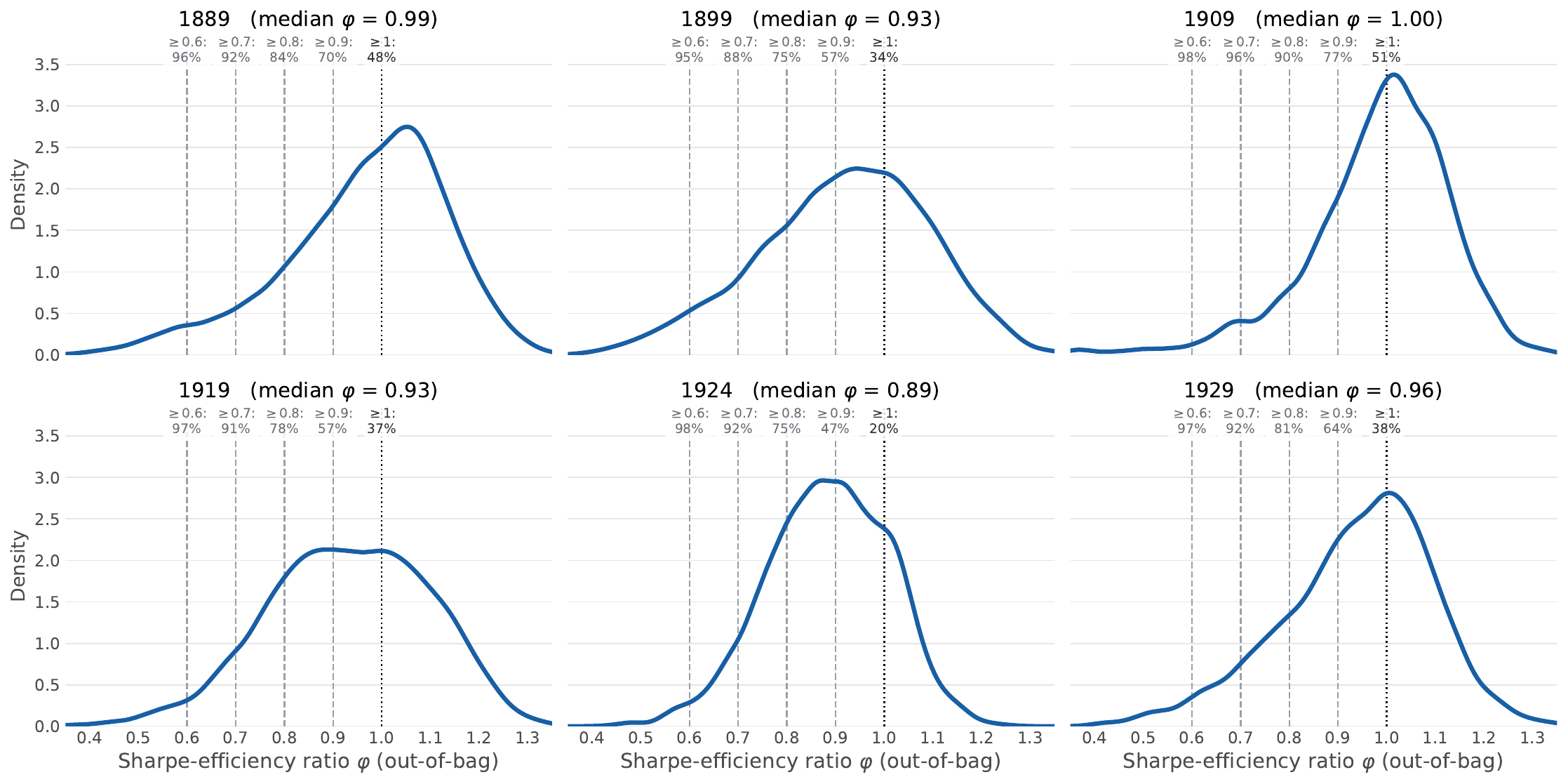}

{\footnotesize\raggedright \textbf{Notes:} Kernel densities of the out-of-bag Sharpe-efficiency ratio $\varphi$ across counties, by census wave (construction as in \Cref{fig: phiefficiency}). The dotted line marks the frontier ($\varphi = 1$). Values are trimmed to $[0.35, 1.35]$ for display.\par}
\end{figure}

\Cref{fig: phiwaves} reports the kernel densities of the out-of-bag Sharpe-efficiency ratio $\varphi$ separately for each census wave, the by-wave companion to the pooled density in \Cref{fig: phiefficiency}. The distributions are stable across the six waves: every wave median lies between $0.89$ and $1.00$, and in each wave the mass concentrates near the frontier.

\subsection{Kansas farm and township risk aversion distributions}

\begin{figure}[H]
\begin{adjustwidth}{-2cm}{-2cm}
\begin{center}
\caption{Distributions of farm- and township-level risk aversion in Kansas \label{fig:histrrakansas}}
\subfloat[Farm-level $\widehat{\rho}$]{\includegraphics[width=0.5\textwidth]{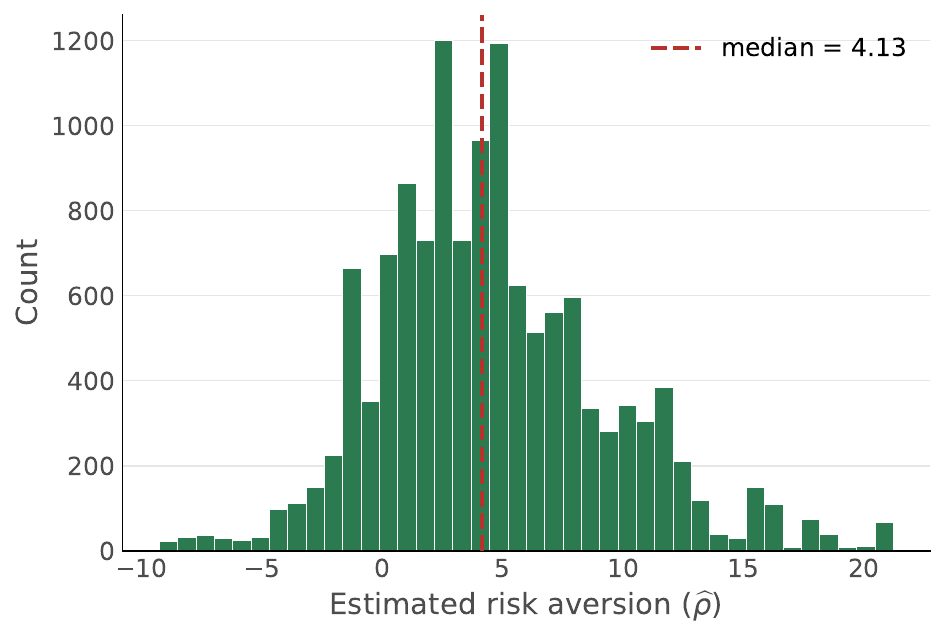}} $\thinspace$
\subfloat[Township-level $\widehat{\rho}$]{\includegraphics[width=0.5\textwidth]{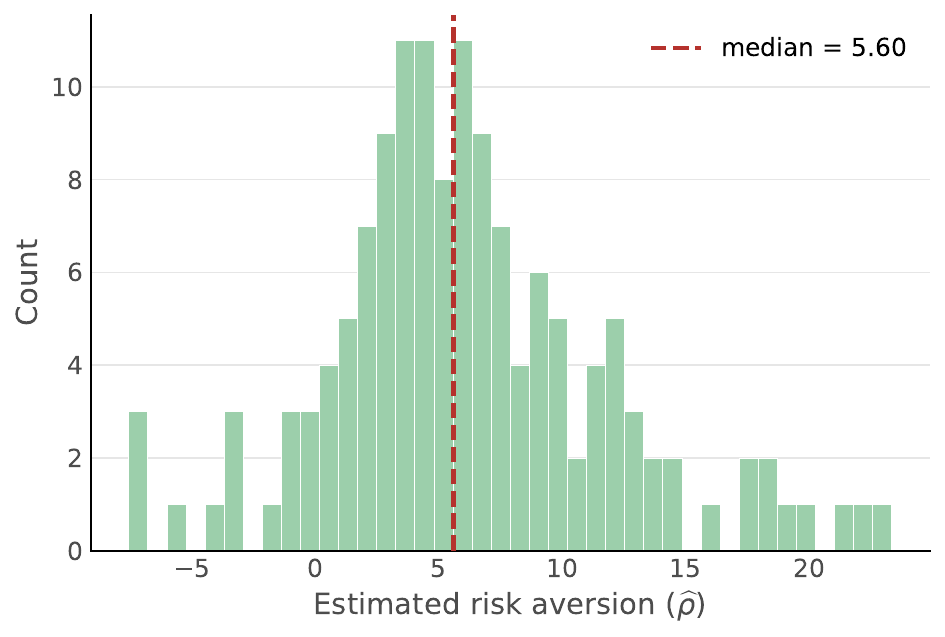}} \\
\end{center}
\end{adjustwidth}
{\footnotesize \textbf{Notes:} These panels plot the distributions of farm- and township-level  $\widehat{\rho}$ across years. The analysis uses the farm-level data of \cite{sylvester2002} for the 24 Kansas townships in 1895, 1905, 1915, 1920, 1925, and 1930.}
\vspace{-0.25cm}
\end{figure}

\begin{figure}[H]
\begin{center}
\caption{Township-level vs.\ farm-level risk aversion in Kansas \label{fig:disttwscatter}}
\includegraphics[width=\textwidth]{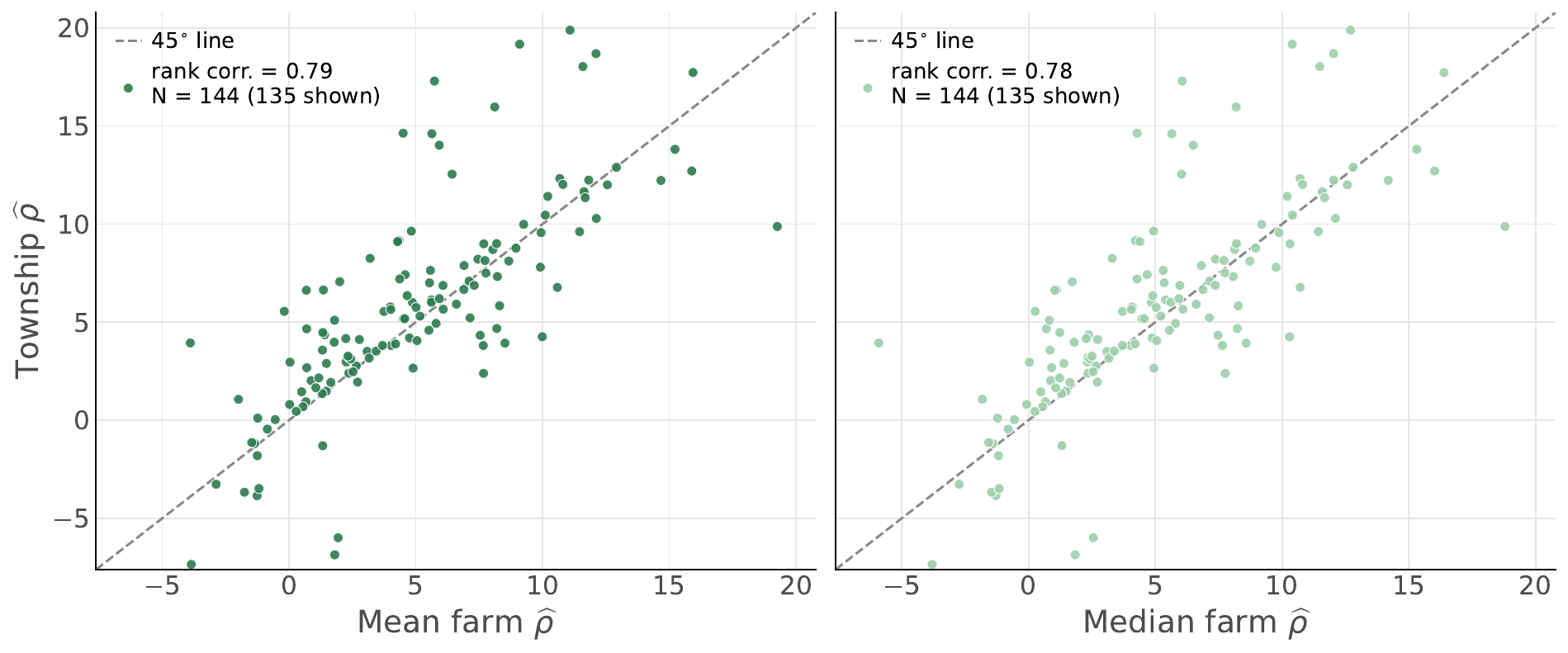}
\end{center}
{\footnotesize \textbf{Notes:} These figures plot the township-level $\widehat{\rho}$ against the mean (left) and median (right) farm-level $\widehat{\rho}$ in the same township-year, with the 45-degree line dashed. The annotated rank correlation is the Spearman correlation across all 144 township-years.}%
\end{figure}

\section{Additional results}
\label{app: results}

\begin{table}[H]
\centering\small
\caption{Summary statistics of the risk-aversion estimates \label{tab:rhosumstats}}
\vspace{.1cm}
\begin{tabular}{l r r r r r r r r r}
\toprule
Wave & $N$ & Mean & SD & p10 & p25 & Median & p75 & p90 & \% $\widehat\rho<0$ \\
\midrule
1889 & 2{,}339 & 4.47 & 6.39 & -0.26 & 1.84 & 3.87 & 6.29 & 9.69 & 11.4 \\
1899 & 2{,}556 & 4.15 & 5.07 & -0.59 & 1.85 & 4.02 & 6.11 & 9.03 & 12.2 \\
1909 & 2{,}692 & 5.11 & 7.07 & -0.88 & 1.01 & 4.37 & 8.29 & 13.51 & 15.6 \\
1919 & 2{,}774 & 3.36 & 2.96 & -0.16 & 1.66 & 3.00 & 4.84 & 7.11 & 10.8 \\
1924 & 2{,}782 & 3.38 & 2.76 & 0.41 & 1.95 & 2.88 & 4.44 & 6.82 & 7.3 \\
1929 & 2{,}752 & 3.38 & 2.67 & 0.89 & 2.03 & 2.79 & 4.17 & 6.24 & 4.7 \\
\midrule
All waves (pooled) & 15{,}895 & 3.95 & 4.81 & -0.05 & 1.81 & 3.26 & 5.52 & 8.83 & 10.3 \\
County average & 2{,}808 & 3.93 & 3.34 & 0.48 & 1.97 & 3.59 & 5.18 & 7.61 & 6.6 \\
\bottomrule
\end{tabular}

\begin{minipage}{0.95\textwidth}
\smallskip\footnotesize{\textbf{Notes:} This table reports the cross-county distribution of the estimated relative risk-aversion coefficient $\widehat{\rho}$, by census wave, pooled across all waves, and as the across-wave county average.}
\end{minipage}
\end{table}

\begin{figure}[H]\begin{center}
\caption{Estimated risk aversion, 1889 -- 1929}\label{fig:rhaoverview}
\vspace{0.15cm}

\subfloat[Mean risk aversion across waves]{\includegraphics[width=0.49\textwidth]{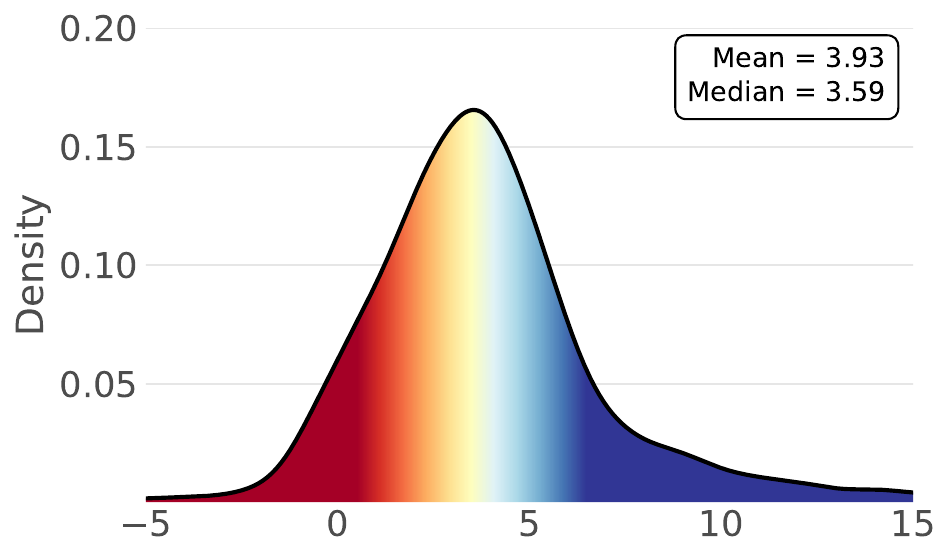}}\hfill
\subfloat[Mean risk aversion over time]{\includegraphics[width=0.49\textwidth]{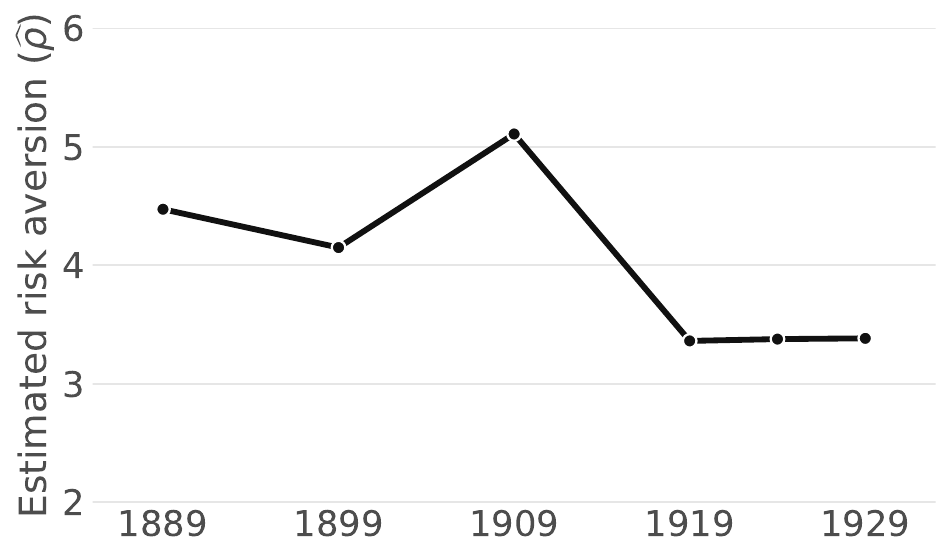}}
\vspace{0.1cm}
\end{center}
{\footnotesize\raggedright \textbf{Notes:} This figure shows the cross-sectional and temporal distribution of the estimated risk aversion coefficient, $\widehat{\rho}$. Panel (a) is the kernel density of each county's mean $\widehat{\rho}$ across the waves in which it is observed. Panel (b) plots the cross-county mean $\widehat{\rho}$ in each wave.\par}
\end{figure}

\begin{figure}[H]\centering
\caption{Distribution of estimated risk aversion by wave, 1889 -- 1929}\label{fig: rhoraw}
\vspace{0.15cm}

\subfloat[1889]{\includegraphics[width=0.45\textwidth]{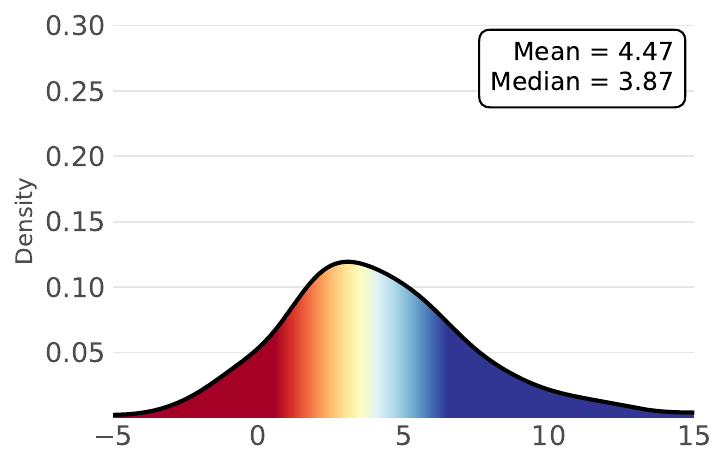}}\hfill
\subfloat[1899]{\includegraphics[width=0.45\textwidth]{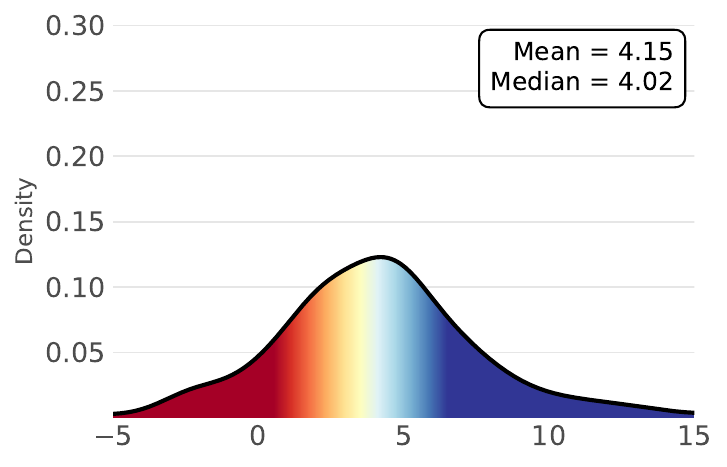}}\\
\subfloat[1909]{\includegraphics[width=0.45\textwidth]{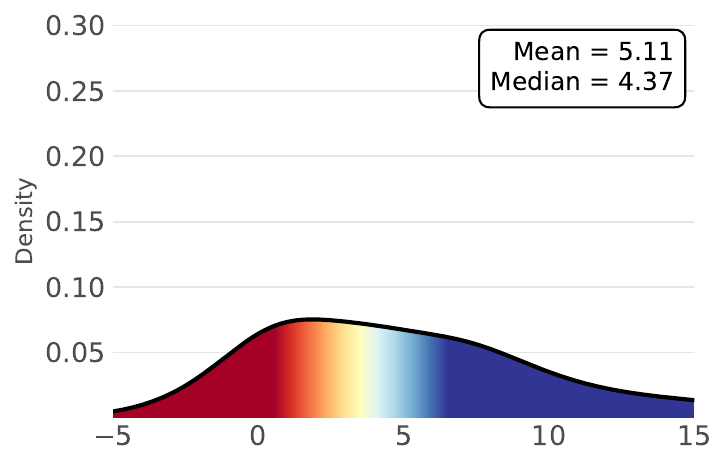}}\hfill
\subfloat[1919]{\includegraphics[width=0.45\textwidth]{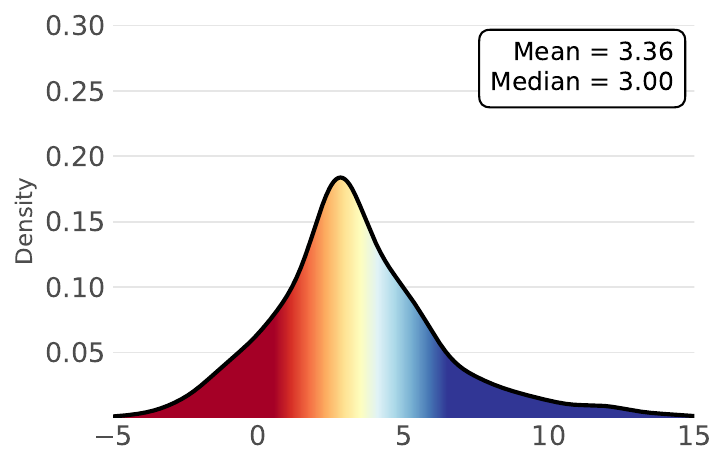}}\\
\subfloat[1924]{\includegraphics[width=0.45\textwidth]{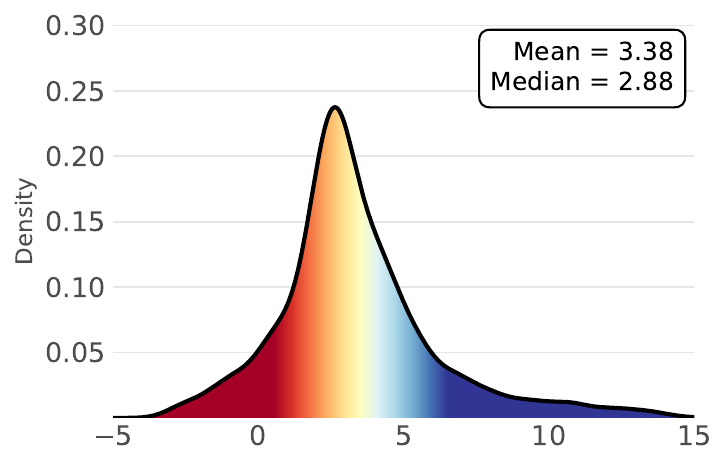}}\hfill
\subfloat[1929]{\includegraphics[width=0.45\textwidth]{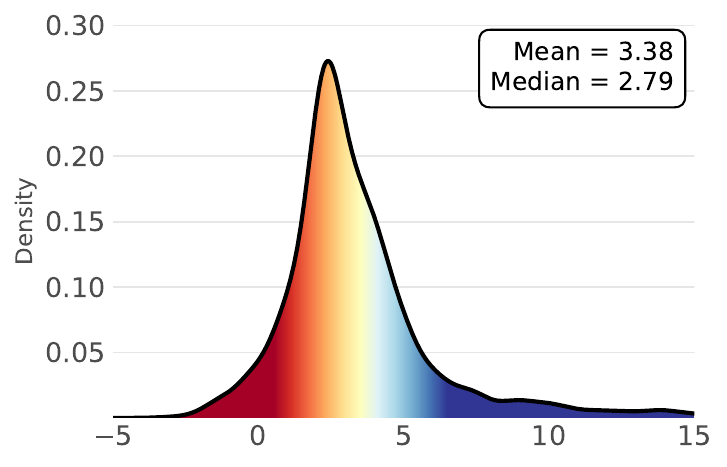}}
\vspace{0.1cm}

{\footnotesize\raggedright \textbf{Notes:} This figure shows the kernel density of the county-level risk-aversion measure
$\widehat{\rho}$ in each census wave.\par}
\end{figure}

\begin{figure}[H]\centering
\caption{Change in estimated risk aversion between waves, 1889 -- 1929}\label{fig:mapchange}
\vspace{0.15cm}

\includegraphics[width=0.49\textwidth]{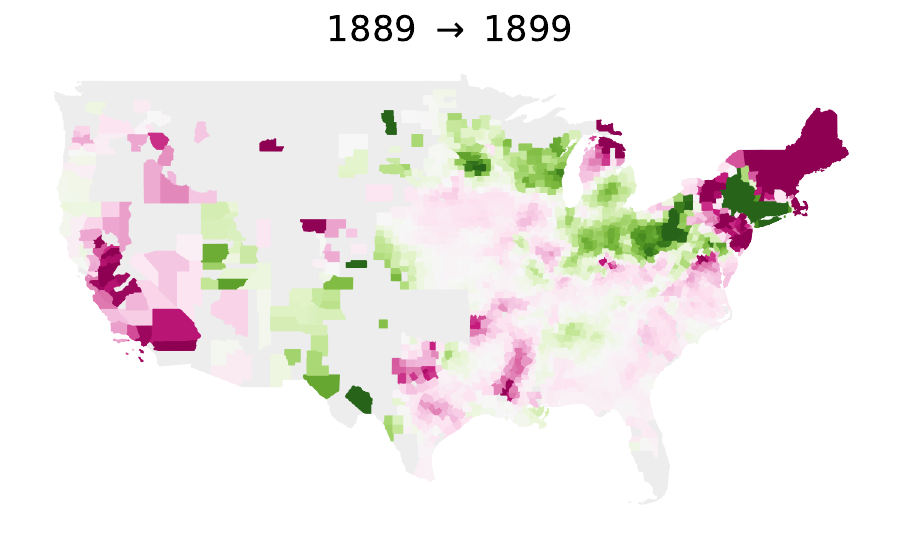}\hfill
\includegraphics[width=0.49\textwidth]{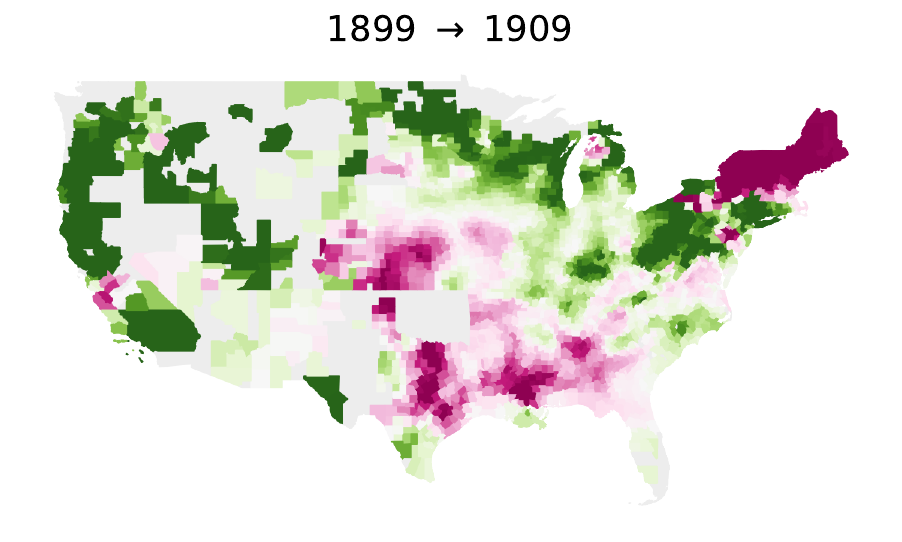}

\includegraphics[width=0.49\textwidth]{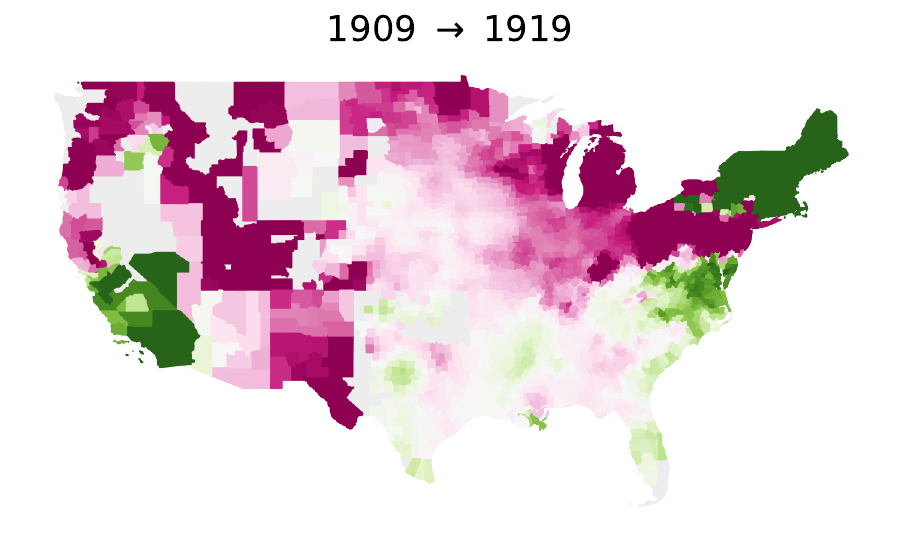}\hfill
\includegraphics[width=0.49\textwidth]{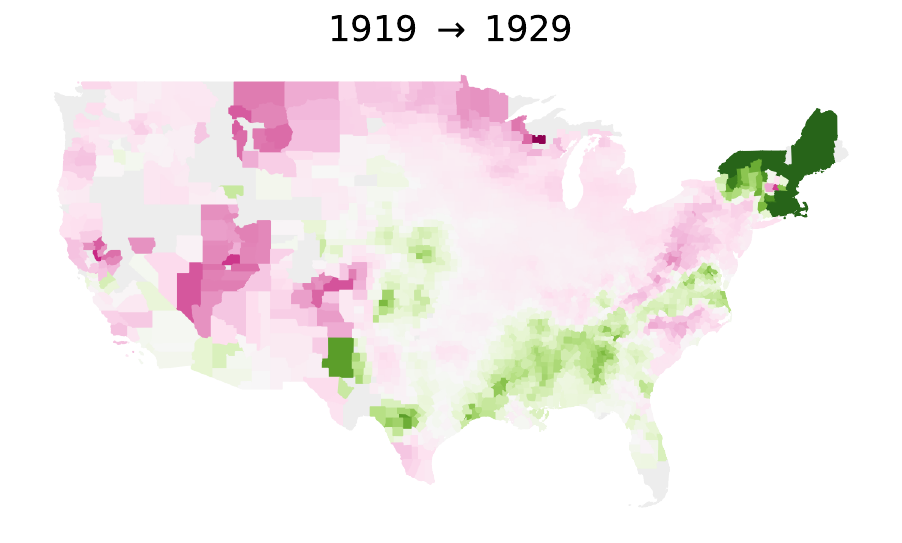}

\vspace{0.2cm}
\includegraphics[width=0.72\textwidth]{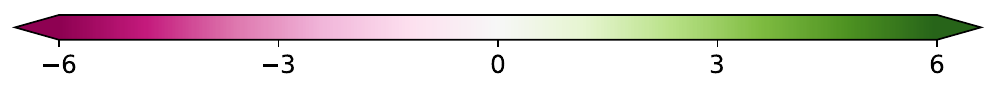}
\vspace{0.1cm}

{\footnotesize\raggedright \textbf{Notes:} This figure shows the between-wave change in the estimated risk aversion coefficient, $\Delta\widehat{\rho} = \widehat{\rho}_{t+1}-\widehat{\rho}_t$, across counties, for each of the four decade-long transitions (the intercensal 1924 wave is skipped). Magenta indicates a decrease in $\widehat{\rho}$ (a county moving toward risk-loving), green an increase (toward risk-averse), and grey no estimate in one or both waves.\par}
\end{figure}

\end{document}